%% file: main.tex
\documentclass[twocolumn]{aastex631}

\pdfoutput=1 

\usepackage{amsmath,amstext}
\usepackage[T1]{fontenc}
\usepackage[figure,figure*]{hypcap}
\usepackage{comment}
\usepackage{graphicx}
\usepackage{natbib}
\usepackage{epsfig}
\usepackage{epstopdf}
\usepackage{enumerate}
\usepackage{longtable} 
\usepackage{multirow, bigdelim}
\usepackage{url}
\usepackage{tabularx}
\usepackage{multirow}
\usepackage{booktabs}
\usepackage{appendix}

\newcommand{\angstrom}{\mbox{\normalfont\AA}}

\received{\today}
\revised{\today}
\accepted{\today}
\submitjournal{ApJ}

\turnoffedit
\turnoffedits

\shorttitle{CCSN light curves and spectra}
\shortauthors{Curtis et al.}

\graphicspath{{./}{figures_revised/}}

\begin{document}

\title{Core-Collapse Supernovae: From Neutrino-Driven 1D Explosions to Light Curves and Spectra}

\correspondingauthor{Sanjana Curtis, Carla Fr\"ohlich}
\email{ssanjan@ncsu.edu, cfrohli@ncsu.edu}

\author[0000-0002-3211-303X]{Sanjana Curtis}
\affiliation{Department of Physics, North Carolina State University, Raleigh NC 27695}
\affiliation{Center for Theoretical Astrophysics, Los Alamos National Lab, Los Alamos NM 87544}
\affiliation{Center for Nonlinear Studies, Los Alamos National Lab, Los Alamos NM 87544}

\author[0000-0003-2540-3845]{Noah Wolfe}
\affiliation{Department of Physics, North Carolina State University, Raleigh NC 27695}

\author[0000-0003-0191-2477]{Carla Fr\"ohlich}
\affiliation{Department of Physics, North Carolina State University, Raleigh NC 27695}

\author[0000-0001-6432-7860]{Jonah M. Miller}
\affiliation{CCS-2, Computational Physics and Methods, Los Alamos National Lab, Los Alamos NM 87544}
\affiliation{Center for Theoretical Astrophysics, Los Alamos National Lab, Los Alamos NM 87544}

\author[0000-0003-3265-4079]{Ryan Wollaeger}
\affiliation{CCS-2, Computational Physics and Methods, Los Alamos National Lab, Los Alamos NM 87544}
\affiliation{Center for Theoretical Astrophysics, Los Alamos National Lab, Los Alamos NM 87544}

\author[0000-0002-0023-0864]{Kevin Ebinger}
\affiliation{GSI Helmholtzzentrum für Schwerionenforschung, D-64291 Darmstadt, Germany}

\begin{abstract}   
We present bolometric and broadband light curves and spectra for a suite of core-collapse supernova models exploded self-consistently in spherical symmetry within the PUSH framework. We analyze broad trends in these light curves and categorize them based on morphology. We find these morphological categories relate simply to the progenitor radius and the mass of the hydrogen envelope. We present a proof-of-concept sensitive-variable analysis, indicating that an important determining factor in the properties of a light curve within a given category is $^{56}$Ni mass. We follow spectra from the photospheric to the nebular phase. These spectra show characteristic iron-line blanketing at short wavelengths and Doppler-shifted Fe $\scriptstyle\mathrm{II}$ and Ti $\scriptstyle\mathrm{II}$ absorption lines. To enable this analysis, we develop a first-of-its-kind pipeline from a massive progenitor model, through a self-consistent explosion in spherical symmetry, to electromagnetic counterparts. This opens the door to more detailed analyses of the collective properties of these observables. We provide a machine-readable database of our light curves and spectra online at \url{go.ncsu.edu/astrodata}.

\end{abstract}

\section{Introduction} 
\label{sec:intro}

Core-collapse supernovae (CCSNe) are the explosive deaths of massive stars (M $\gtrsim$ 8--10 $M_{\odot}$). The CCSN problem is complex and, despite considerable efforts, the explosion engine is not yet fully understood 
\citep[c.f.][and references therein for a discussion of the status of explosion engine simulations]{Mueller.LRCA:2020}. 
However, due to the vastly different timescales of the gravitational collapse of the core and of the electromagnetic display of the SN, we can still investigate the properties of the SN light curves for different progenitor stars. This is particularly timely and useful as ongoing and upcoming transient surveys---for example ASAS-SN \citep{asassn-2017}, The Zwicky Transient Factory \citep[ZTF;][]{ztf:2019}, or the Vera Rubin Observatory \citep{rubin-lsst}---continue to increase the size and diversity of the sample of SNe.

The light curves of CCSNe are quite heterogeneous but typically fall into the Type Ib/c or Type II spectral classes. Type I supernovae are distinguished from Type II supernovae by the absence of hydrogen lines in their spectra. The further distinction between Type Ib and Type Ic supernovae is made based on the presence or absence (respectively) of helium lines in the spectrum.

Canonical Type II supernovae are subdivided into two main groups based on the shape of the light curve: plateau (IIP) and linear (IIL). Another subgroup is Type IIb, similar in shape to a Type IIL but characterized by significant spectral evolution including the appearance of strong He-lines reminiscent of Type Ibs. A small fraction of Type II events are classified as 1987A-like, named after the famous light curve of SN 1987A. Here, the plateau is replaced by a broad peak that reaches a maximum at around 85 days post-explosion, followed by the exponential decay typical of a Type IIP. Yet other sub-classes exist, such as Type IIn, which show narrow H emission lines, and Type II superluminous supernovae, which are hundreds of times more luminous than typical Type II supernovae. The latter of these are classified as Type II if they show hydrogen lines in their spectra, however, the origin and energy source(s) of these events are unclear and under debate \citep{GalYam2019}. 

The evolution of the supernova luminosity contains important information about the explosion and the amount and distribution of the radioactive $^{56}$Ni synthesized during the explosion. Supernova spectra are even more revealing and provide information about the chemical composition of the star, the physical conditions of the ejecta at the photosphere (velocity and continuum temperature), and their evolution with time. The photospheric phase spectra, used to constrain kinematics and composition, are a useful test for explosion models and the radiation hydrodynamics modeling of the ejecta. During the nebular phase, the line strengths and line profiles provide information about the physical conditions, expansion velocity, hydrostatic and explosive nucleosynthesis, ejecta morphology, mixing, and dust formation. 

Together, CCSN light curves and spectra can help us determine progenitor properties for observed supernovae, test stellar evolution and hydrostatic nucleosynthesis, constrain the explosion mechanism and the nuclear equation of state, test explosive nucleosynthesis, and understand the formation of neutron stars and black holes. However, interpreting these observables correctly is a formidable challenge, one that requires detailed and accurate theoretical modeling. The apparent similarities within the different subgroups hide a range of diverse behaviors and trends. For example, all Type IIP light curves feature a plateau but the details, such as the luminosity of the plateau and radioactive tail as well as the plateau duration, vary between the different observed events in this class. See \cite{Zampieri2017}, \cite{Sim2017} and \cite{Jerkstrand2017} for recent reviews of the topic. 

One approach towards this problem is to focus on what is required, in terms of the progenitor and the explosion, to match the observations of a specific CCSN. This was done in a number of studies for SN 1987A \citep{Arnett1988, Shigeyama1988, Woosley1988, Utrobin1993}, SN 1993J \citep{Nomoto1993, Bartunov1994, Shigeyama1994, Young1995, Blinnikov1998, Dessart.Yoon.ea:2018} and SN 1999em \citep{Baklanov2005, Utrobin2007a, Bersten2011, Utrobin.ea:2017}, among many other supernovae \citep[e.g.,][]{Goldberg.Bildsten:2020}. Recently, \cite{Utrobin2019} presented light curves from three-dimensional explosion simulations of a sample of blue supergiant models and compared their predictions to SN 1987A. Nebular spectral models have also been presented, for example, for SN 1987A \citep{Fransson1987,Kozma1998a, Kozma1998b, deKool1998, Jerkstrand2011}, SN 2004et \citep{Jerkstrand2012}, SN 2012aw \citep{Jerkstrand2014}, and SN 2012ec \citep{Jerkstrand2015}. 

A complementary approach is to start with specific models, analytic or numerical, and calculate synthetic observables. Usually, the goal of these studies is to develop a general understanding of the imprint of progenitor characteristics (like mass, radius, radioactive debris, mixing) on the morphology of CCSN light curves. Many investigations of Type II light curves have adopted this approach, including analytical studies \citep{Arnett1980, Chugai1991, Popov1993} as well as detailed numerical works \citep{Litvinova1983, Nadyozhin:2003, Chieffi2003, Young2004, Kasen2009, Dessart2010, Dessart2013, Morozova.IIp:2016, sukhbold16, Kozyreva.IIp:2019}. 
Spectra for the photospheric as well as the nebular phase were presented in \citet{Dessart2011} and \citet{Dessart.ea:2020} for piston-driven explosions of 12--25 $M_{\odot}$ stars.  \cite{Jerkstrand.9msun:2018} performed light curves and nebular spectra calculations from a single 9 $M_{\odot}$ neutrino-driven explosion. 

In this study, we adopt the latter approach but with some crucial improvements. 
We start from self-consistent effective explosion simulations using the PUSH method \citep{push1,push2}, instead of using traditional methods which artificially induce the explosion and its desired strength. These traditional methods for inducing explosions in 1D, like thermal bomb or piston, treat the mass cut, explosion energy, and/or $^{56}$Ni yields effectively as free parameters and therefore have limited predictive power. 
In contrast, the PUSH setup is entirely free of hand-tuning beyond the initial calibration process of \emph{one} model against SN~1987A. The key aspects of the PUSH method (and major differences to methods such as the piston or thermal bomb) are: 
(i) The simulation follows the formation and evolution of the compact central object, thus including a consistent neutrino and antineutrino emission from the forming neutron star. 
(ii) The transport of electron-type neutrinos and antineutrinos is not modified (unlike early attempts at effective models such as \cite{cf06a,fischer2010}), 
thus preserving a consistent $Y_e$-evolution including the changes throughout collapse, bounce, and onset of the explosion. 
And (iii), the mass cut (bifurcation between the proto-neutron star and ejected matter) emerges naturally from the simulations, thus preserving an ejecta mass consistent with the other explosion properties. Thus, with the PUSH method, we can predict $^{56}$Ni yields, explosion energies, and ejecta masses that are entirely self-consistent to each other.

The consistency of the $^{56}$Ni yields is especially important observationally, as the amount and distribution of $^{56}$Ni is a key input for light curve predictions. The radioactive decay of $^{56}$Ni heats the ejecta and affects the morphology of the entire light curve. The late-time radioactive tail, in particular, is entirely powered by the decay of $^{56}$Co, which is the product of $^{56}$Ni decay.

The nucleosynthesis is sensitive to how and where the explosion is launched as well as to the structure and composition of the layers that undergo explosive burning. This is especially relevant for the iron-group (including $^{56}$Ni) elements that originate from the innermost stellar layers. 
These layers are most affected by the multi-dimensional nature of CCSNe, resulting in thermodynamic trajectories that are more complex than in the spherically symmetric case. However, at the present time, only very few nucleosynthesis calculations from multi-dimensional simulations exist \citep{Eichler.Nakamura.ea:2017, Harris.Hix.ea:2017, yoshida2017, wanajo2d}, and they have limitations of their own (e.g., problems with convection due to the assumption of axisymmetry). 
Some uncertainties of postprocessing simulations for abundances are discussed in \citet{Harris.Hix.ea:2017}.

We calculate bolometric and broadband light curves as well as spectra for a suite of 62 progenitor stars spanning a range of masses and three different metallicities, representing normal Type IIP, stripped-envelope like, and SN~1987A-like supernovae. The explosion energy and the mass cut are predictions from \texttt{Agile} simulations within the PUSH framework. The amount of $^{56}$Ni is a predictions from detailed explosive nucleosynthesis calculations. 
The observables we present can be compared to past and future observations of CCSNe and used to assess the effectiveness of our models. To our knowledge, this is the first work to self-consistently predict both light curves and spectra of a broad set of CCSNe starting from effective explosion simulations and using yields from detailed nuclear reaction network calculations. 

To summarize, our approach has several crucial strengths for the present study:
\begin{itemize}
    \item We have predictions for the explosion energy of different progenitor stars from \texttt{Agile} simulations. The explosion energy is expected to have a strong influence on the supernova light curve. 
    \item The mass cut emerges as a prediction from the \texttt{Agile} explosion simulations. This is relevant for estimating the total ejected mass, which is an important quantity for light curve predictions, and for nucleosynthesis in the inner stellar layers.
    \item Finally, the amount of $^{56}$Ni is predicted from detailed explosive nucleosynthesis calculations rather than input by hand as is still done in many studies of CCSN light curves. 
\end{itemize}

The remainder of this paper is organized as follows. In Section~\ref{sec:snec}, we describe the \texttt{SNEC} code and the process of mapping from \texttt{Agile} to \texttt{SNEC}. We present light curves for all our models computed using \texttt{SNEC}, comment on qualitative differences, and discuss the physical behavior that leads to different light curve morphologies. In Section~\ref{sec:supernu}, we describe the \texttt{SuperNu} code, the process of mapping from \texttt{SNEC} to \texttt{SuperNu}, and discuss light curves as well as spectra of selected models. In Section ~\ref{sec:discussion} we discuss implications and limitations of the present study. Section~\ref{sec:summary} summarizes this work and presents our conclusions as well as future directions.

\section{Input Models and Method}

\begin{figure}
\begin{center}
        \includegraphics[width=\columnwidth]{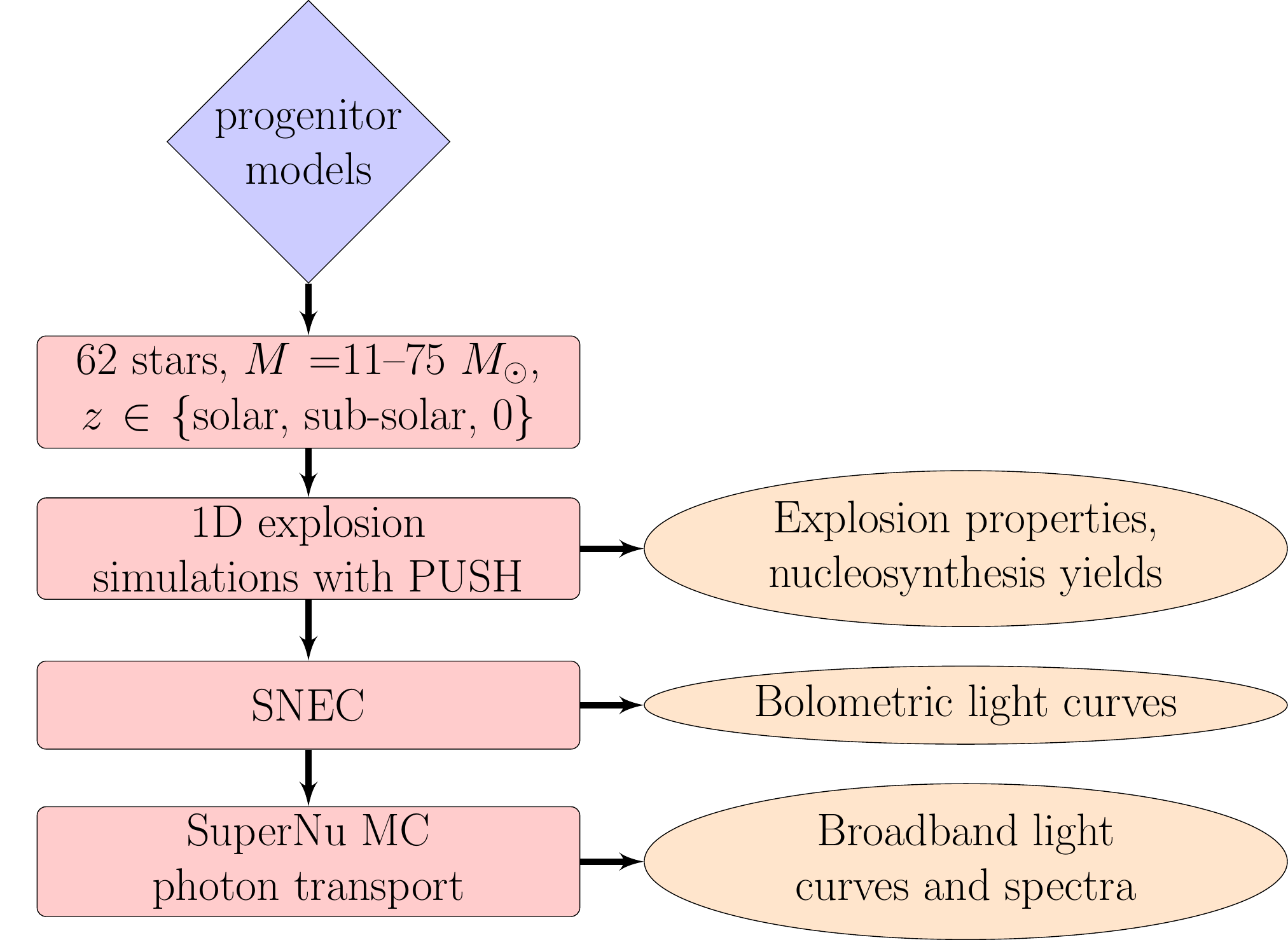}
        \caption{Flowchart showing the process of computing light curves and spectra starting with hydrodynamic explosion simulations. 
        \label{fig:flowchart}
        }
\end{center}
\end{figure}

\begin{table*}
        \begin{center}
                \caption{Progenitor models included in this study
                \label{tab:explmodels}
        }
                \begin{tabular}{lccccc}
                        \tableline \tableline
                        Series & Label & Metallicity  & $M_{\rm{ZAMS}}$\\
                        (-) & (-) & ($Z_{\odot}$) & ($M_{\odot}$)\\
                        \tableline
                        s-series & s & 1  &  11.0 -- 18.0, 18.8, 19.0 -- 22.0, 26.0 -- 38.0, 40.0, 75.0 \\
                        u-series & u & 10$^{-4}$  &  11.0 -- 20.0, 24.0 -- 28.0, 30.0\\

                        z-series & z & 0 & 11.0 -- 23.0, 27.0 -- 31.0\\
                        \tableline
                \end{tabular}
        \end{center}
    \tablecomments{Progenitors from \cite{Woosley.Heger:2002}, exploded using PUSH in \cite{push2} and \cite{push4}. We only include exploding models with integer ZAMS mass.}
\end{table*}

We compute light curves and spectra for 28 solar metallicity, 16 sub-solar metallicity and 18 zero metallicity progenitor models from \cite{Woosley.Heger:2002}, exploded using the PUSH method in \cite{push2} and \cite{push4}. The complete list of models included in this study is given in Table \ref{tab:explmodels}. We will label models by their zero-age main sequence (ZAMS) masses, preceded by the letters ``s'', ``u'' or ``z'' indicating their metallicity (Z) as shown in Table \ref{tab:explmodels}. 

All progenitor models are non-rotating single stars from the stellar evolution code \texttt{KEPLER}. The explosion simulations for the solar metallicity (s-series) progenitors  were performed in \cite{push2}, and those for the sub-solar (u-series) and zero metallicity (z-series) progenitors were performed in \cite{push4}. Both studies employed the PUSH method \citep{push1, push2} to induce explosions in spherical symmetry using parametrized neutrino heating. Here, we will study the integer-mass exploding models only, with the exception of s18.8 ($Z=Z_{\odot}$, $M_{\rm{ZAMS}}$=18.8 $M_{\odot}$), which we include as an interesting case since it reproduces the observed explosion energy and $^{56-58}$Ni yields of SN 1987A. Our models constitute a representative subset of the $\sim$150 explosion simulations presented in the two works mentioned above.

The explosion simulations were done in spherical symmetry (1D) with the general-relativistic hydrodynamics code \texttt{Agile}, which is coupled to neutrino transport. It uses the isotropic diffusion source approximation (IDSA) \citep{Liebendoerfer.IDSA:2009} for electron-flavor neutrino transport and the advanced spectral leakage (ASL) scheme  for heavy-flavor neutrino transport \citep{perego16}. For matter in nuclear statistical equilibrium (NSE), the HS(DD2) \citep{Hempel.SchaffnerBielich:2010} equation of state (EOS) is used, while matter outside of NSE is described by an ideal gas EOS coupled with an approximate alpha-network. 

The \texttt{Agile} simulations follow the collapse, bounce and explosion of \texttt{KEPLER} progenitors and were originally run for a total simulation time of 5~s. For our study, we have extended the end time of these simulations to 15~s or until the shock leaves the computational domain in order to follow the shock propagation (and resulting heating) with \texttt{Agile} for as long as possible before mapping to a code capable of computing light curves. Note that the hydrodynamical simulations do not include the entire progenitor model since the outer layers of the star are not affected by the explosion dynamics until much later. For most models, only matter from the center of the progenitor up to the helium layer is included. The exceptions are a few s-series models with ZAMS masses above 30 $M_{\odot}$ that experience large pre-explosion mass loss.

The exploding models were post-processed with the nuclear reaction network \texttt{CFNET} to predict detailed isotopic nucleosynthesis yields for 2902 isotopes, presented in \cite{push3} for the s-series and in \cite{push4} for the u- and z-series. This is especially important for determining the amount and distribution of the $^{56}$Ni synthesized in the supernova explosion.

To construct input models for our light curve calculations, we extract the relevant hydrodynamic quantities and explosion properties from \texttt{Agile} simulations and the corresponding composition details from the abundances predicted by \texttt{CFNET}. This information constitutes our input profiles for \texttt{SNEC} \citep{Morozova2015}
which is capable of Lagrangian hydrodynamics and equilibrium-diffusion radiation transport. We follow the evolution of the explosion in \texttt{SNEC} (see section \ref{sec:snec}), and then map to \texttt{SuperNu} \citep{Wollaeger2013, Wollaeger2014}, a Monte Carlo (MC) radiative transfer code, once the outflow becomes homologous i.e. $ v \propto r$ (see section \ref{sec:supernu}). 

Mapping the \texttt{Agile} explosion simulations first to \texttt{SNEC} and then to \texttt{SuperNu} is necessary because \texttt{SuperNu} cannot handle any non-trivial hydrodynamics and assumes that the outflow is homologous. This assumption is not satisfied at the end of our explosion simulations. Additionally, the diffusion scheme used by \texttt{SNEC} is a good approximation during the first tens of days post-explosion, when the ejecta are still optically thick. However, as the ejecta become optically thin, this scheme becomes less reliable and a more detailed approach like MC transport is better suited for simulating radiative transfer in this regime. Therefore, we evolve the explosion in \texttt{SNEC} and map to \texttt{SuperNu} at a suitable point in the evolution. This mapping usually happens on the timescale of tens of days after the explosion but differs from model to model, occurring as early as $\sim$ 0.5 days for the more compact (in terms of radius) progenitors. Figure~\ref{fig:flowchart} shows a schematic describing our workflow.

\section{Synthetic Light Curves from SNEC}
\label{sec:snec}
\subsection{The SNEC code}
\texttt{SNEC} solves the equations of Lagrangian hydrodynamics in spherical symmetry and includes radiation transport via flux-limited diffusion \citep{Mihalasbook}. It also follows the basic physics relevant for predicting supernova light curves, such as ionization/recombination of elements and radioactive heating by $^{56}$Ni. Given a suitable input model, \texttt{SNEC} is capable of computing the bolometric light curve as well as light curves in different broad bands under the assumption of blackbody emission. The code is open-source and described in detail in \cite{Morozova2015}.

In our \texttt{SNEC} simulations, we use the analytic EOS of \cite{Paczynski1983}, which contains contributions from radiation, ions and electrons and accounts for electron degeneracy in an approximate way. \texttt{SNEC} supplements the EOS with a routine that solves the Saha equations in the non-degenerate approximation. In principle, we can track the recombination of all the elements included in our input composition. However, for the sake of computational efficiency, we choose to track the recombination of elements from hydrogen up to oxygen only. For most models, with the exception of heavily stripped stars, hydrogen and helium together make the dominant contribution to energy release via recombination.

\texttt{SNEC} allows the user to induce an explosion using either a thermal bomb or a piston. Our models do not require this capability since we can provide the velocity of the outflow from the \texttt{Agile} simulations to \texttt{SNEC}. We therefore disable this feature by running \texttt{SNEC} with a thermal bomb with effectively zero energy input.

One-dimensional explosion models cannot capture any mixing of the chemical composition during shock propagation due to Rayleigh-Taylor and Richtmyer-Meshkov instabilities, found to occur in multi-dimensional simulations \citep{Kifonidis2006, Wongwathanarat2015, Utrobin2019, Stockinger2020}. They retain sharp gradients in the composition profile known to produce artificial features in the predicted light curves, such as an abrupt decrease of bolometric luminosity \citep{Utrobin2007} or bump/spike/knee-like features \citep{Utrobin.ea:2017} at the end of the plateau (in Type IIP), which are not observed in nature. To mimic mixing in spherical symmetry, \texttt{SNEC} applies boxcar smoothing to the input composition profile as done in other works \citep{Kasen2009, Dessart2012, Dessart2013}. We perform boxcar smoothing of the composition profile using the same prescription as \cite{Morozova2015}. Note that the boxcar mixing only changes the distribution of nuclear species in velocity space, but leaves their total amount as predicted from the detailed nucleosynthesis calculations.
See Appendix~\ref{app:mixing} for the input composition for one of our models with and without boxcar smoothing and a comparison of the resulting light curves that illustrates the effect of mixing on our results.

We use non-uniform gridding in mass similar to that used in \cite{Morozova2015}. The proto-neutron star is excised from the grid and we cannot track any fallback of material onto the remnant. The grid resolution is concentrated in the interior, where the explosion originates,
and near the surface, where the photosphere is located at/after shock breakout. We performed extensive convergence tests to determine the minimum number of grid cells required to adequately resolve our input models. See Appendix~\ref{app:snec_convergence} for an example and additional details about the grid resolution used here.

\subsubsection{Opacities}

The opacities used in \texttt{SNEC} are Rosseland mean opacities and the opacity tables included with the code are valid at solar metallicity ($Z=0.02$). The tables are a combination of OPAL Type II opacity tables \citep{Iglesias1996} and those from \cite{Ferguson2005}, both for solar compositions. The OPAL Type II tables are used in the high temperature regime (10$^{3.75}$ K $<  T  <$ 10$^{8.7}$ K) while tables from Ferguson et al. are used at low temperatures (10$^{2.7}$ K $<  T  <$ 10$^{4.5}$ K). In the overlap region, the Ferguson et al. opacities are preferred since they account for contribution from molecular lines, missing in OPAL tables. In regions where opacity values are not available from either table, the opacity (being most sensitive to temperature) is set to the nearest value available at the same temperature. We use these opacity tables in our \texttt{SNEC} simulations of the s-series models.

To simulate the u-series and z-series models with \texttt{SNEC}, we need opacity tables that are valid at low and zero metallicities. Following the same prescription as the one outlined above for the solar-metallicity opacities, we constructed an equivalent set of tables for zero metallicity matter. We use this new set of tables to simulate both the u-series and the z-series models. Since the metallicity of u-series models is extremely low (nearly zero), we assume that the relevant opacities can be approximated by the corresponding values at zero metallicity. To determine the extent to which our approximation could affect the u-series light curves we compute, we simulated a u-series model using the solar-metallicity opacity tables included with \texttt{SNEC} as well as the zero-metallicity opacity tables constructed here. We found negligible differences between the resulting light curves, which suggests that using approximate opacity values introduces very little uncertainty into our results. See Appendix~\ref{app:opacities} for details about the construction of new opacity tables and the sensitivity of u-series light curves to the opacities used. 

As is common practice, we impose an opacity floor to account for additional effects missing from the Rosseland mean opacities. In \texttt{SNEC}, the opacity floor is set to be linearly proportional to the metallicity at each grid point, ranging from 0.01 cm$^2$g$^{-1}$ for $Z=0.02$ to 0.24 cm$^2$g$^{-1}$ for $Z=1$. We leave this prescription unchanged.

\subsubsection{Radioactive Nickel heating}

Since \texttt{SNEC} does not include a nuclear reaction network, it allows the user to specify an amount and distribution of $^{56}$Ni by hand. For our models, we can extract this information from the explosive nucleosynthesis calculations performed in \cite{push3} and \cite{push4}. We supply the $^{56}$Ni mass-fractions in different zones in our input profile to \texttt{SNEC} and set the \texttt{Ni\_by\_hand} flag to zero, so \texttt{SNEC} reads mass-fractions from the profile to determine the total $^{56}$Ni mass for radioactive decay.

Thermalization of gamma-rays emitted in the $^{56}$Ni $\rightarrow$ $^{56}$Co $\rightarrow$ $^{56}$Fe decay contributes radioactive heating that shapes the supernova light curve. \texttt{SNEC} uses the gray transfer approximation of \cite{Swartz1995} to track the gamma-ray transport. The effective gamma-ray opacity is assumed to be absorptive and energy-independent ($\kappa_{\gamma} = 0.06 \ Y_{e} \rm{\ cm}^2 \rm{\ g}^{-1}$). The rate of energy release per gram of $^{56}$Ni is calculated as:

\begin{equation}
    \begin{aligned}
    \epsilon_{rad} = 3.9 \times 10^{10} e^{-t/\tau_{Ni}} + 6.78 \times 10^9\\
    [e^{-t/\tau_{Co}} - e^{-t/\tau_{Ni}}] \mathrm{\ erg \ g}^{-1}\mathrm{\ s}^{-1},
    \end{aligned}
\end{equation}

where $\tau_{Ni}$ and $\tau_{Co}$ are the mean lifetimes of $^{56}$Ni and $^{56}$Co, equal to 8.8 and 113.6 days respectively. \texttt{SNEC} does not account for positron energies in the $^{56}$Co $\rightarrow$ $^{56}$Fe decay, which translates into a small (3-4\%) error in the overall energetics of $^{56}$Ni decay. Finally, we note that though the mass-fractions of $^{56}$Ni, $^{56}$Co and $^{56}$Fe evolve as a result of radioactive decay, \texttt{SNEC} does not track this evolution. This becomes relevant when mapping from \texttt{SNEC} to \texttt{SuperNu} -- we must separately compute the new (decayed) mass-fractions of relevant isotopes at the time of mapping (see section \ref{sec:supernu}).

\subsection{Mapping from Agile to SNEC}
For every model, \texttt{SNEC} requires a structure file as well as a composition file. The structure files follow the \texttt{.short} format and contain hydrodynamic and thermodynamic quantities, such as mass, radius, temperature, density, velocity, electron fraction and angular velocity. The composition files contain mass-fractions of the elements and isotopes that constitute the input model. 

We start by importing the supernova ejecta profile from \texttt{Agile} along with the composition information from \texttt{CFNET}. The mapping from \texttt{Agile} is done shortly before the shock reaches the edge of the simulation domain, typically between $\sim$4--15~s depending on the model and the amount of mass included in the explosion simulation.

\begin{figure*}
\begin{center}
        \begin{tabular}{cc}
        \includegraphics[width=0.48\textwidth]{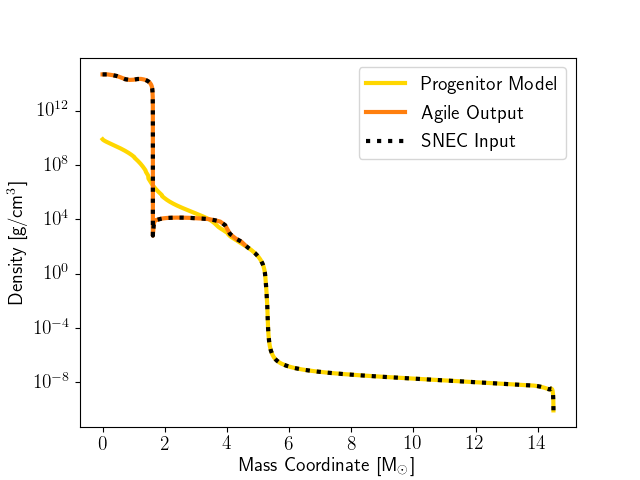}
        \includegraphics[width=0.48\textwidth]{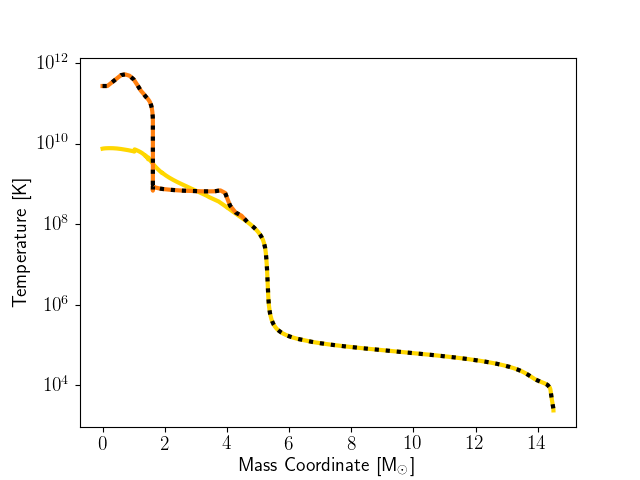}\\
        \includegraphics[width=0.48\textwidth]{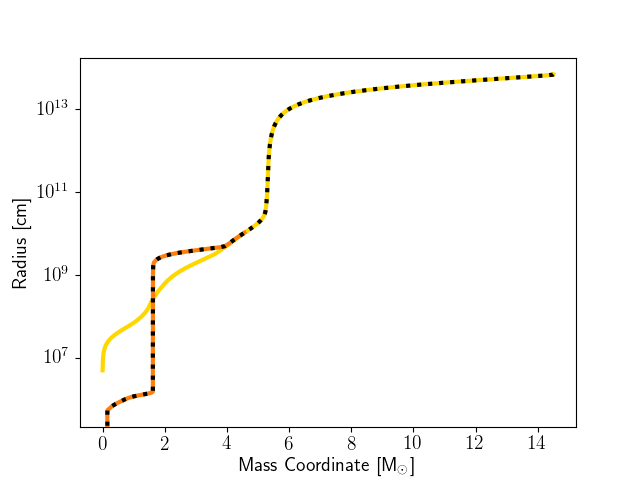}
        \includegraphics[width=0.48\textwidth]{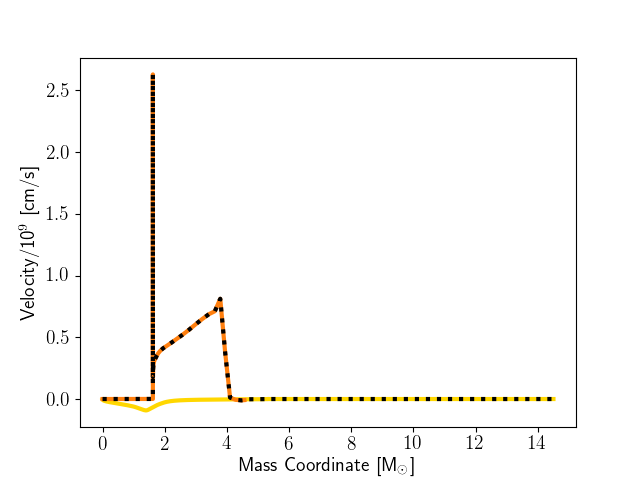}
        \end{tabular}
        \caption{Interpolated profiles of density (top left), temperature (top right), radius (bottom left), and velocity (bottom right) for s18.0 used as input for \texttt{SNEC}.
        \label{fig:interp_hydro}
        }
\end{center}
\end{figure*}

Since the explosion simulations do not include the entire progenitor star (typically including matter up to the helium layer only), we need to append the excised envelope back on to the profile from \texttt{Agile} to obtain the complete hydrodynamic profile of the ejecta. This is done by finding the mass coordinate of the edge of the trimmed progenitor, locating the corresponding zone in the progenitor model, and linearly interpolating desired quantities over a chosen number of mass zones. We do this for all quantities required by the structure files except for angular velocity, which is simply set to zero since our models are non-rotating. See Figure~\ref{fig:interp_hydro} for a plot of some of the hydrodynamical quantities in the input structure file for model s18.0. 

\begin{figure*}
\begin{center}
        \begin{tabular}{cc}
        \includegraphics[width=0.48\textwidth]{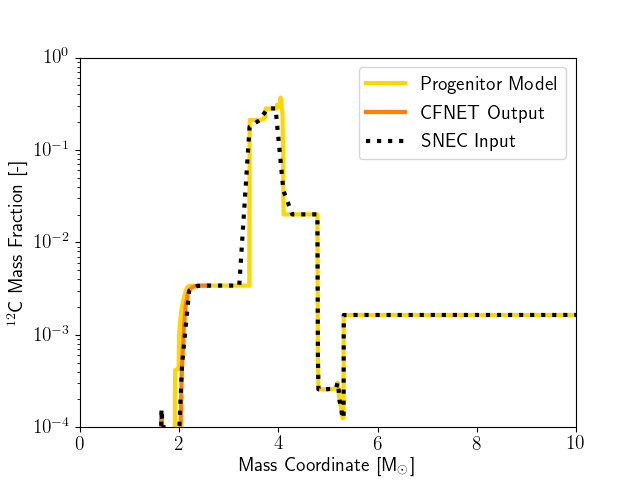}
        \includegraphics[width=0.48\textwidth]{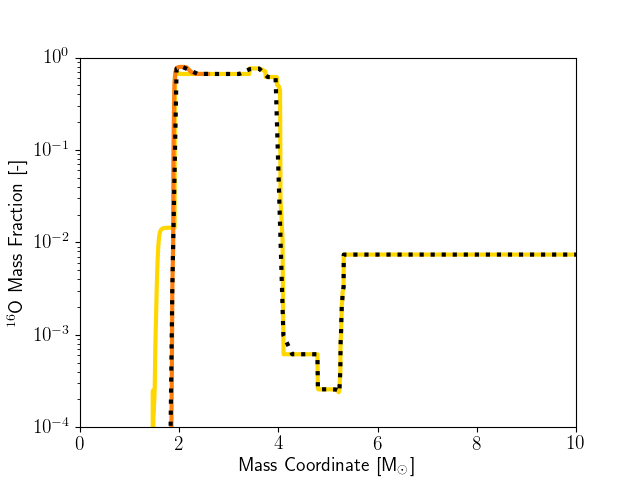}\\
        \includegraphics[width=0.48\textwidth]{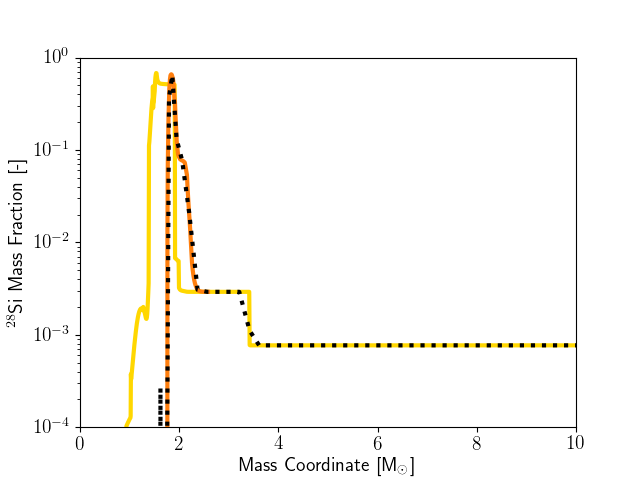}
        \includegraphics[width=0.48\textwidth]{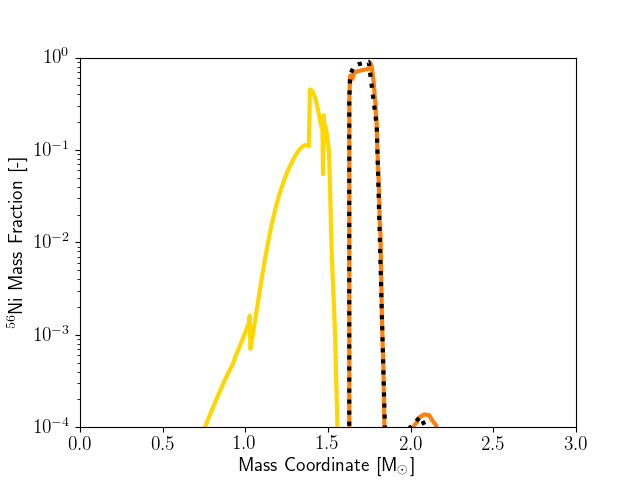}
        \end{tabular}
        \caption{Interpolated mass fractions of $^{12}$C (top left), $^{16}$O (top right), $^{28}$Si (bottom left), and $^{56}$Ni (bottom right) for s18.0 used as input for \texttt{SNEC}.
        \label{fig:interp_mfs}
        }
\end{center}
\end{figure*}

Only a fraction of the mass included in the \texttt{Agile} simulations (matter that attains peak temperatures $\geq 1.75$ GK) is post-processed for nucleosynthesis yields with \texttt{CFNET}. Although yields for 2902 isotopes are available for the post-processed material, for our study, it is sufficient to include the following isotopes in the composition files: $^{1}$H, $^{3}$He, $^{4}$He, $^{12}$C, $^{14}$N14, $^{16}$O, $^{20}$Ne, $^{24}$Mg, $^{28}$Si, $^{32}$S, $^{36}$Ar, $^{40}$Ca, $^{44}$Ti, $^{48}$Cr, $^{52}$Fe, $^{54}$Fe and $^{56}$Ni. Together, these isotopes provide a good description of the composition of our input models. Wherever available, we use the abundances predicted by \texttt{CFNET} rather than those from the approximate alpha-network within \texttt{Agile}. Using predictions from the nuclear reaction network lends greater confidence to our input values of $^{56}$Ni, a key quantity for the supernova light curve. Employing the same procedure as described above for the hydrodynamical quantities, we append mass-fractions from the progenitor model for any unprocessed material to get the full composition of our ejecta. However, we have to interpolate network abundances over both time and space to produce mass-fractions at the same point in time and on the same mass grid as the corresponding structure files. The mass-fractions are normalized to add up to unity in each zone.  The composition profile of model s18.0 supplied as input to \texttt{SNEC} is shown in Figure~\ref{fig:interp_mfs}.

We generate light curves for the first 300  days after explosion. Shock breakout is not well-resolved in \texttt{SNEC} since the photosphere is located in the outermost grid cell, and the light curves during this phase are unreliable. Once the photosphere begins to move inward into the expanding ejecta, the \texttt{SNEC} light curves become robust. \texttt{SNEC} also computes band light curves by assuming emission from the photosphere and using bolometric corrections from \cite{Ofek2014}. However, the $U$- and $B$-band light curves are  strongly influenced by iron-group line blanketing at around tens of days and \texttt{SNEC} cannot accurately reproduce them. The $IR$-band and $V$-band light curves, which can be adequately described by a blackbody spectrum, are captured more accurately. At late times, an increasingly large fraction of the ejecta become optically thin and the luminosity has a large contribution from the radioactive decay of $^{56}$Ni/$^{56}$Co.
\texttt{SNEC} terminates the band light curves when this contribution amounts to more than 5\% of the total luminosity.

\subsection{Results: SNEC Light Curves}

\begin{figure}
\includegraphics[width=0.48\textwidth]{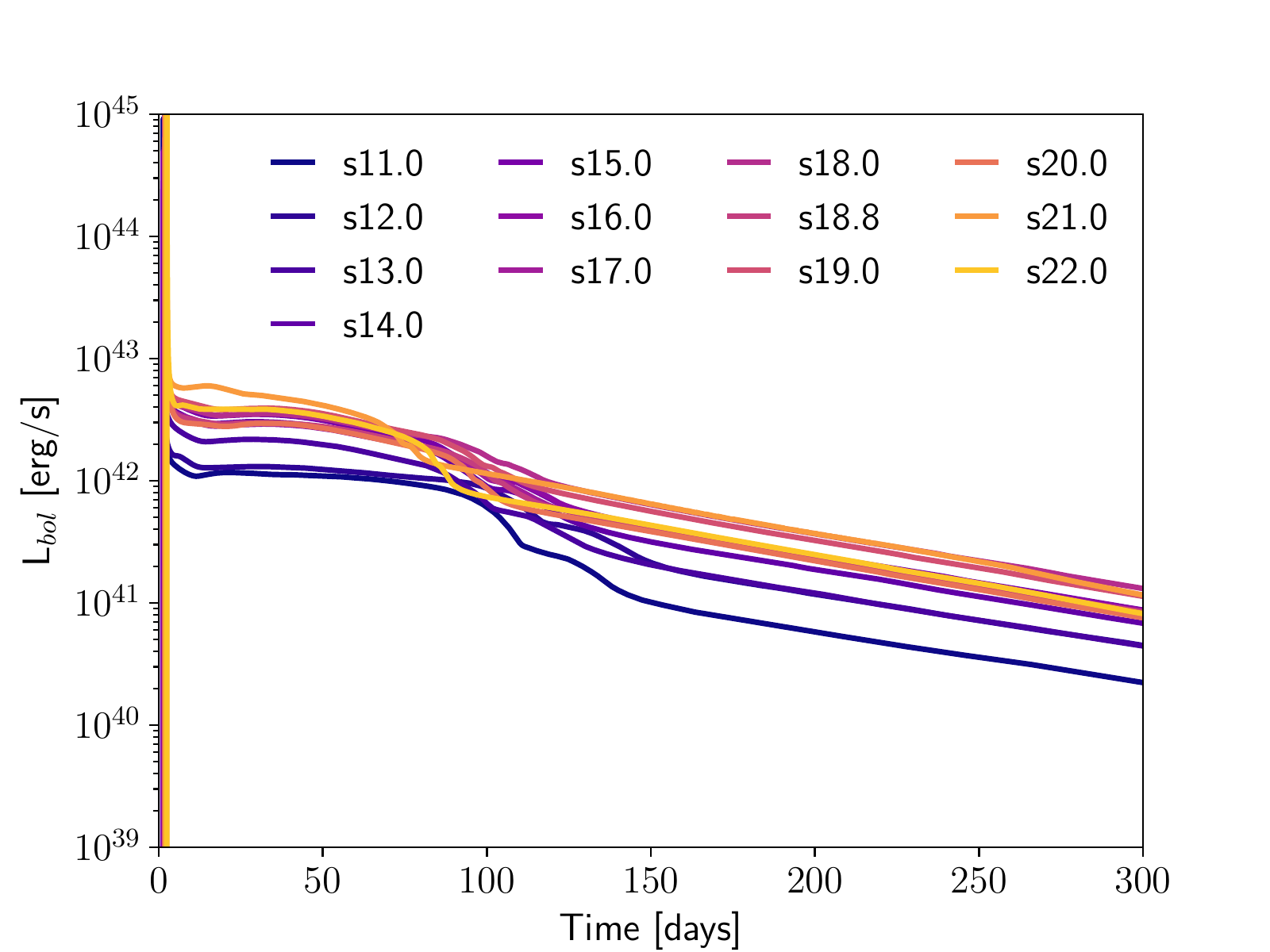}
\includegraphics[width=0.48\textwidth]{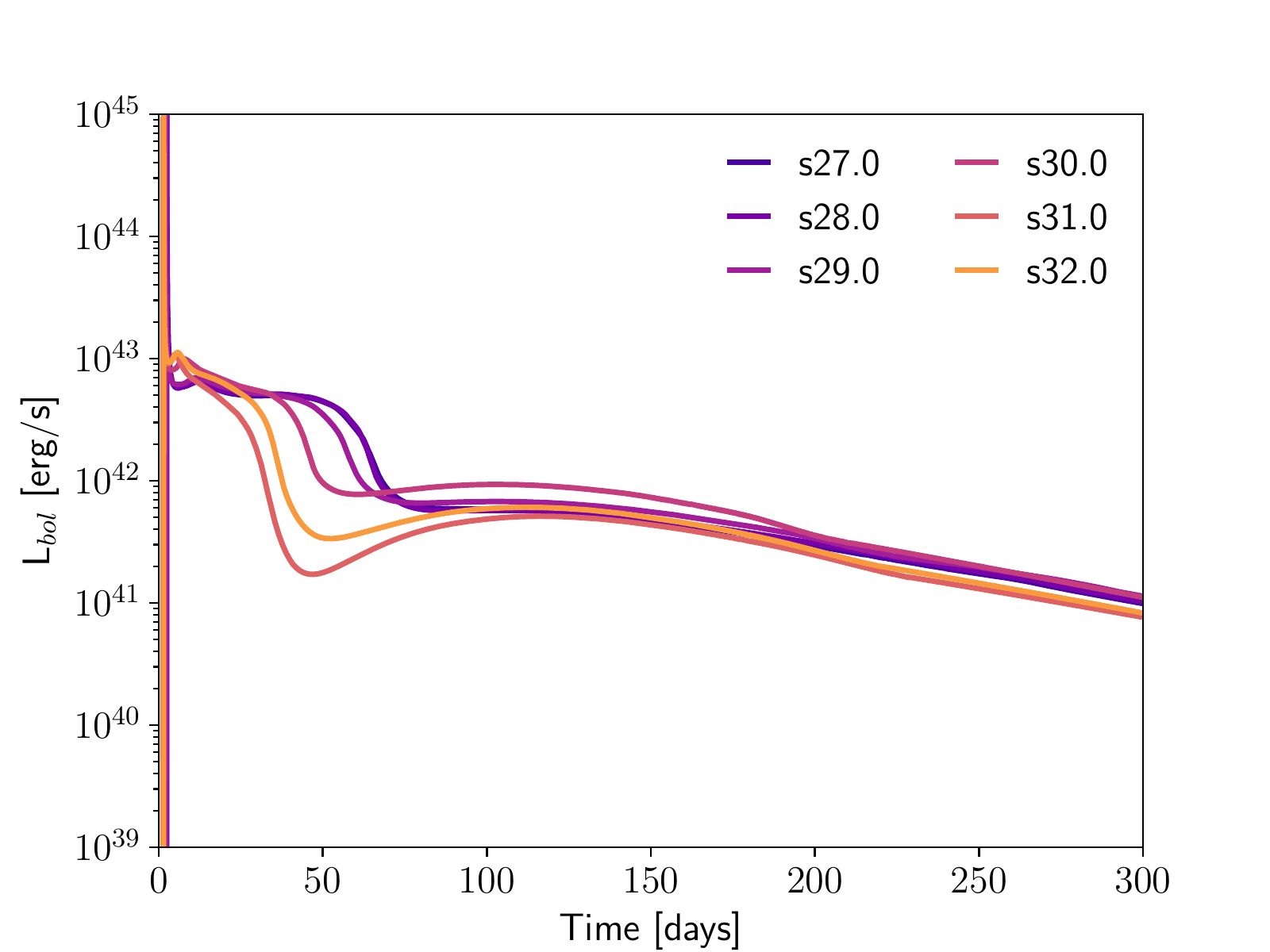}
\includegraphics[width=0.48\textwidth]{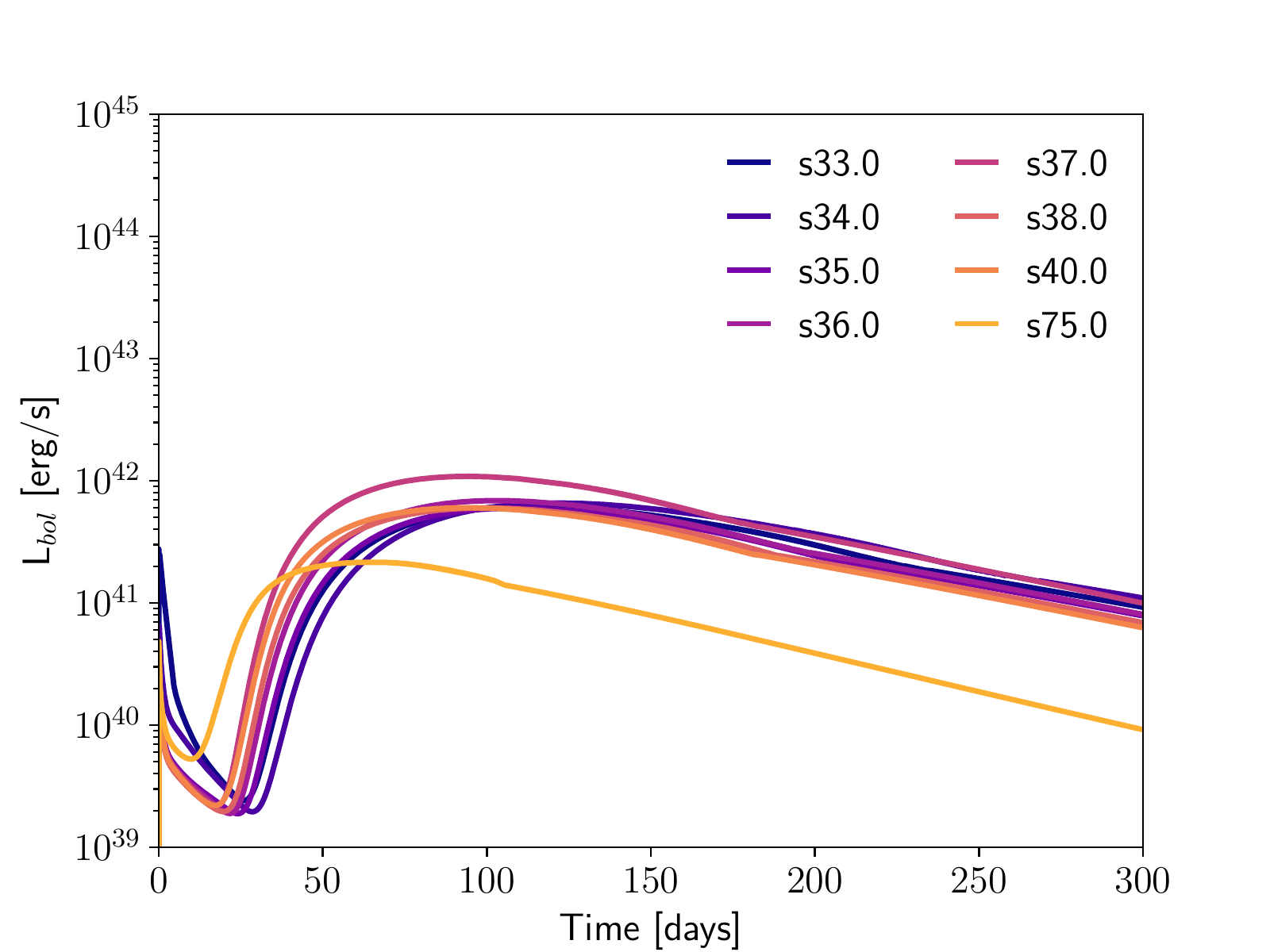}
\caption{SNEC bolometric light curves for s-series models.
\label{fig:s02_lcs}
        }
\end{figure}

The \texttt{SNEC} bolometric light curves for the s-series models are shown in Figure~\ref{fig:s02_lcs}. We observe a systematic change in the morphology of these light curves as we go from low to high ZAMS masses. Models s11.0--s22.0 produce light curves that resemble a normal Type IIP, featuring an extended plateau that drops off at $\sim$100 days to an exponentially-declining radioactive tail. For models s26.0 through s32.0, we find the plateaus getting shorter, the decrease in luminosity at the end of the plateau becoming more and more pronounced, and the radioactive tail giving way to what looks like a very broad, late-time luminosity peak. Finally, models s33.0--s75.0 have no distinguishable plateaus at all. Instead, the bolometric luminosity falls quickly (by a few orders of magnitude) and rises to peak quickly, in a manner qualitatively similar to stripped-envelope supernovae. The main feature of these light curves is a clear late-time luminosity peak.

We note that model s18.8, which lies in the 18--21 $M_{\odot}$ range estimated for the progenitor of SN~1987A, and found to match the observed explosion energy and $^{56-58}$Ni yields of this SN in \cite{push2}, does not produce a 1987-like light curve. This is not unexpected since this model is a red supergiant while the progenitor of SN 1987A is known to be a blue supergiant \citep{Blanco.ea:1987,Walborn.ea:1987}.

\begin{figure}
        \includegraphics[width=0.48\textwidth]{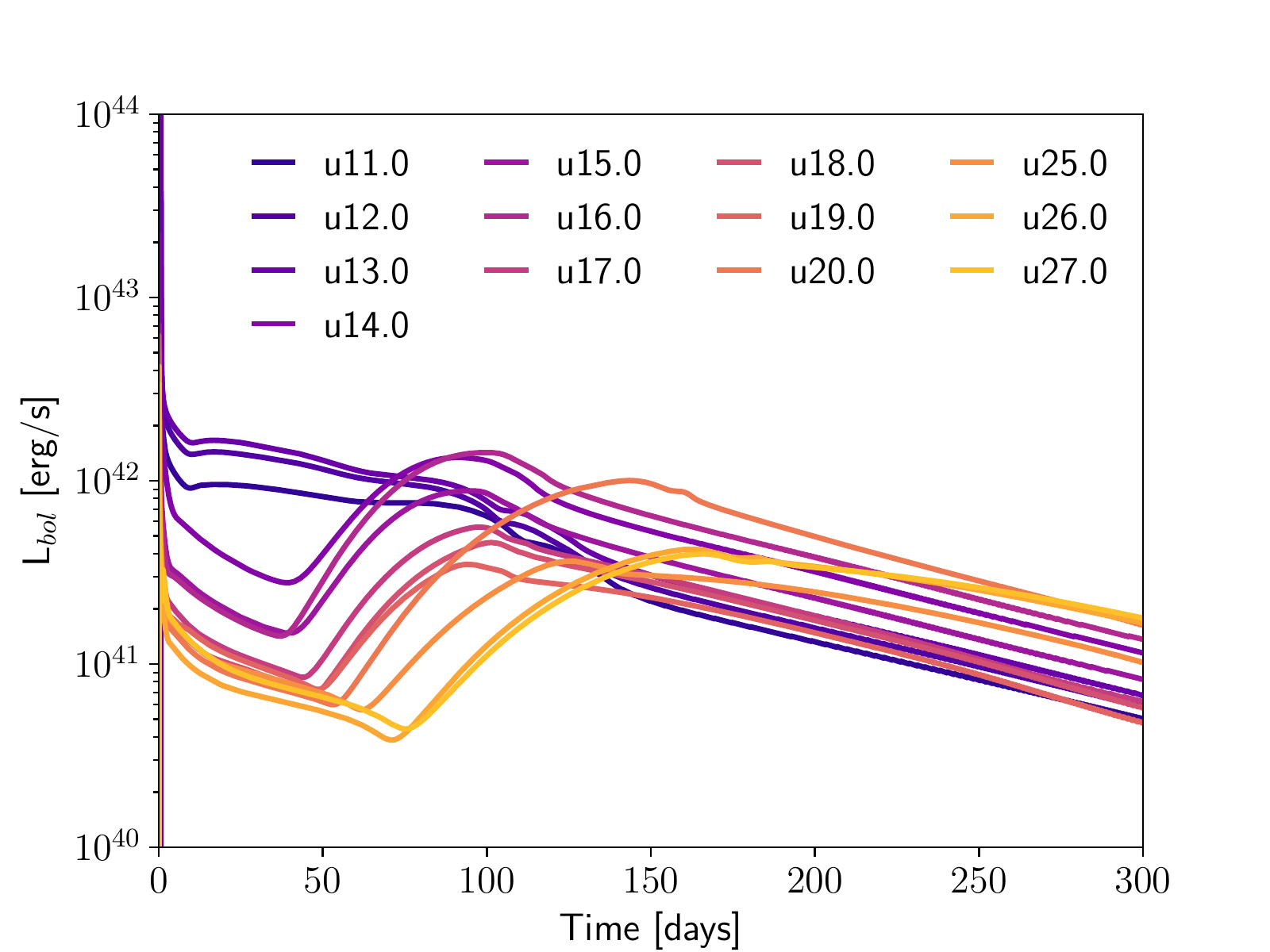}
        \includegraphics[width=0.48\textwidth]{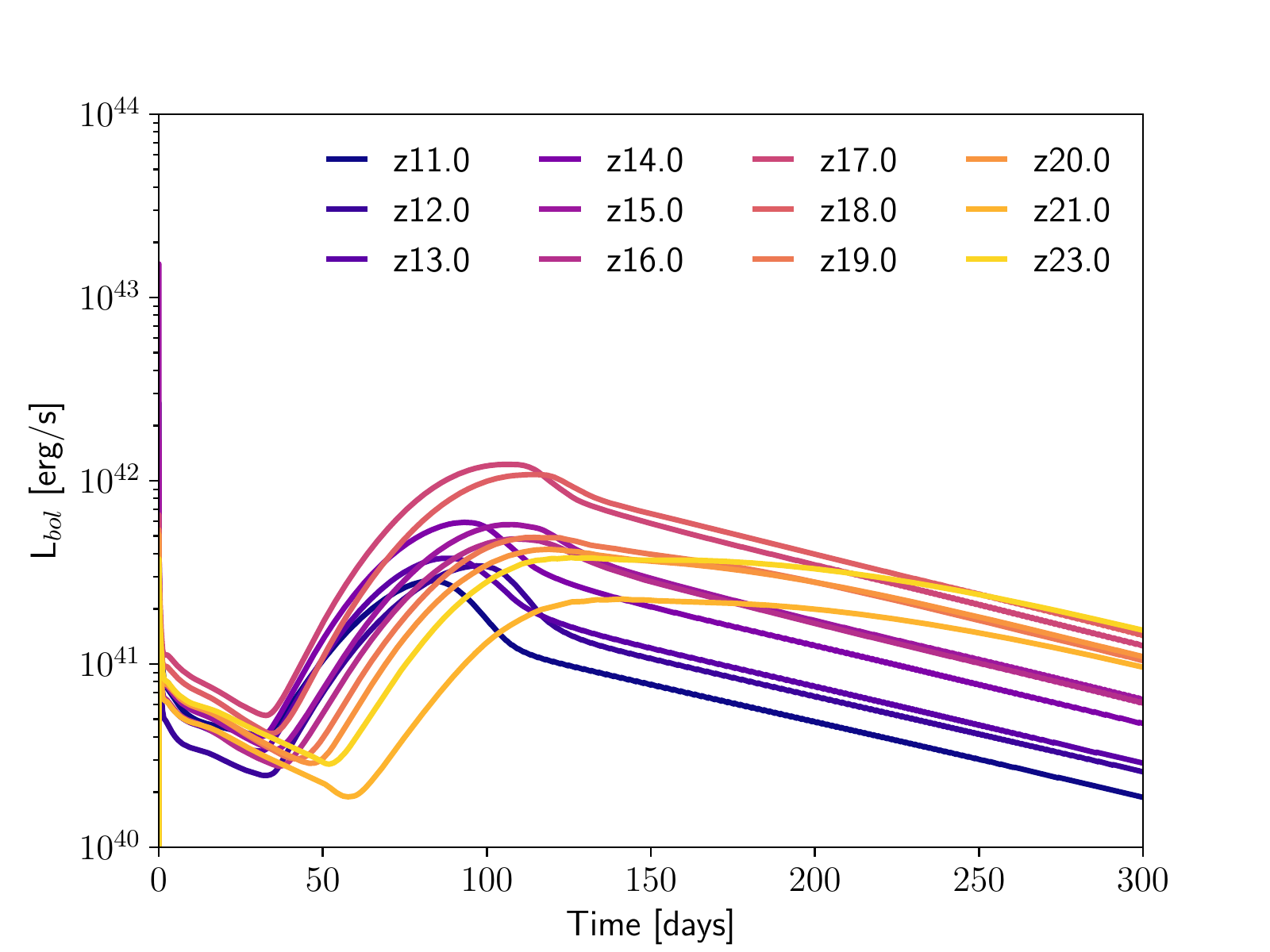}
        \caption{SNEC bolometric light curves for the u-series and z-series models.
        \label{fig:u02_z02_lcs}
        }
\end{figure}

The \texttt{SNEC} light curves for the u-series and z-series models are shown in the top panel and bottom panel of Figure~\ref{fig:u02_z02_lcs} respectively. Here, only the lowest mass models from the u-series, u11.0--u13.0, produce Type IIP light curves with a clear plateau. While models u14.0--u27.0 and z11.0--z21.0 do show a constant or slowly declining bolometric luminosity until $\sim$50 days, which could be interpreted as an under-luminous plateau, the luminosity rises rather than falls at the end of this phase. It continues rising until it attains a peak value roughly equal to the typical plateau luminosity of normal Type IIP supernovae (a few times 10$^{42}$erg/s). Soon after, the light curve drops by a small amount and settles on the radioactive tail, in a manner reminiscent of the plateau-to-tail transition seen in Type IIP. 

For 10 of the 62 models in our study, the light curves we compute with \texttt{SNEC} do not resemble typical CCSN light curves, nor those presented above for similar models. We discuss these peculiar light curves separately in Appendix~\ref{app:weird} and exclude them from our analysis of broader trends in the next subsection.

The bolometric luminosities computed with \texttt{SNEC} are available as machine-readable tables included in our online database\footnote{\url{go.ncsu.edu/astrodata}}. The corresponding broadband light curves from \texttt{SNEC} (produced under the assumption of blackbody emission) are also available online but we will not examine them in any detail here. We expect the broadband light curves predicted by \texttt{SuperNu} to be more accurate and hence postpone that discussion to Section~\ref{sec:supernu}. Additionally, we provide upon request machine-readable tables containing the evolution of the photospheric radius, photospheric velocity and effective temperature for all our models.

\subsection{Discussion of Light Curve Morphologies}
\label{subsec:morphologies}

Having described the \texttt{SNEC} light curves for different progenitor sets separately, we now discuss the three broad classes of qualitative behavior we observe across our entire sample of light curves: 
\begin{enumerate}
    \item Normal Type IIP: seen for models with a large radius and a massive hydrogen envelope. Most of the low mass s-series models and the lowest mass u-series models fall into this category. The light curve shows an extended plateau powered by hydrogen recombination that drops slightly to a radioactive tail.
    \item Stripped-envelope like: seen for models with a small radius and almost no hydrogen and/or helium envelope. The highest mass s-series models fall into this category. Here the bolometric luminosity falls quickly and rises quickly, mostly powered by heating due to $^{56}$Ni decay.
    \item SN 1987A-like: seen for models with a small radius but a massive hydrogen envelope. Most of the u-series and z-series models fall into this class. The bolometric luminosity is low and slowly declining for a few tens of days, eventually rising to typical plateau luminosities, followed by an end-of-plateau style drop and transition to a radioactive tail. These light curves are powered by a combination of hydrogen recombination and $^{56}$Ni decay.
\end{enumerate}

\begin{figure}
        \includegraphics[width=0.48\textwidth]{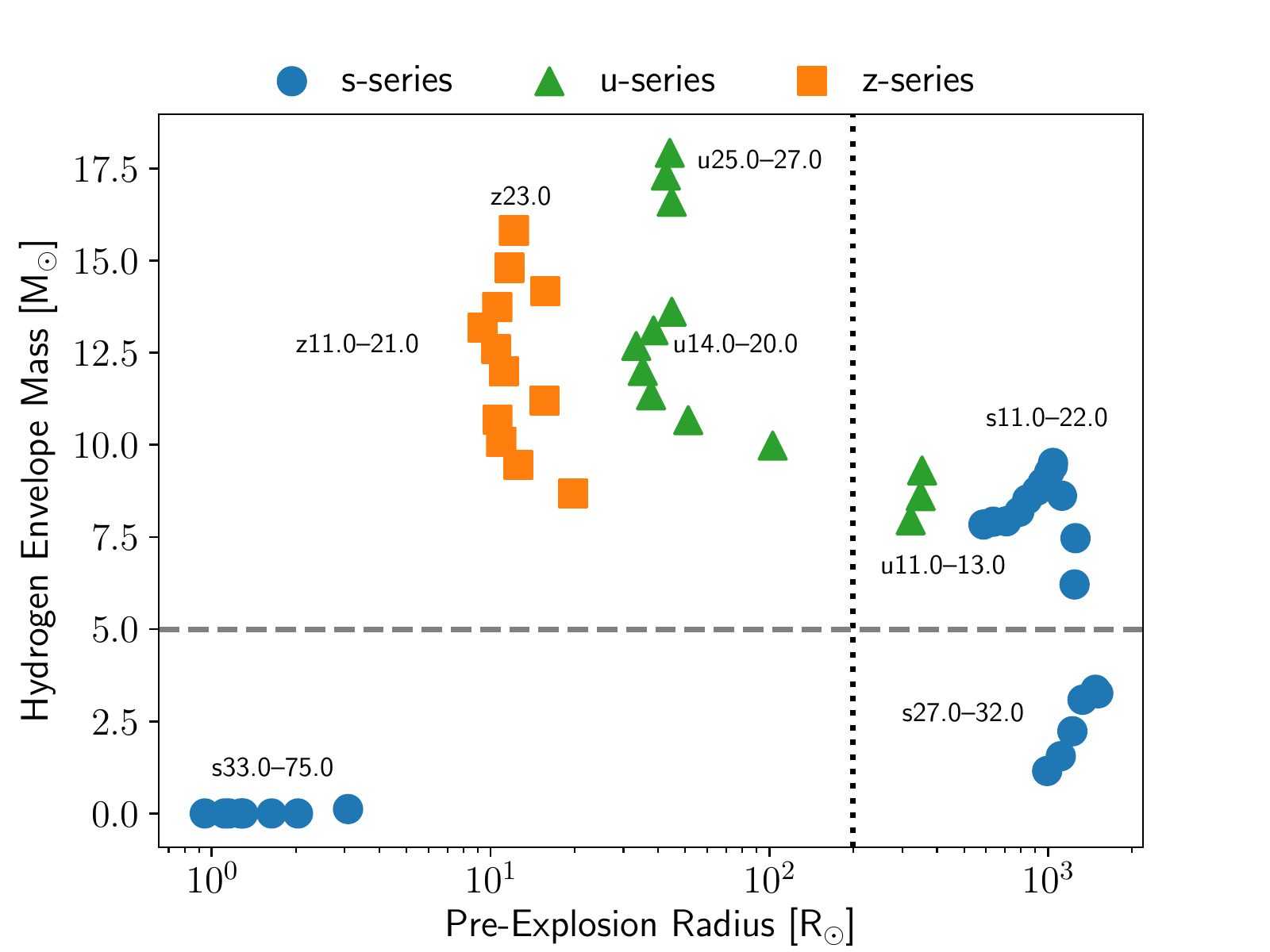}
        \caption{Hydrogen envelope mass as a function of the pre-explosion radius for all models of this study. The clusters correspond to different classes of qualitative behavior of the light curves. The vertical dotted line is at 200$R_{\odot}$ and the horizontal dotted line is at 5$M_{\odot}$. 
        \label{fig:quadrants}
        }
\end{figure}

 The initial size of the progenitor, the explosion energy, the amount of radioactive debris ($^{56}$Ni), the degree of chemical mixing, the total ejected mass, and the mass of the hydrogen envelope are all important quantities with respect to supernova light curves. In Figure~\ref{fig:quadrants}, we plot the progenitor radii and H-envelope masses for all our models and find that the different qualitative behaviors we identify correspond to different regions of this parameter space. Interestingly, models s26.0--s32.0 also appear to occupy their own sub-space. As we remarked earlier, the light curves of these models represent the transition case, lying somewhere between Type IIP light curves and stripped-envelope like light curves. 

For Type IIP light curves, both the explosion energy and $^{56}$Ni mass influence the luminosity during the plateau phase. More energetic explosions tend to have brighter but faster evolving light curves, while the energy deposited by $^{56}$Ni decay can decrease the rate of decline of the plateau and increase its length, depending on the extent of outward mixing of $^{56}$Ni \citep{Kasen2009}. 

\begin{figure*}
\begin{center}
        \begin{tabular}{cc}
        \includegraphics[width=0.48\textwidth]{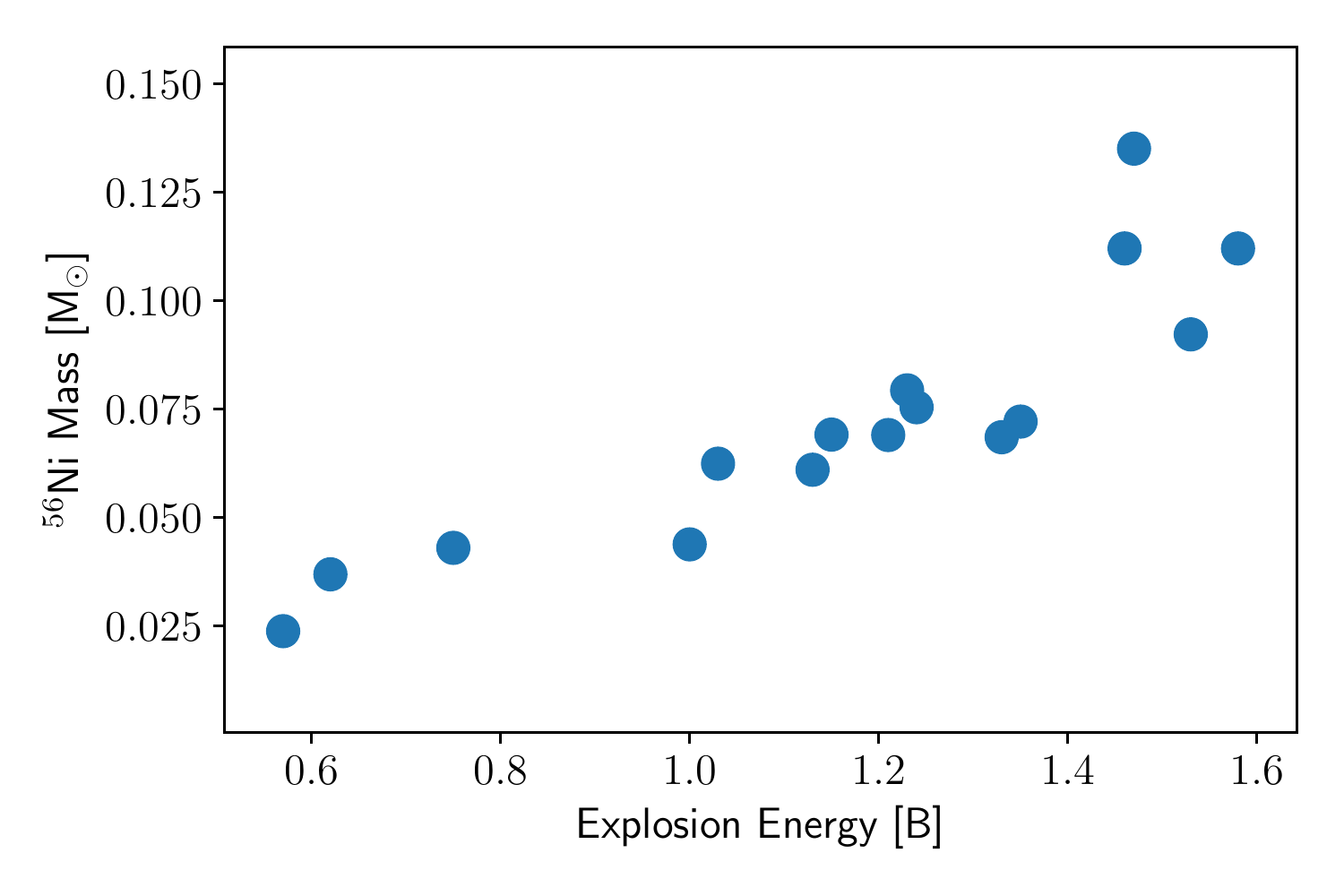}
        \includegraphics[width=0.48\textwidth]{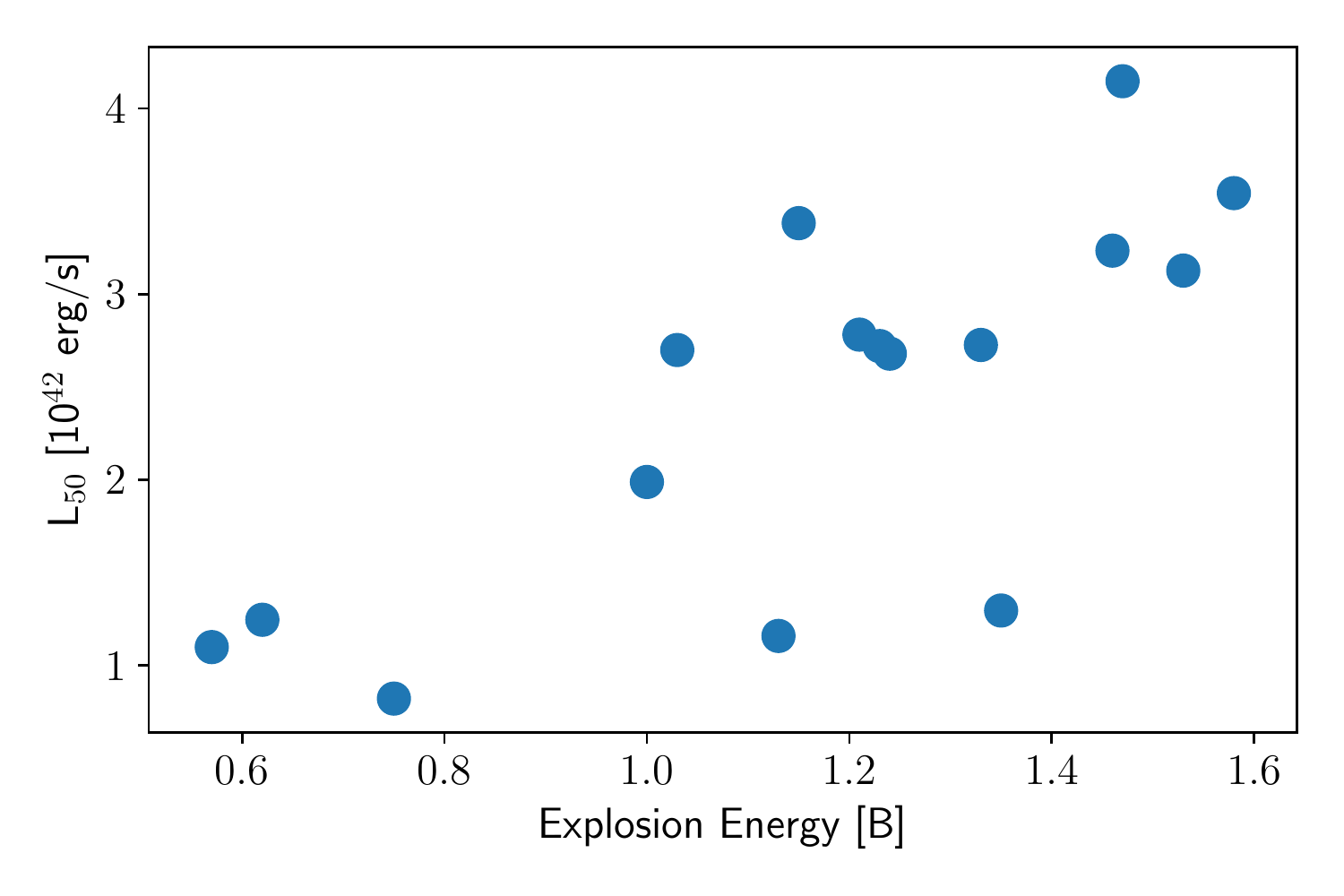}\\
        \includegraphics[width=0.48\textwidth]{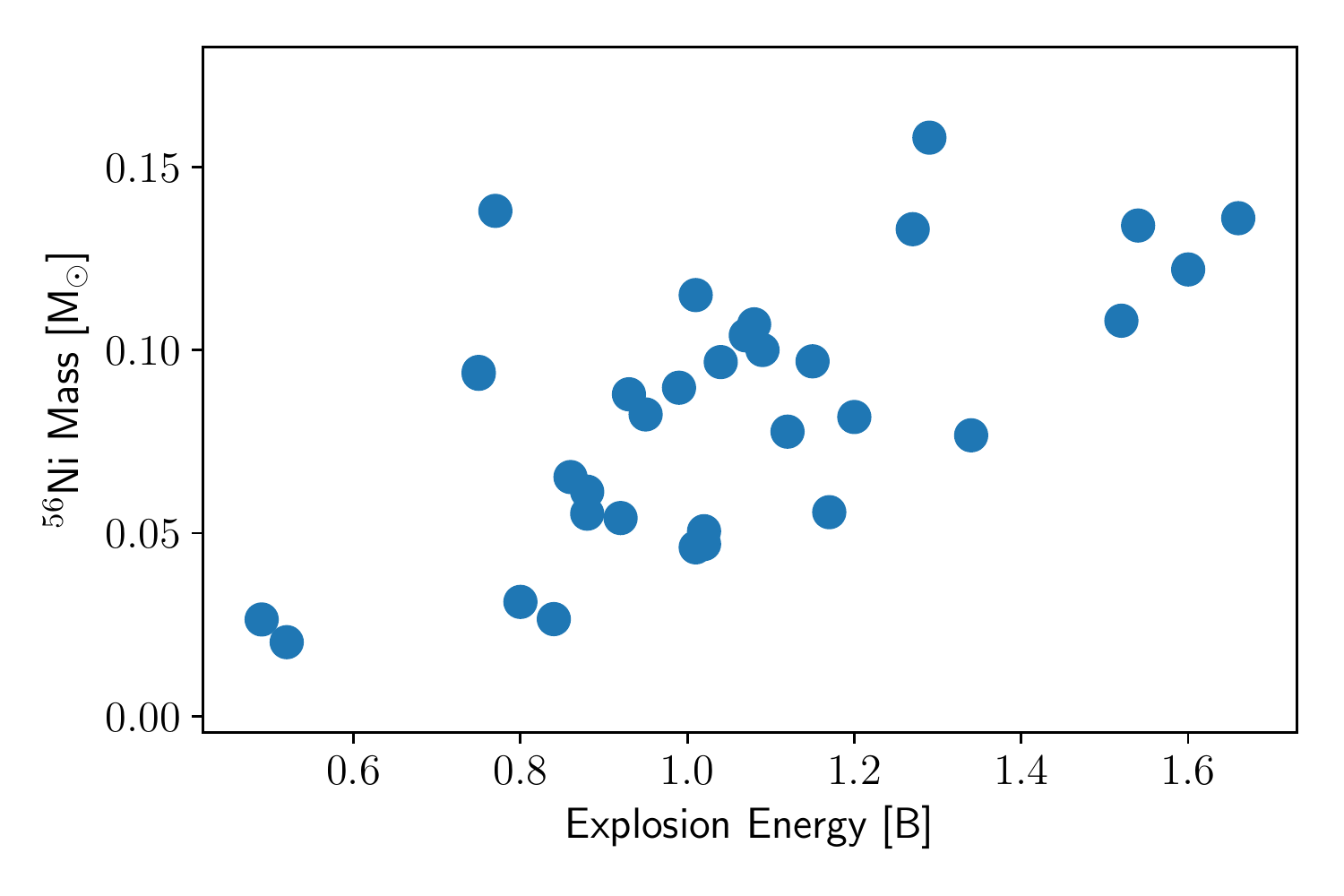}
        \includegraphics[width=0.48\textwidth]{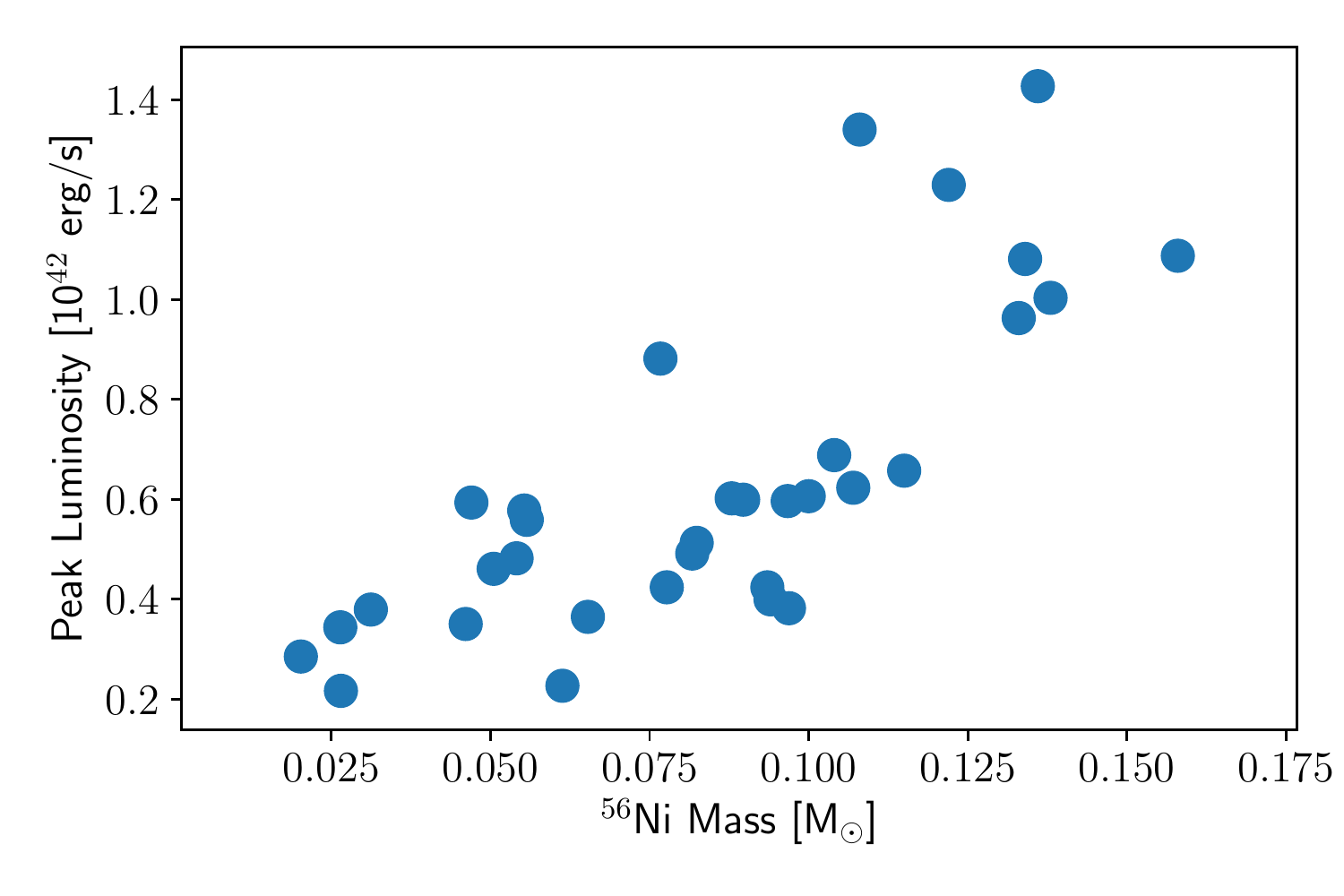}
        \end{tabular}
        \caption{The top two panels show the \deleted{luminosity at 50 days for Type IIP light curves, as a function}\replaced{of the $^{56}$Ni mass (left) and explosion energy (right) predicted using PUSH models}{ $^{56}$Ni mass (left) and luminosity at 50 days (right) as a function of explosion energy for Type IIP light curves}. The bottom panels show the \replaced{peak luminosity for light curves with broad peaks (both SN1987A-like and stripped-envelope like), as a function of $^{56}$Ni mass (left) and explosion energy (right)}{$^{56}$Ni mass (left) as a function of explosion energy and peak luminosity (right) as a function of $^{56}$Ni mass for light curves with broad peaks i.e. both SN1987A-like and stripped-envelope like models}.
        \label{fig:lin_corr}
        }
\end{center}
\end{figure*}

When drawing correlations from observational data, an ensemble of objects is collected and looked at. Our analysis mimics this process with a synthetic ensemble. In the top two panels of Figure~\ref{fig:lin_corr}, we plot \deleted{the plateau luminosity at 50 days ($L_{50}$) for our} \deleted{Type IIP light curves as a function of} \replaced{the $^{56}$Ni mass and}{the $^{56}$Ni mass and the plateau luminosity at 50 days ($L_{50}$) for our Type IIP light curves, as a function of the} explosion energy of the models \deleted{as} predicted by PUSH. \replaced{We find}{Our results show} a linear correlation between the amount of $^{56}$Ni produced in the explosion and \added{the explosion energy, as well as a weaker correlation between} $L_{50}$ \added{and the explosion energy.} \replaced{, as reported in previous studies for}{Previous studies using} observational samples of Type IIP supernovae \citep{Nakar2016, Hamuy2003} \added{report linear correlations between the $^{56}$Ni mass and $L_{50}$, which can also be reproduced using our results}. \deleted{There is also a positive but weaker correlation of} \deleted{$L_{50}$ with the explosion energy.}

The bottom two panels of Figure~\ref{fig:lin_corr} show the \replaced{same correlations}{$^{56}$Ni mass as a function of explosion energy and the peak luminosity as a function of $^{56}$Ni mass,} \deleted{but} for \deleted{the peak luminosities attained by} models producing stripped envelope-like or SN 1987A-like light curves. Here, we \added{once again} find \added{a linear correlation between the $^{56}$Ni and explosion energy. There is also} a strong correlation of the peak luminosity with \deleted{both} the $^{56}$Ni mass \deleted{and the explosion energy}. Such a correlation \deleted{of the peak luminosity with $^{56}$Ni mass} was reported for stripped-envelope supernovae by \citet{Prentice2016}. We note that the trends we present here arise naturally across the sample of models in our study, without any systematic hand-tuning of the relevant explosion properties and $^{56}$Ni yields. A quantitative description of the correlations between for various progenitor, explosion and light curve properties is presented in Section~\ref{subsec:correlations}.

Before we discuss the detailed physics behind some of our light curves, we need to understand the general physical evolution of the ejecta\added{,} which we summarize here. The nature of supernova light curves is roughly determined by five aspects of the event: shock breakout, expansion, radiative diffusion, radioactive heating, and recombination. The shock wave heats and accelerates matter so that it expands and radiates. As it expands, the internal energy is converted to kinetic energy as the pressure does work upon expanding matter. This conversion of internal energy to kinetic energy is relevant for the observed luminosity. The radius of the star after the passage of the shock wave is expected to be comparable to or larger than the radius of the pre-supernova star. If the star cools significantly by the time it expands to the large radius typical of supernovae (being cooled by that very expansion), the amount of thermal energy that can escape and be seen is greatly reduced and the supernova appears less luminous. The energy released by radioactive decay, however, is unaffected by expansion. 

The subsequent evolution of normal Type IIP supernovae is divided into three canonical epochs: the diffusive phase, the recombination phase, and the radioactive tail phase. The first two phases are together called the `photospheric' phase and the radioactive tail phase is often referred to as the `nebular phase'.
\begin{itemize}
\item Diffusive phase: This phase lasts for the first tens of days post-explosion. The ejecta are ionized and optically thick and the luminosity is primarily due to the release of internal energy which diffuses outward.
\item Recombination phase: This phase begins once the ejecta have expanded and cooled enough to allow for hydrogen recombination. As the ejecta recombine, the photosphere moves inward in mass, accompanied by release of energy. The typical duration of this phase is up to 100--120 days post-explosion. 
\item Radioactive tail phase: Once the ejecta are recombined and optically thin, the luminosity comes from thermalization of $\gamma$-ray photons produced in the decay of $^{56}$Ni and is completely dominated by radioactive heating. 
\end{itemize}

In the remainder of this Section, we focus on two models, one producing a normal Type IIP light curve and the other a 1987A-like light curve. We illustrate the differences in their physical evolution that lead to differences in their luminosity evolution, in particular, the formation of an extended plateau or a broad peak.

\subsubsection{Normal Type IIP: the case of s18.0}

\begin{figure}
\includegraphics[width=0.48\textwidth]{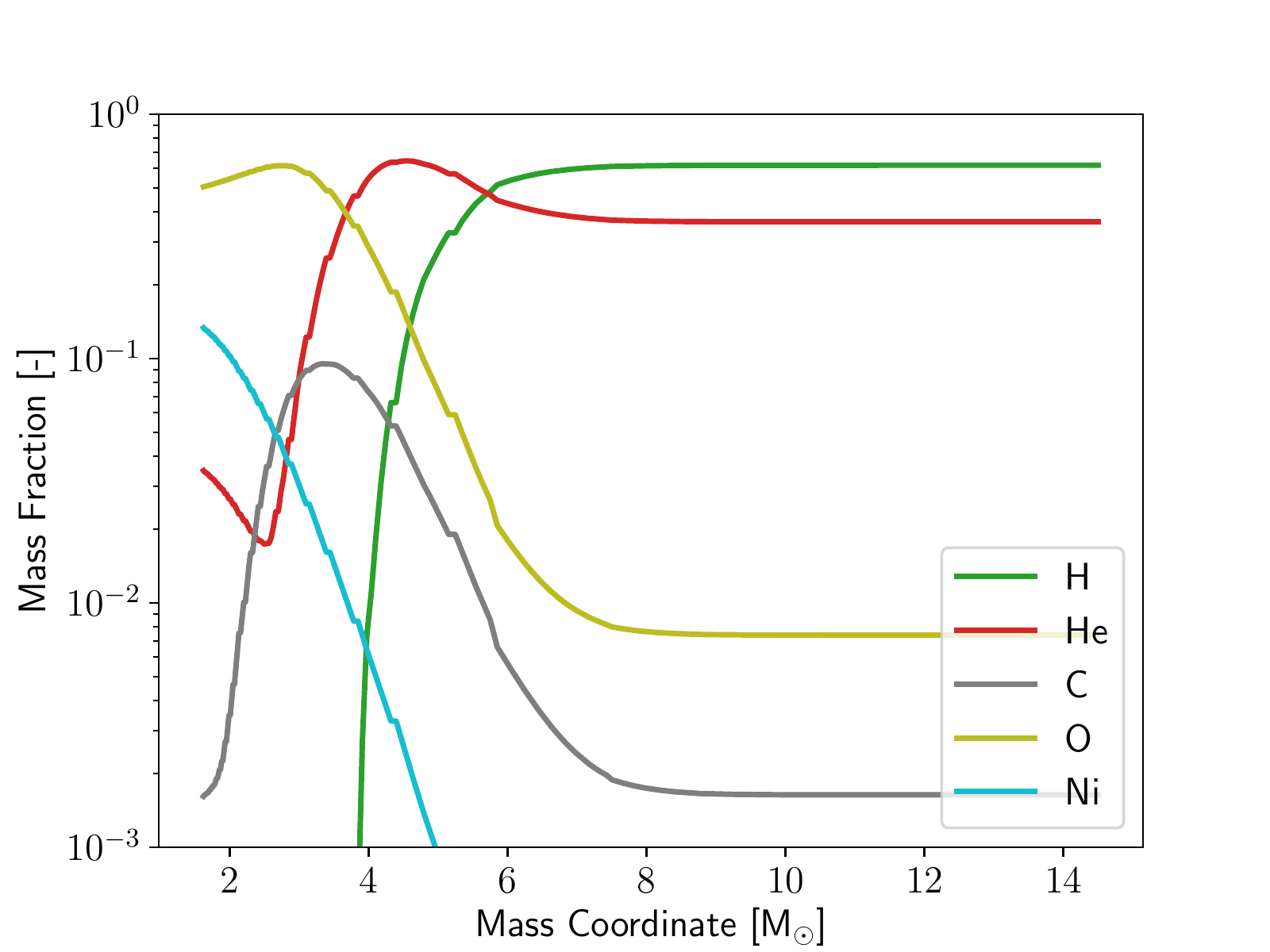}
\includegraphics[width=0.48\textwidth]{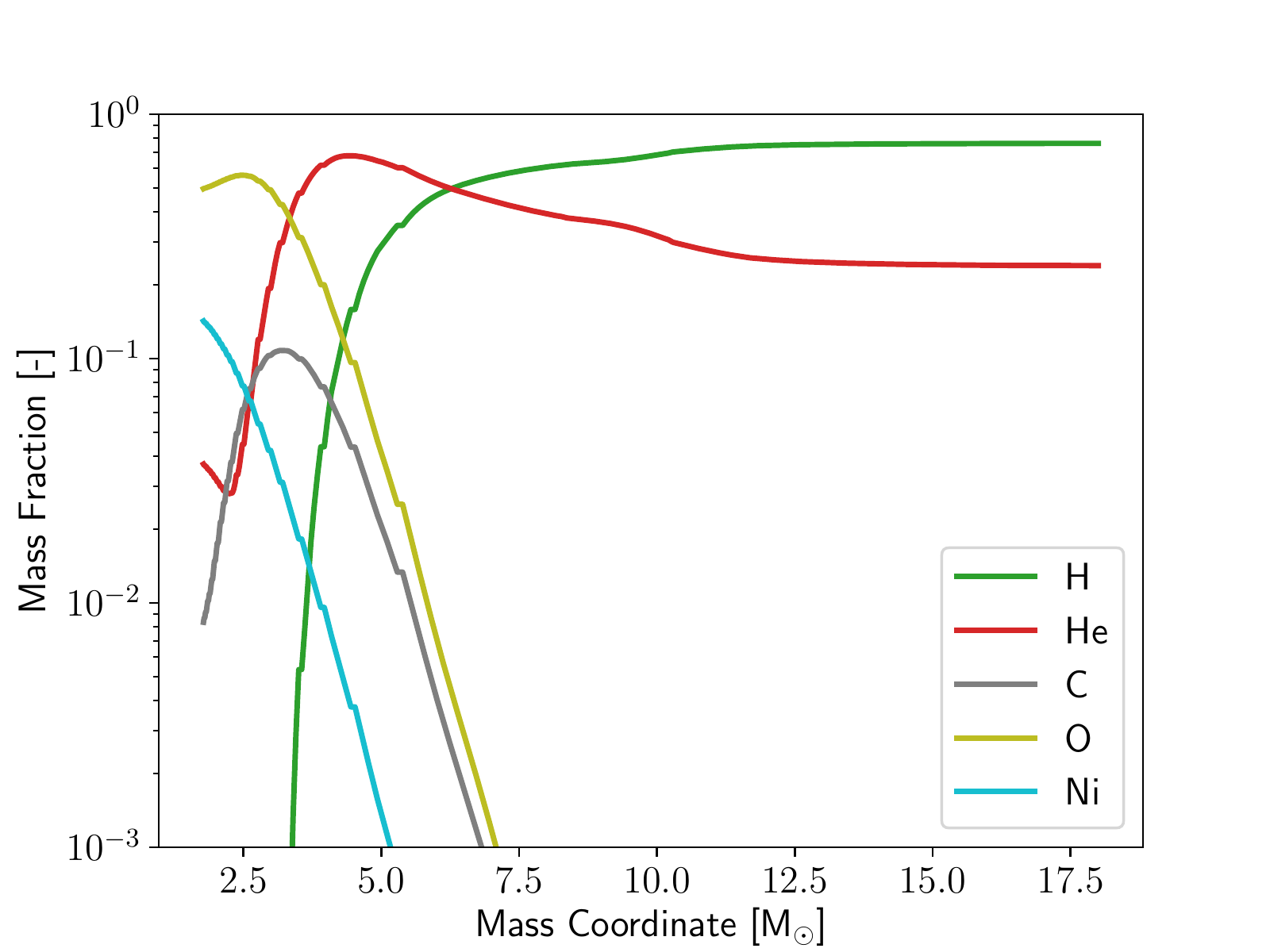}
\includegraphics[width=0.48\textwidth]{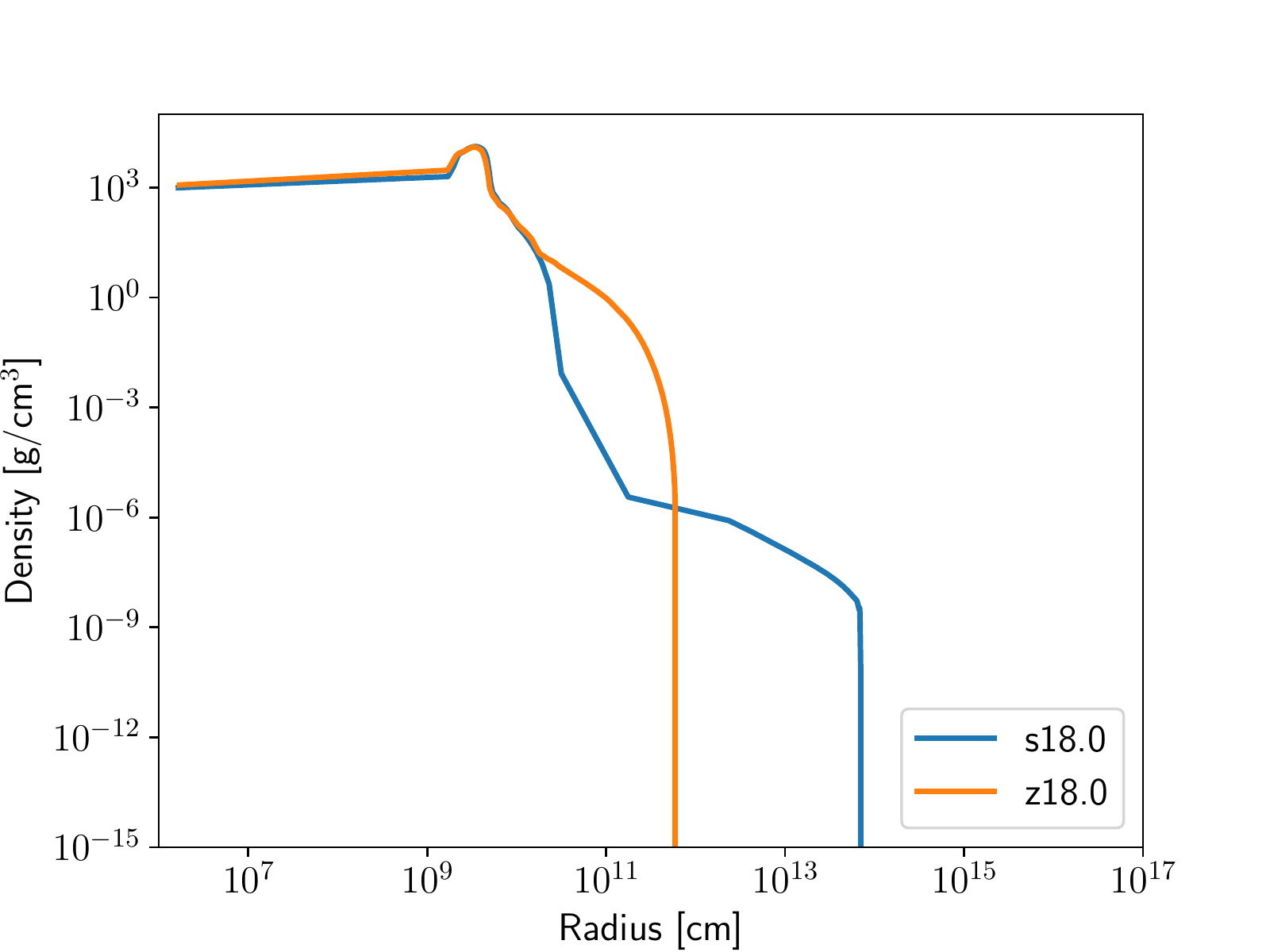}
\caption{Composition of models s18.0 (top) and z18.0 (middle) and density as a function of radius for both models (bottom). Only the ejecta above the mass-cut are shown. The slight discontinuities in the composition profiles are an artifact of smoothing in \texttt{SNEC} via boxcar averaging. 
\label{fig:compare_profiles}
}
\end{figure}

Model s18.0 is a solar-metallicity star with a ZAMS mass of 18 $M_{\odot}$. Although it has experienced some mass loss during its lifetime, it retains a massive hydrogen envelope of 9.25 $M_{\odot}$ and has a radius of $\sim$1010 $R_{\odot}$.  The explosion energy of this model is 1.45 Bethe and the $^{56}$Ni yield is 0.112 $M_{\odot}$. The initial composition of this model (post-smoothing in \texttt{SNEC}) is shown in Figure~\ref{fig:compare_profiles}.

\begin{figure*}
\begin{center}
        \begin{tabular}{cc}
        \includegraphics[width=0.48\textwidth]{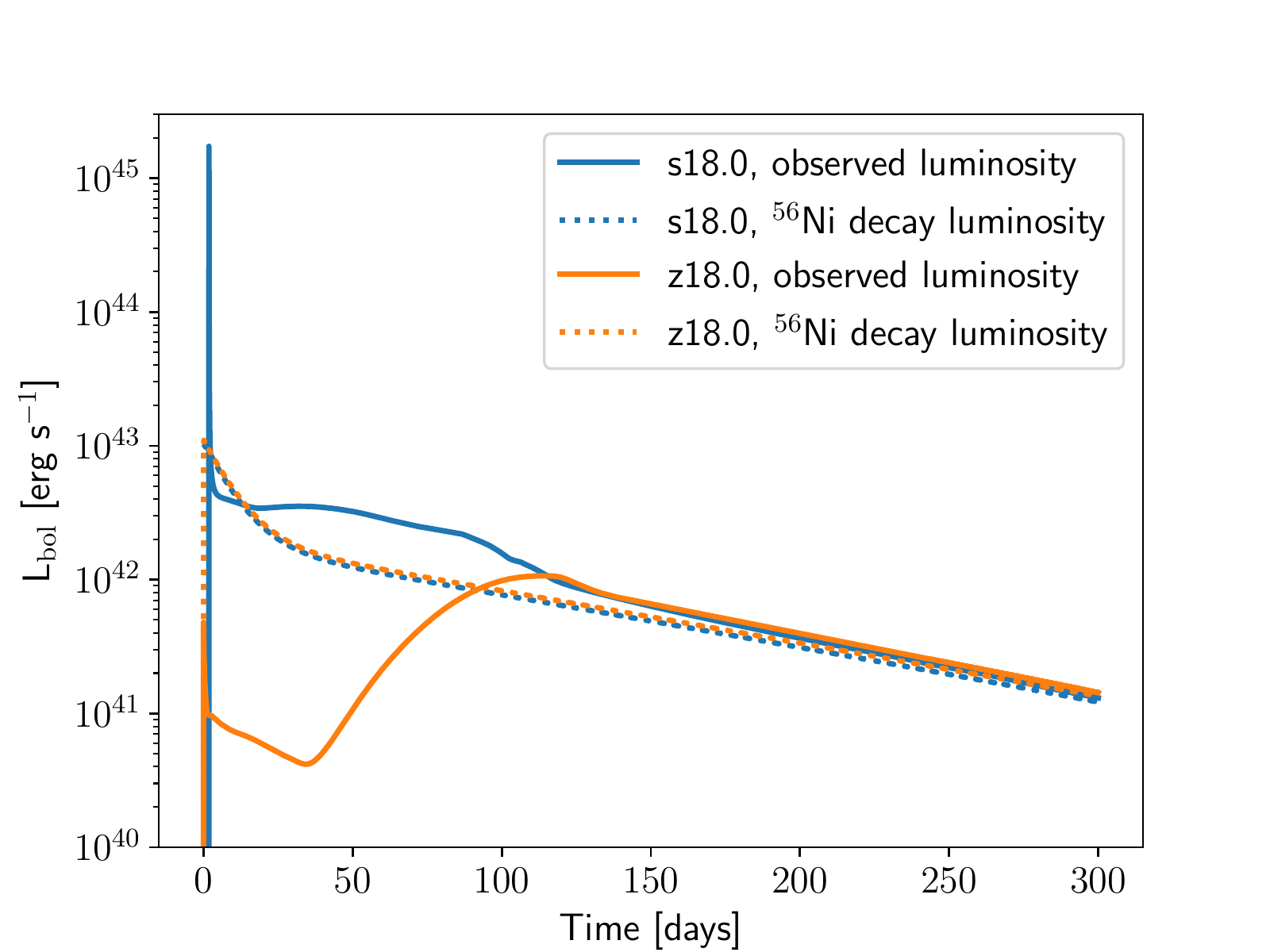}
        \includegraphics[width=0.48\textwidth]{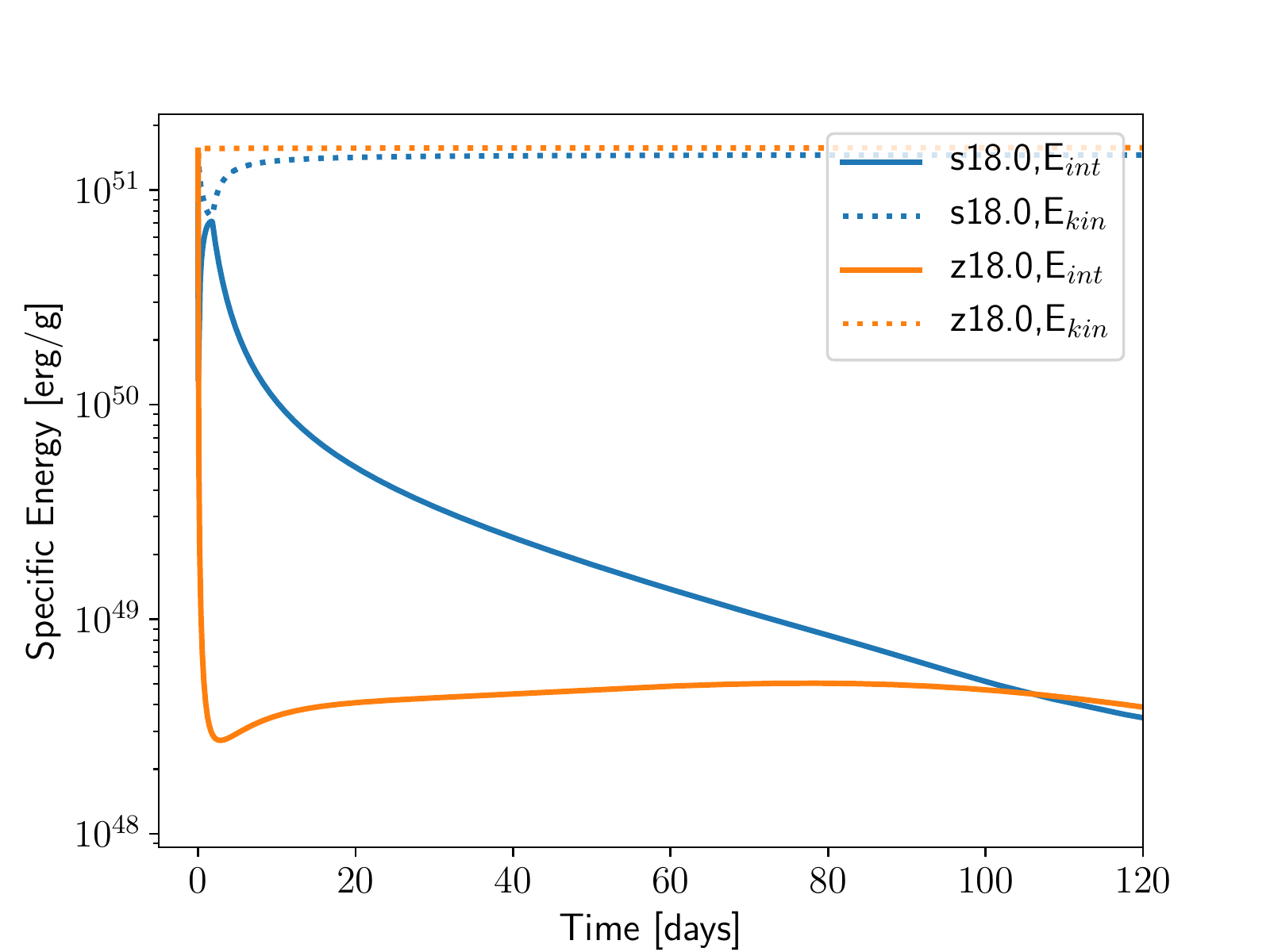}\\
        \includegraphics[width=0.48\textwidth]{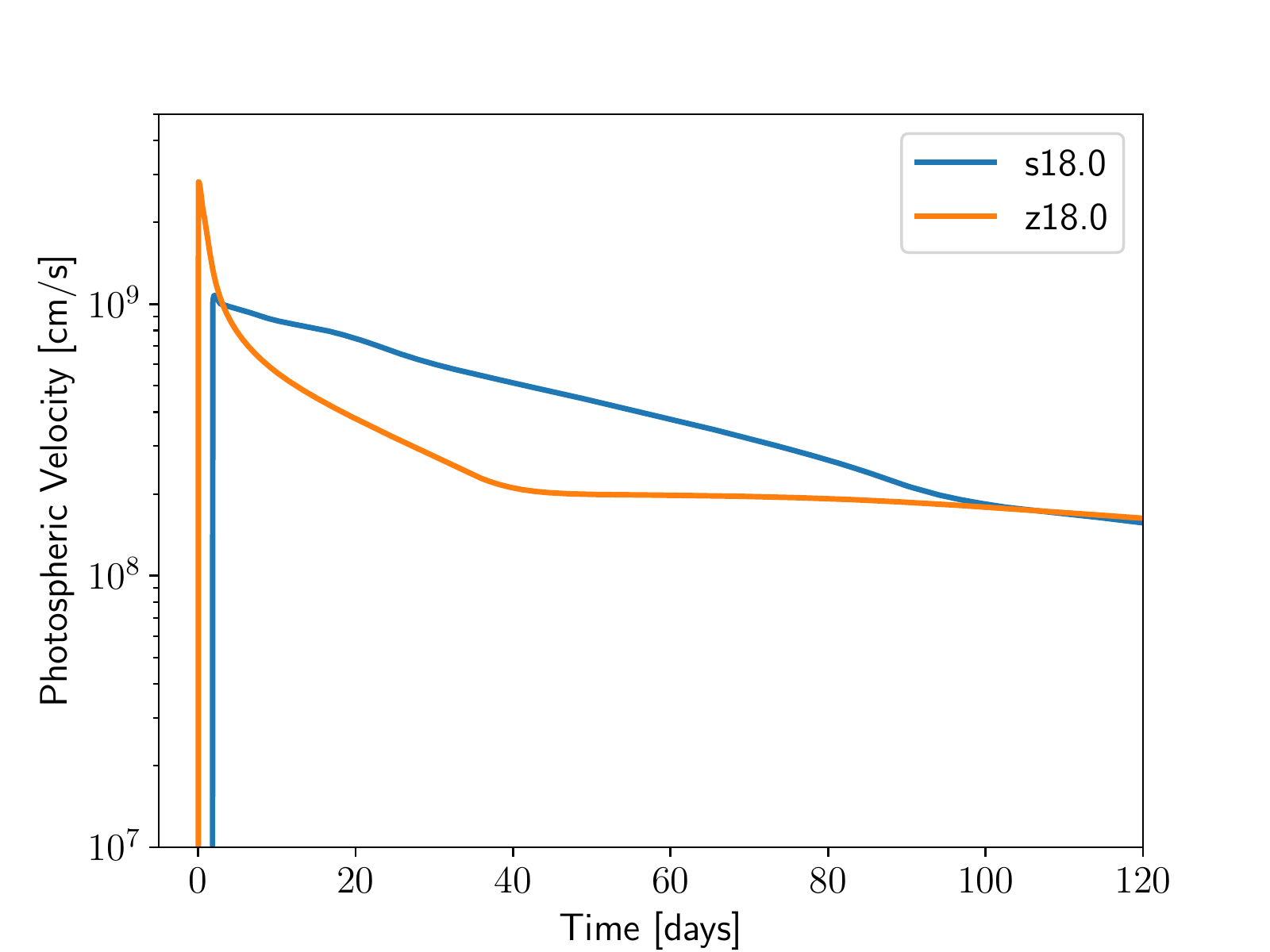}
        \includegraphics[width=0.48\textwidth]{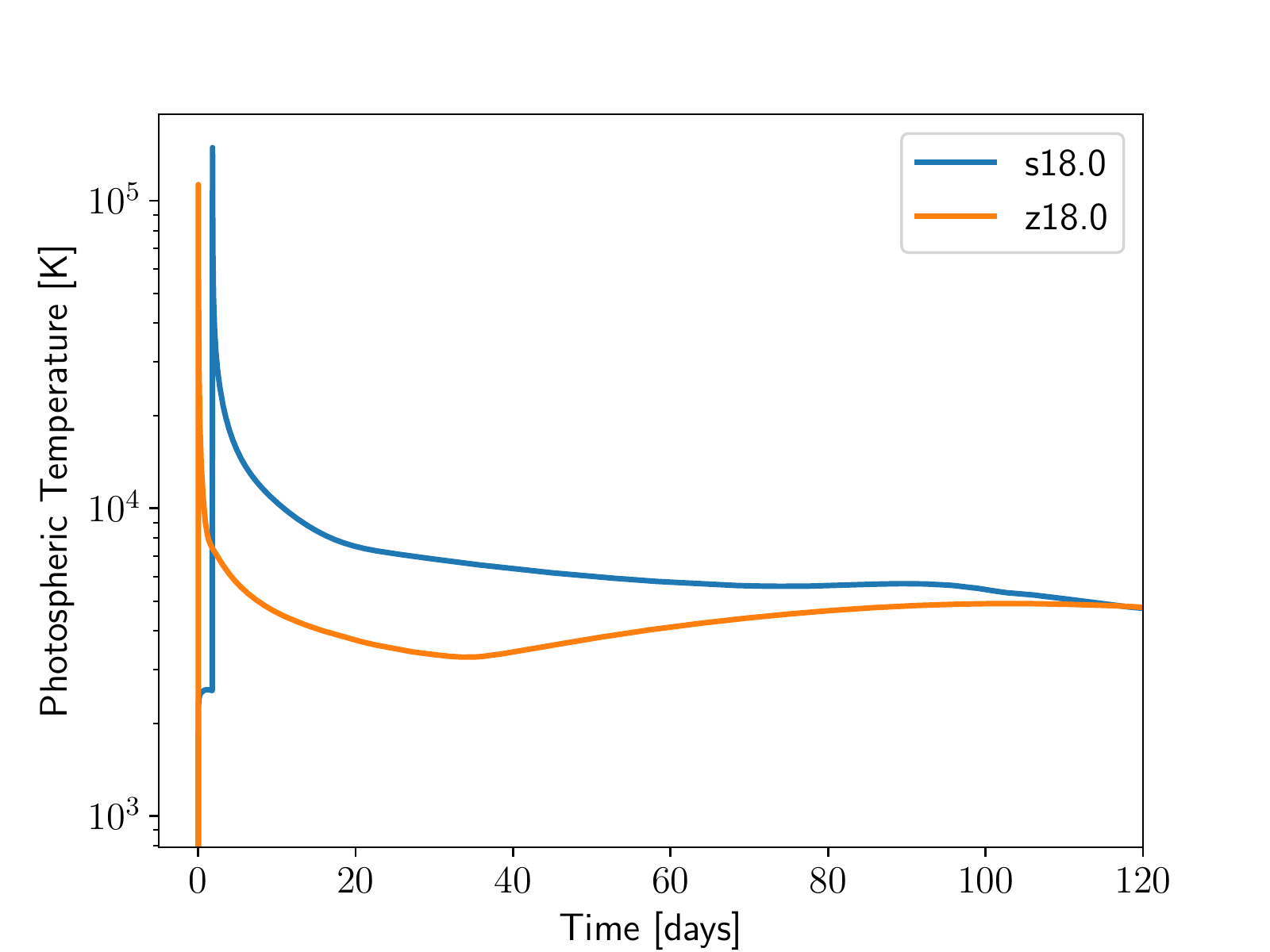}
        \end{tabular}
        \caption{Bolometric light curves (top left), energy density evolution (top right),  photospheric velocity (bottom left), and effective temperature(bottom right) for models s18.0 (blue) and z18.0 (orange).
        \label{fig:compare_lcs}
        }
\end{center}
\end{figure*}

The bolometric light curve of s18.0 looks much like that of a normal Type IIP supernova, as evident in the first panel of Figure~\ref{fig:compare_lcs}. The second panel shows the gradual conversion of internal energy to kinetic energy for this model. The bottom two panels show the evolution of the photospheric velocity and effective temperature. 

Shock breakout occurs at t $\sim$ 1.8 days when the optical depth\footnote{typically dominated by electron scattering} of the plasma lying ahead of the shock front is less than $\sim c/v$, where $v$ is the shock velocity. The photospheric radius increases and the bolometric luminosity peaks at $\sim$1.73 $\times 10^{45}$ ergs/s. The corresponding effective temperature is roughly 1.49 $\times$ $10^{5}$ K. This is followed by a short cooling phase that lasts until $\sim$20 days, before hydrogen recombination begins at a temperature of $\sim$7500 K. As the hydrogen envelope recombines, the electron-scattering opacity drops dramatically and an ionization front develops. The matter inside the front is ionized and opaque while the matter outside is neutral and optically thin. Thus, the location of the photosphere essentially coincides with that of the ionization front, both receding inward with respect to mass-coordinate. However, the photospheric radius stays roughly constant due to the combined effect of the ejecta expanding and the recombination front moving inward in mass. This hydrogen recombination powers the well-known plateau phase and we see very little variation in bolometric luminosity until t$\sim$104 days.

The plateau ends when the photosphere reaches the helium core. Since helium recombines at $T\gtrsim 10^4$K, which is much hotter than the photospheric temperature at this point, the recombination wave begins going through the helium layer at an increased rate, causing a drop in the bolometric luminosity as the photospheric radius decreases. The duration of the plateau is typically defined as the time from shock breakout to the luminosity drop when the photosphere reaches the helium core. As the photosphere sweeps through the helium layer, it may uncover some additional luminosity input from $^{56}$Co, depending on the degree to which $^{56}$Ni was mixed outward into this layer. This leads to the formation of a small, knee-like feature during the decline from the plateau to the radioactive tail. Finally, we enter the nebular phase at $\sim$110 days and the radioactive tail of the light curve is powered exclusively by the decay of $^{56}$Co. The hydrodynamical evolution of this model in \texttt{SNEC} is described in more detail and contrasted with that of model z18.0 in the next subsection.

\subsubsection{SN 1987A-like: the case of z18.0}

Model z18.0 is a zero-metallicity star with a ZAMS mass of 18 $M_{\odot}$. It has experienced almost no mass loss during its lifetime and retains its entire massive hydrogen envelope of $\sim$13.2 $M_{\odot}$. The pre-explosion radius of this model is $\sim$ 9.34 $R_{\odot}$, the explosion energy is 1.54 Bethe, and the $^{56}$Ni yield is 0.134 $M_{\odot}$. The initial composition of z18.0 is shown in the second panel of Figure~\ref{fig:compare_profiles}. 

The light curve of this model is SN 1987A-like, as opposed to that of s18.0, although both models have the same ZAMS mass and similar hydrogen envelope masses, explosion energies and $^{56}$Ni yields. The primary difference between them (apart from metallicity), and key to their diverging evolution, is the initial radius of the progenitor model. While the solar-metallicity model has a radius typical of red supergiants (R $\gtrsim$200 $R_{\odot}$), the zero-metallicity model is extremely compact ($\lesssim$ 10$R_{\odot}$). 

Figure~\ref{fig:compare_lcs} also shows the evolution of the bolometric luminosity, internal energy, photospheric velocity and effective temperature for model z18.0. The luminosity of s18.0 is higher during the diffusive and recombination phases since the observed luminosity depends on the photospheric radius, which is larger for models with large radii. While the expansion of s18.0 occurs slowly and more thermal energy remains available, for z18.0, the work done to expand the ejecta is large and the initial expansion velocity is higher. Thus, the ejecta cool down faster, quickly exhausting the internal thermal energy. This can be seen in the top right panel of Figure~\ref{fig:compare_lcs}.

\begin{figure*}
\begin{center}
    \begin{tabular}{cc}
        \includegraphics[width=0.45\textwidth]{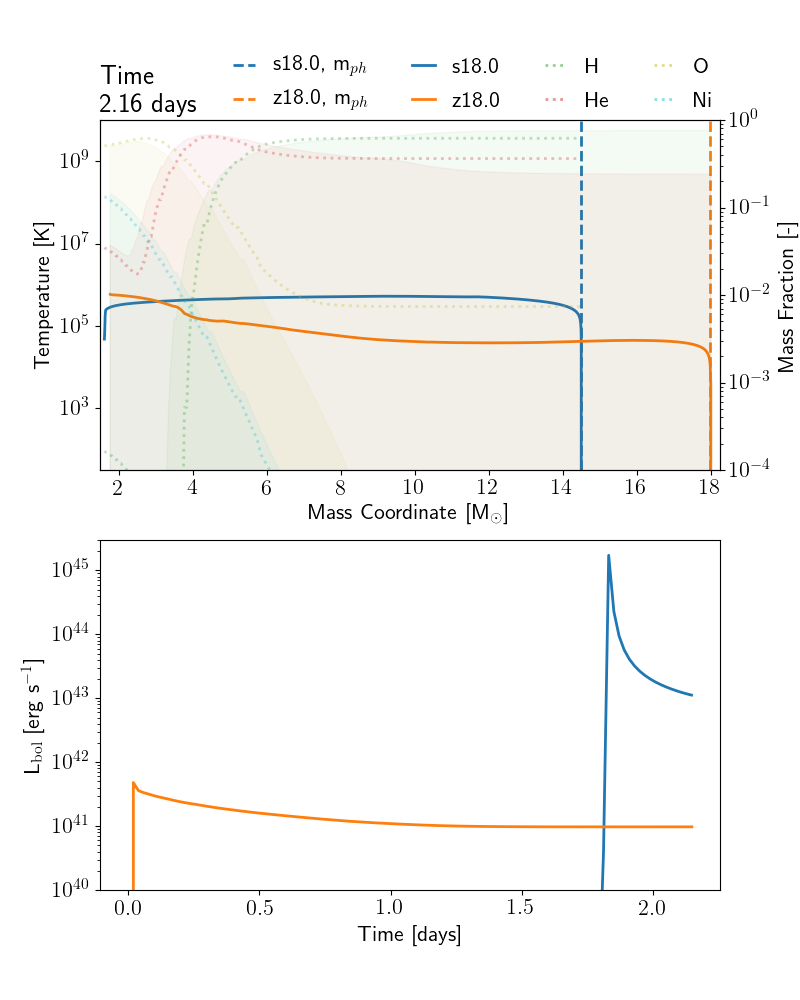}
        \includegraphics[width=0.45\textwidth]{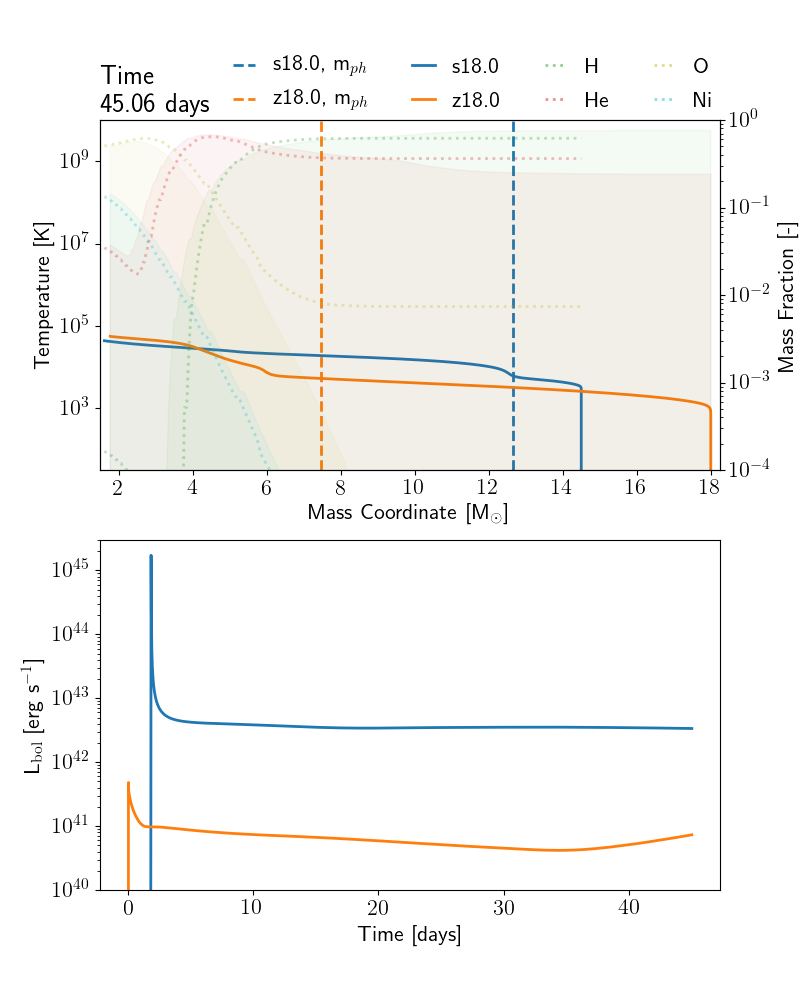}\\
        \includegraphics[width=0.45\textwidth]{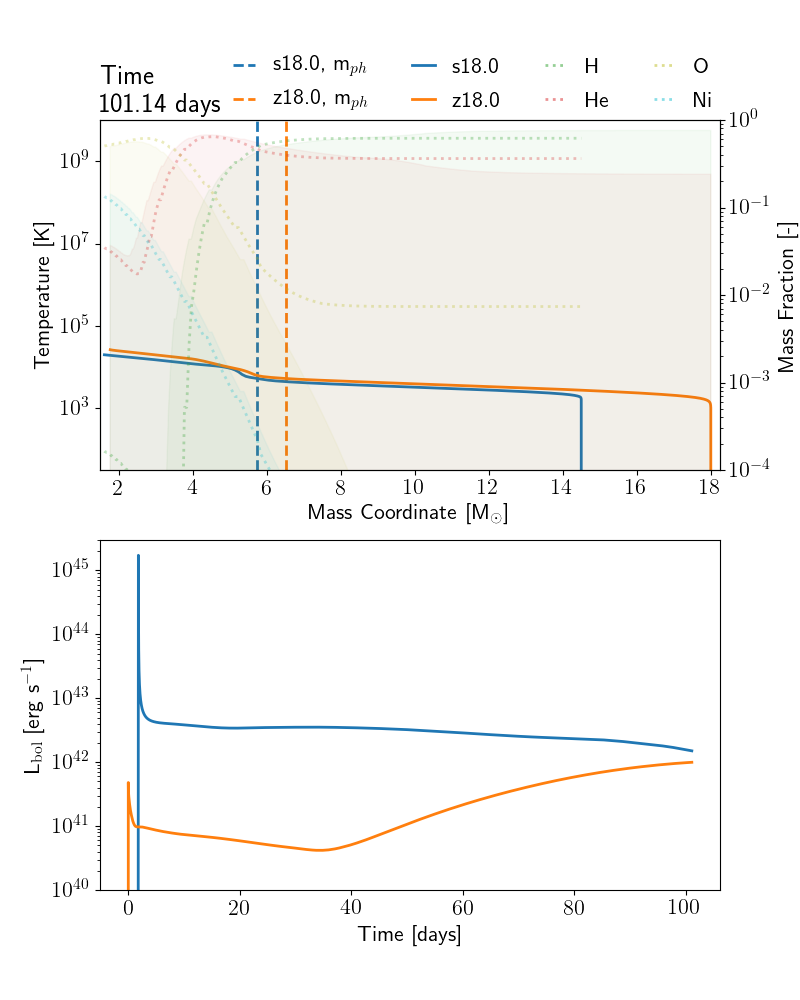}
        \includegraphics[width=0.45\textwidth]{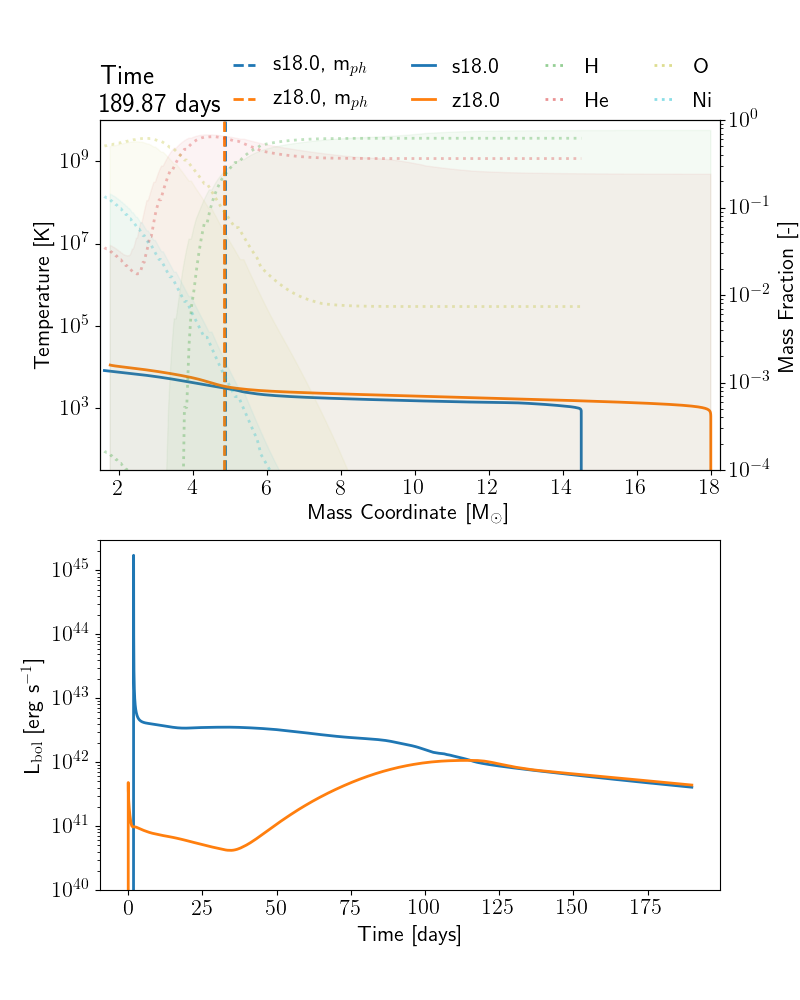}
    \end{tabular}
    \caption{Evolution of models s18.0 and z18.0 with respect to temperature and the resulting bolometric light curves. The location of the photosphere is indicated by the dashed vertical lines. The shaded region reflects the composition of model z18.0 while the dotted lines correspond to the mass fractions of the same elements for model s18.0.
    \label{fig:compare_evol}
    }
\end{center}
\end{figure*}

We can see the same behavior reflected in the evolution of the temperature profiles of these models in \texttt{SNEC}, plotted at selected times in Figure~\ref{fig:compare_evol}. The inflection point in the temperature profile roughly corresponds to the location of the photosphere. For z18.0, recombination starts quite early around day 5 and by day 45, almost all the hydrogen has already recombined. The subsequent luminosity is governed by radioactive decay, which is not diluted by expansion work. Nickel heating slows the motion of the recombination wavefront but it eventually begins moving deeper into the helium core, producing the small drop in luminosity after day 100. Finally, the late-time tail declines in much the same way for both models regardless of initial radius, since it is powered almost exclusively by radioactive decay.

\subsubsection{Correlations with Progenitor and Explosion Properties}
\label{subsec:correlations}
\begin{figure*}[t!]
\begin{center}
    \includegraphics[width=0.48\textwidth]{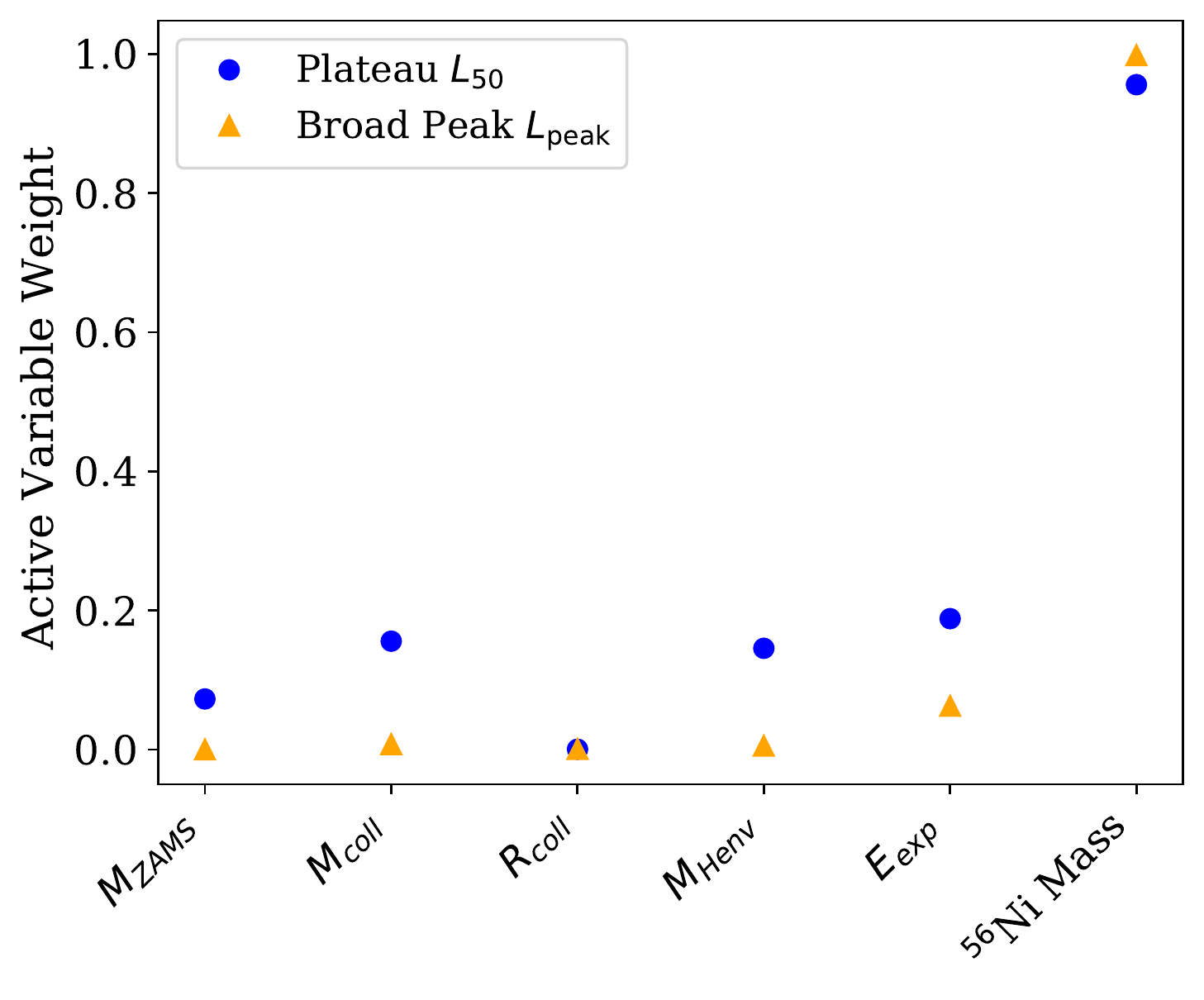}
    \includegraphics[width=0.48\textwidth]{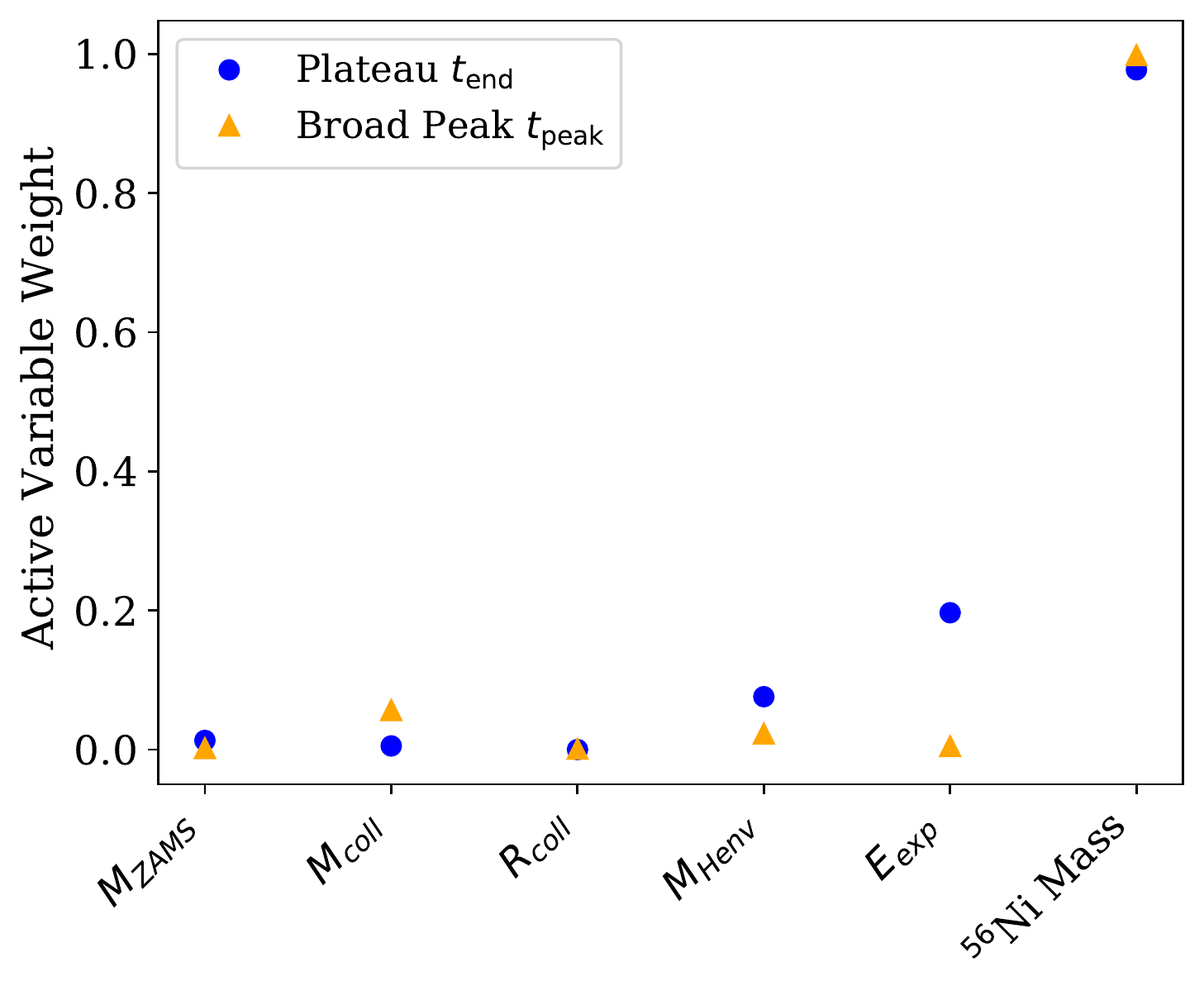}
    \caption{Input parameter weights (absolute values) for the characteristic luminosity (left) and characteristic timescale (right)
     for models with plateau (blue circles) and broad peak (orange triangles) light curves. The broad peak category includes both SN1987A-like and stripped-envelope like light curves. The input parameters are the ZAMS mass ($M_\mathrm{ZAMS}$), mass at collapse ($M_\mathrm{coll}$), radius at collapse ($R_\mathrm{coll}$), hydrogen envelope mass ($M_{\mathrm{Henv}}$), explosion energy ($E_\mathrm{exp}$), and $^{56}$Ni mass). }
    \label{fig:as_active_variable_weights}
\end{center}
\end{figure*}

Here, we use Active Subspace Sampling (\citet{as_textbook_constantine} and also Appendix \ref{app:uq}) to identify the most important progenitor and explosion properties that impact the light curve properties. We use six parameters (ZAMS mass, mass at collapse, radius at collapse, hydrogen envelope mass, explosion energy, and $^{56}$Ni mass) in our analysis. Figure \ref{fig:as_active_variable_weights} shows the active variable weights for our sample of Type IIP light curves with a plateau (blue circles) and separately for our sample of `broad peak' light curves (orange triangles) which includes both stripped-envelope like and SN1987A-like light curves.  From this, we can see that in both cases the $^{56}$Ni mass emerges as the single-most important parameter in determining the characteristic luminosity of the light curves ($L_{50}$ for plateau models, $L_{peak}$ for broad peak models, left panel) and for the characteristic timescale of the light curves ($t_{end}$ for plateau models, $t_{peak}$ for broad peak models, right panel).
Due to the small sample size (for this technique) we cannot make strong statements about the weaker correlations of the other active parameters included in the analysis. 

Important quantities relating to the progenitor, the explosion and the nucleosynthesis, along with the corresponding light curve properties are summarized in Tables~\ref{tab:summary_plateau} and \ref{tab:summary_peak}.

\input{summary_table_plateau}
\input{summary_table_peak}

\section{Synthetic light curves and Spectra from SuperNu}
\label{sec:supernu}
\subsection{SuperNu}

\texttt{SuperNu} \citep{Wollaeger2013, Wollaeger2014} is a time-dependent radiation transport code that uses Implicit Monte Carlo (IMC; \citet{Fleck1971, Wollaber2016}) and Discrete Diffusion Monte Carlo (DDMC; \citep{Gentile2001, Densmore2007,Densmore2012,Abdikamalov2012} to solve the radiative transfer equations, under the assumption of local thermodynamic equilibrium (LTE). DDMC is used to accelerate IMC in optically thick regions where it becomes inefficient. 

\texttt{SuperNu} is capable of computing bolometric and broadband light curves as well as synthetic spectra, but does not handle hydrodynamic coupling between matter and radiation. Hence, there is no momentum transfer between radiation and matter, although the radiation does affect the temperature of the material. This approximation is valid after sufficient time has passed since the explosion and the ejecta are expanding homologously. \texttt{SuperNu} uses the fact that the outflow is homologous to formulate the method over a velocity grid. 

The MC approach is a probabilistic method where the radiation energy is discretized as packets (or particles), each representing a bundle of photons. The packets are emitted wherever the energy sources are non-zero and transported over the computational domain. As they move, they interact (absorb and scatter) with matter according to the local energy-dependent opacities. The interactions are carried out stochastically through random number sampling of the corresponding probability density distributions.

The opacities in \texttt{SuperNu} include Thomson scattering and multi-group absorption opacities. The leakage and Planck opacities are calculated from the scattering and absorption opacities. The multi-group absorption opacities include bound-bound, bound-free and free-free data for elements from hydrogen up to cobalt. The line data for bound-bound opacities are taken from \cite{Kurucz1994}. \texttt{SuperNu} also tracks radioactive heating by the alpha-chain isotopes $^{56}$Ni, $^{52}$Fe and $^{48}$Cr, and their decay products. The electron fraction of the material is updated with radioactive decay in every time step. 

\subsection{Mapping from SNEC to SuperNu}

\begin{figure*}
        \includegraphics[width=0.46\textwidth]{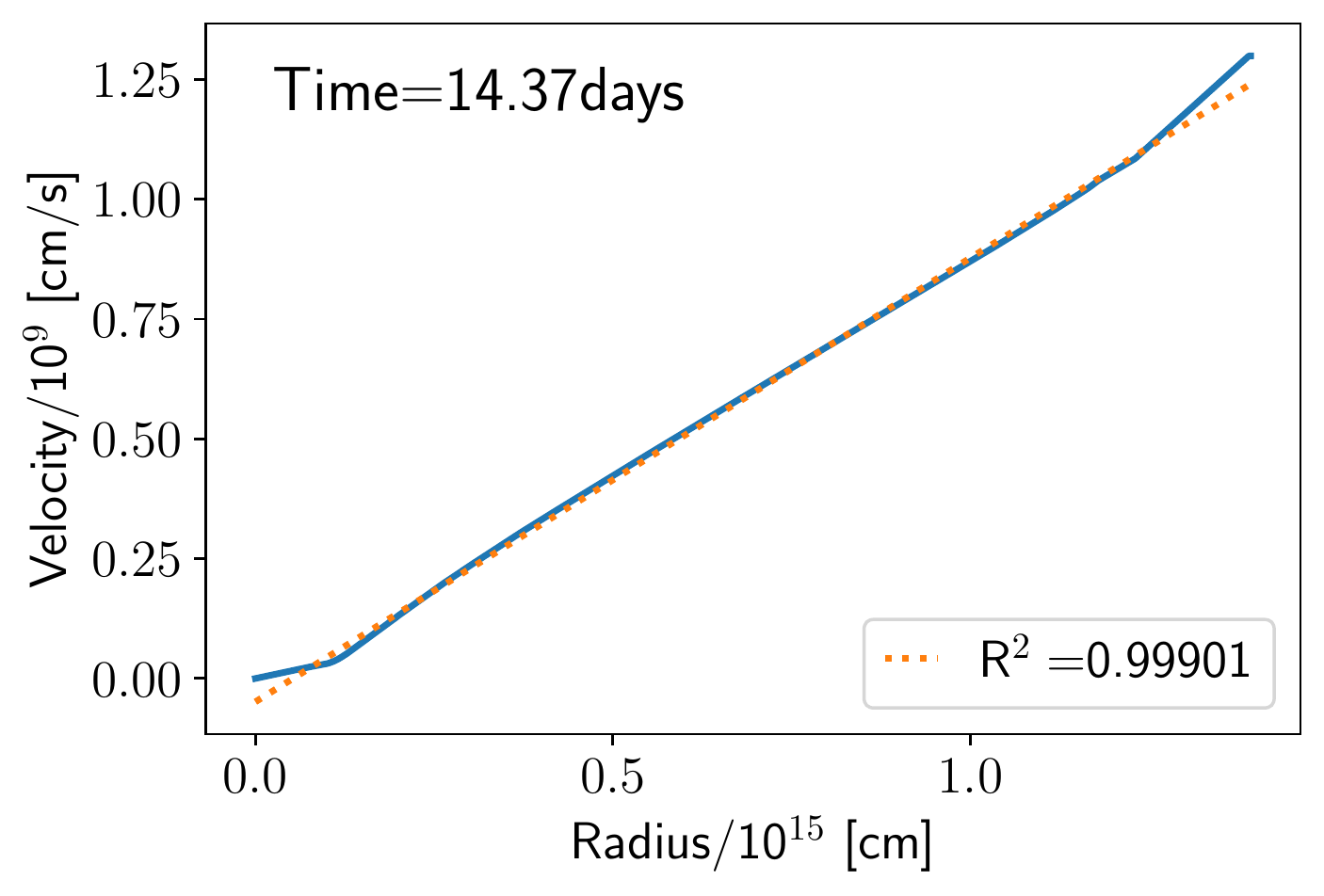}
        \includegraphics[width=0.54\textwidth]{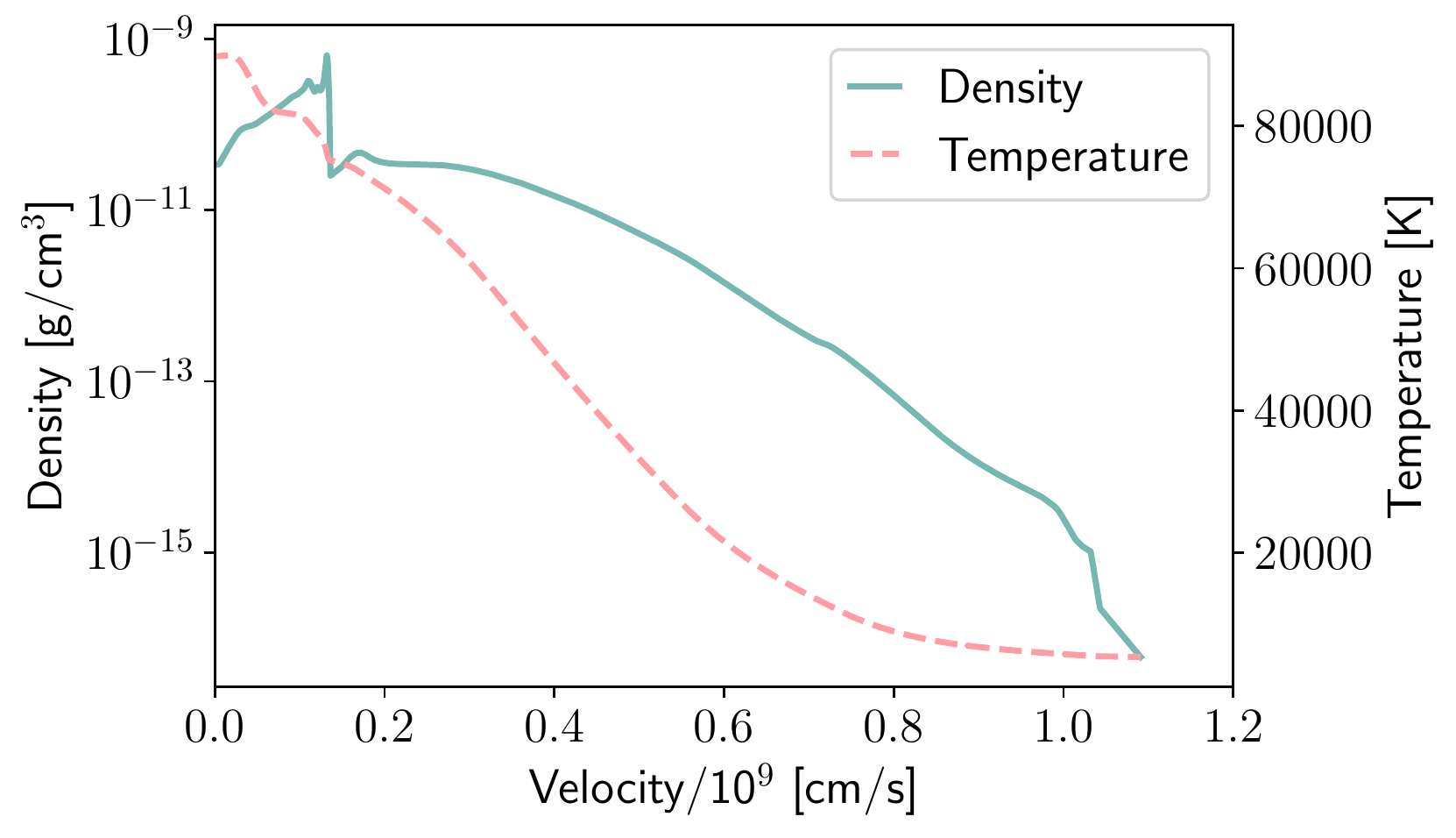}
        \caption{Left: Actual (solid line) and fitted (dotted line) velocity profile at the time of mapping from \texttt{SNEC} to \texttt{SuperNu} for model s18.0. 
        Right: Density (red, left axis) and temperature (blue, right axis) profiles in velocity space corresponding to the fitted velocity profile on the left.
        \label{fig:input_str}
        }
\end{figure*}

We cannot map our explosion simulations with \texttt{Agile} directly to \texttt{SuperNu} since the homologous expansion assumption is not satisfied at the end of our simulations. Instead, we evolve the outflow in \texttt{SNEC} until it becomes homologous, at which point we map from \texttt{SNEC} to \texttt{SuperNu}. We determine the time step for this mapping by performing a simple linear fit of velocity with respect to radius and checking whether the coefficient of determination ($R^2$) has met a chosen threshold value. Here, we choose to map when the $R^2$ value exceeds 0.999. In the left panel of Figure~\ref{fig:input_str}, we plot our velocity fit and the corresponding mapping time for model s18.0.

Once the appropriate time step has been identified, we extract the relevant quantities like mass, velocity, temperature, electron fraction, and the mass fractions of different elements from the \texttt{SNEC} output. Since \texttt{SNEC} does not evolve the mass-fractions of the radioactive isotopes supplied to it in the input composition profile (here, $^{56}$Ni, $^{52}$Fe and $^{48}$Cr), these mass-fractions are still set to their values at a few seconds post-explosion. We calculate the post-decay mass fractions of these isotopes at the chosen mapping time using the Bateman equation \citep{Bateman1910}.

Finally, we map from the mass-based grid used by \texttt{SNEC} to the velocity-based grid required by \texttt{SuperNu}. For our spherically-symmetric models, this is a simple procedure that involves computing the mass contained in each grid cell and assigning the appropriate velocity value to it. To ensure that the velocity grid is monotonically increasing, as would be the case for truly homologous outflows, we use the linear fit to the velocity profile of our nearly homologous outflow as the input velocity grid for \texttt{SuperNu}. The right panel of Figure~\ref{fig:input_str} shows the density and temperature of model s18.0 as a function of velocity, used as input for \texttt{SuperNu}.

At sharp gradients, SNEC produces the characteristic ringing known as the Gibbs phenomenon \citep{wilbraham1848certain, du1874ueber, michelson1898HarmonicAnalyser, michelson1898letter1, love1898fourier1, michelson1898fourier2, love1899fourier2, Gibbs1, Gibbs2, Bocher1906, hewitt1979gibbs}, as is typical for continuum codes \citep{toro2013riemann}. These artifacts are clearly visible in the density profile shown in Figure~\ref{fig:input_str}. We experimented with smoothing over the Gibbs-ringing by convolving the density with a gaussian kernel. However, this had minimal effect on the light curves computed by \texttt{SuperNu} and we use the profiles produced by \texttt{SNEC} without any additional smoothing.

We carried out extensive convergence tests to determine the number of groups, time step resolution and MC particle count required to produce converged light curves and spectra for our models. We present the results of these tests for one model in Appendix~\ref{app:supernu_convergence}. 

\subsection{Results: SuperNu Light Curves and Spectra}

We compute bolometric light curves, \textit{UBVRI} magnitudes as well as spectra using \texttt{SuperNu} for all models in our sample except the peculiar light curves presented in Appendix \ref{app:weird}. The bolometric light curves and spectral evolution data are published in machine-readable format with this paper (see Appendix~\ref{app:data} for sample tables) and are also available online \footnote{\url{go.ncsu.edu/astrodata}}. 

\begin{figure*}
\begin{center}
        \begin{tabular}{cc}
\includegraphics[width=0.48\textwidth]{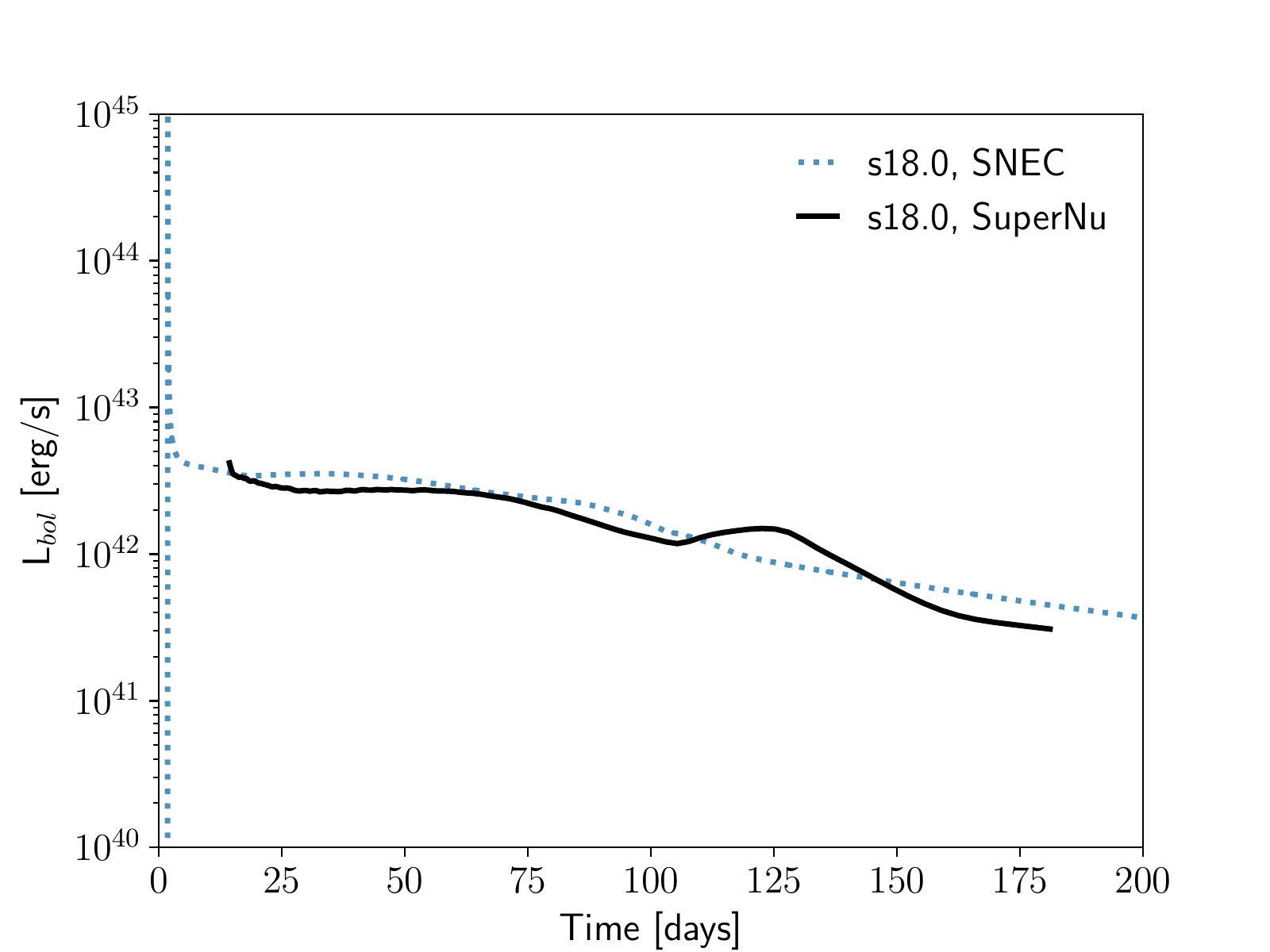}
\includegraphics[width=0.48\textwidth]{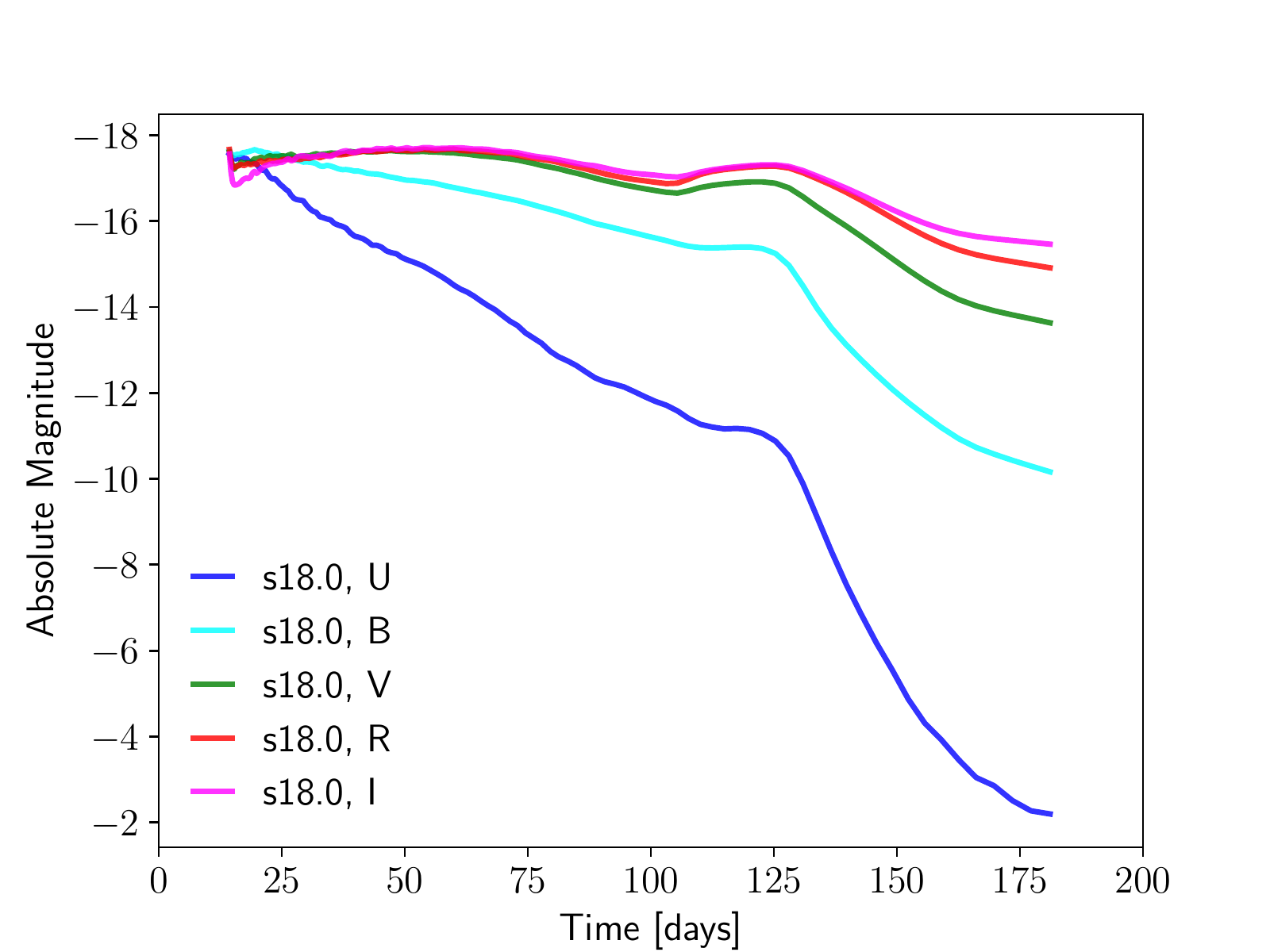}\\
\includegraphics[width=0.48\textwidth]{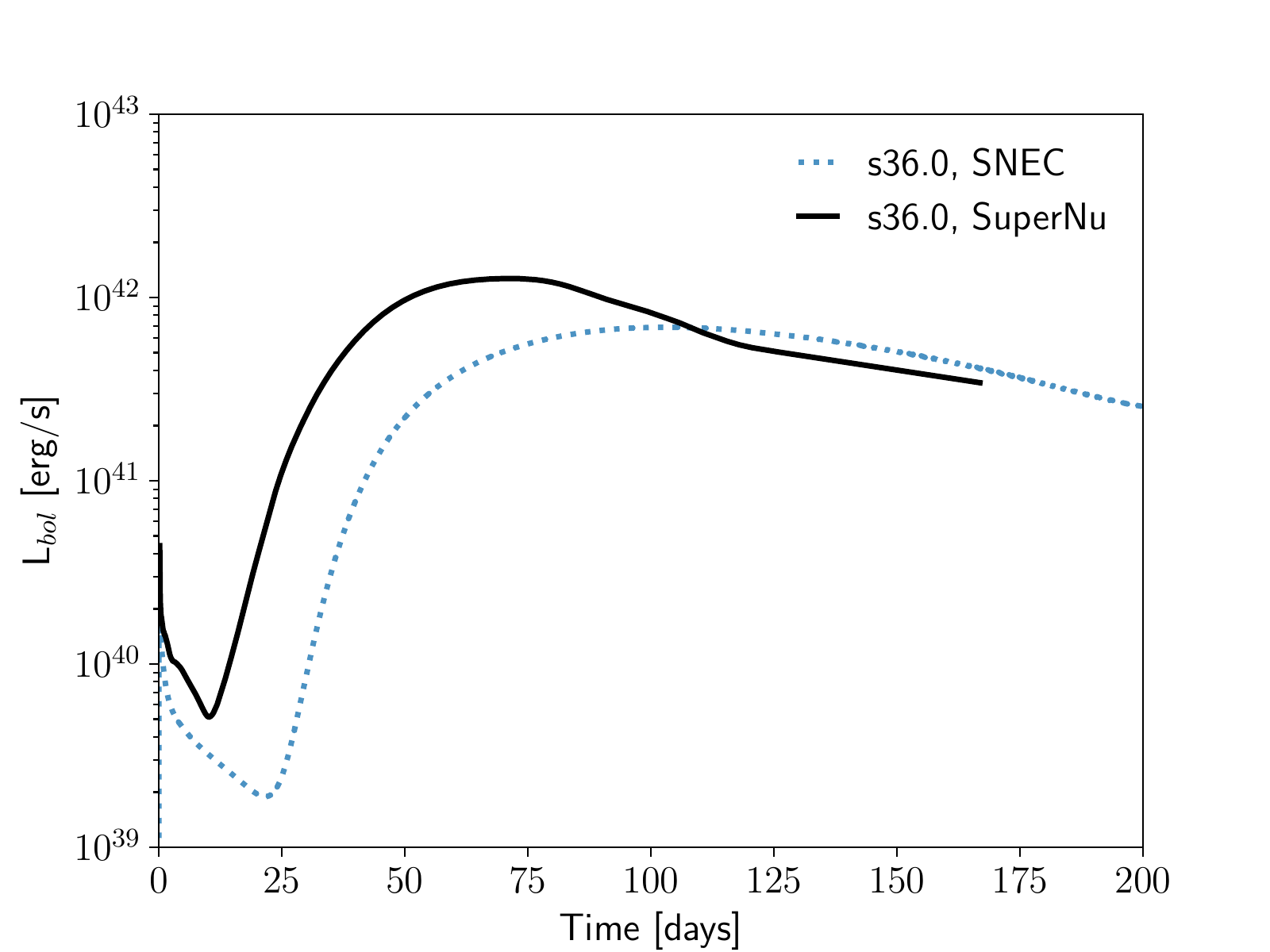}
\includegraphics[width=0.48\textwidth]{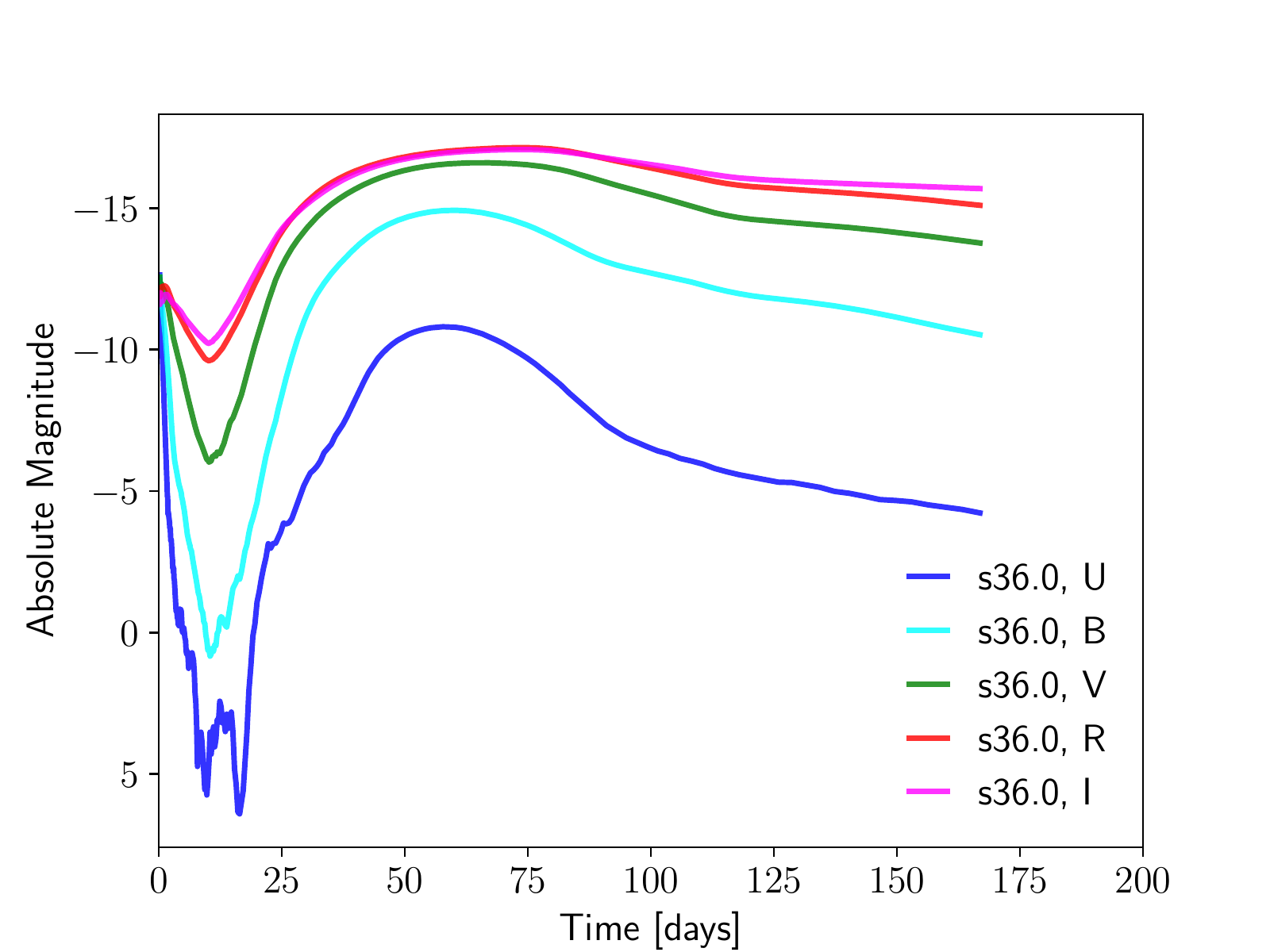}\\
\includegraphics[width=0.48\textwidth]{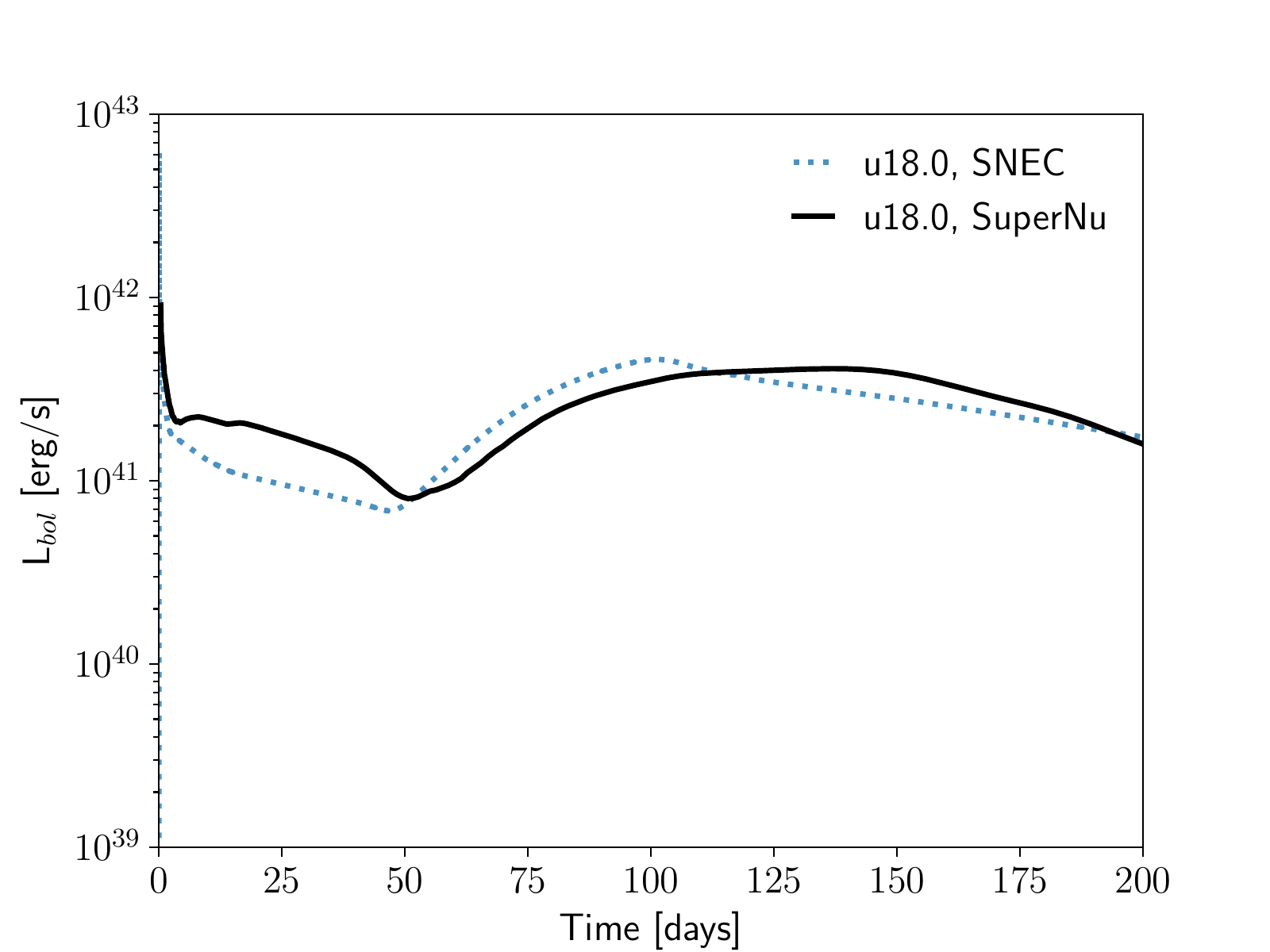}
\includegraphics[width=0.48\textwidth]{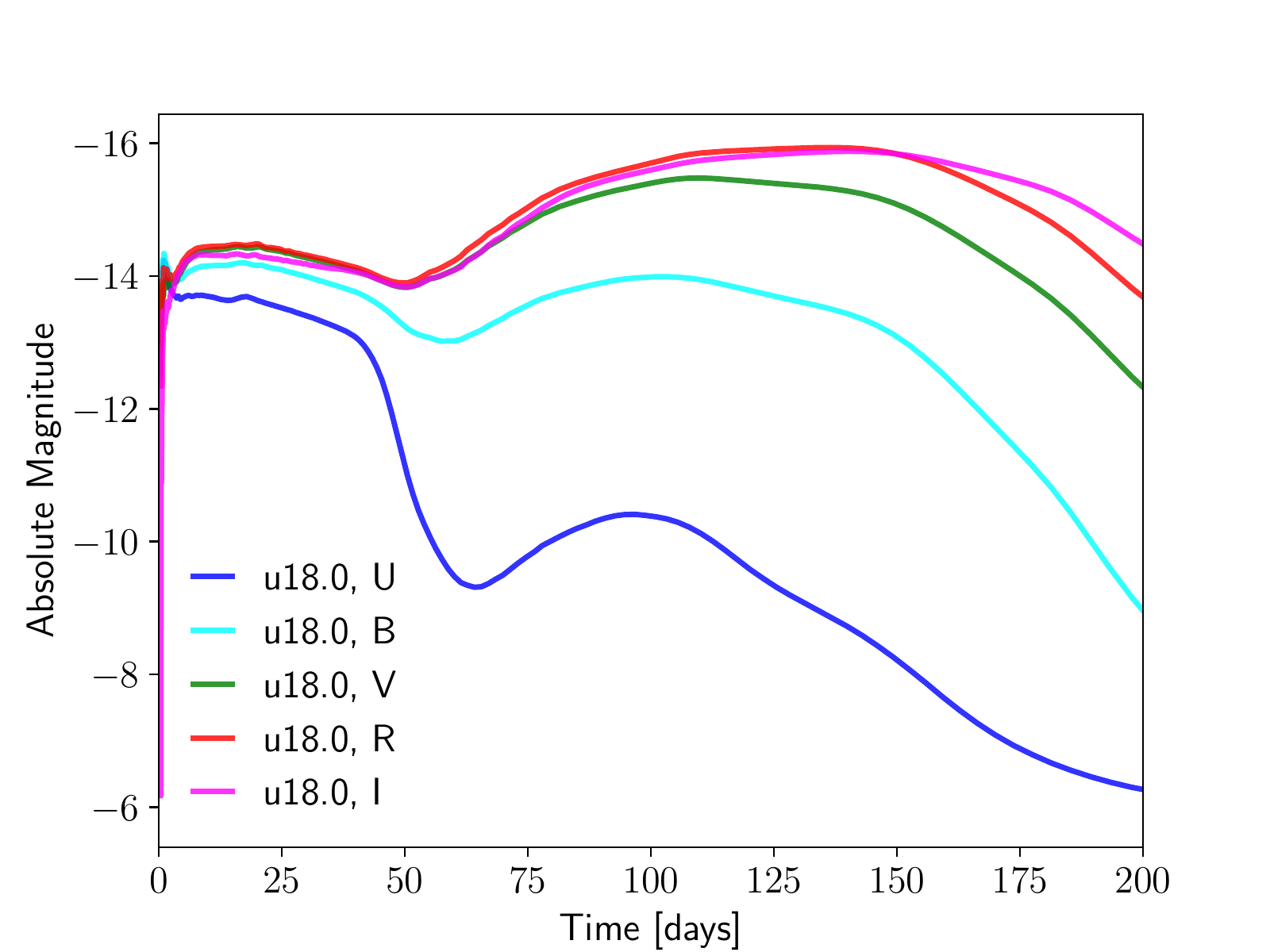}
        \end{tabular}
\end{center}
\caption{Models s18.0 (top), s36.0 (middle) and u18.0 (bottom) belong to the normal Type IIP, stripped-envelope like and SN 1987A-like qualitative classes respectively. The panels on the left show the bolometric light curves for these models computed using \texttt{SuperNu} (solid line) and \texttt{SNEC} (dotted line). The panels on the right show the corresponding absolute magnitudes in different bands computed using \texttt{SuperNu}. 
\label{fig:supernu_lcs}
        }
\end{figure*}

In Figure~\ref{fig:supernu_lcs} (left panel) we show the \texttt{SuperNu} bolometric light curves for three models -- s18.0, s36.0 and u18.0 -- one from each of the three qualitative classes discussed in subsection~\ref{subsec:morphologies}. For comparison, we also plot the \texttt{SNEC} bolometric light curves for these models. While the quantitative results from the two codes show differences, as is expected due to the different treatments of radiative transfer and associated approximations, the qualitative behavior of the models remains unchanged. However, we note the appearance of a small bump at the end of the plateau for the Type IIP light curve of model s18.0, also visible in the different broadbands on the right. Light curves predicted using averaged 3D explosion models do not show this feature, instead reproducing the observed monotonic decline from plateau to radioactive tail. This artifact is a known limitation of light curves calculated from one-dimensional explosion models \citep{Chieffi2003, Young2004, Utrobin.ea:2017} and is related to the development of a density step at the H/He interface during shock propagation. We experimented with artificially smoothing the density profile extracted from \texttt{SNEC} in addition to the boxcar mixing already employed, but this did not remove the feature completely in our simulations. As an extreme case, we tested a model with a fully-mixed composition and a heavily smoothed density profile and found that the feature does disappear, indicating that it is related to both the degree of chemical mixing and the density structure of the model, as previously noted by \cite{Utrobin.ea:2017}.

Figure~\ref{fig:supernu_lcs} also shows the \textit{UBVRI} broadband light curves of our models (right panel). As expected, the \textit{U} and \textit{B} band light curves show a relatively fast decline while the \textit{VRI} bands are flat until day 150. This is due to the blanketing of the spectrum at shorter wavelengths ($\lambda <$ 5000$\angstrom$) by millions of blended iron group lines. For model s18.0, there is no plateau in the \textit{U} band and only the hint of one in the \textit{B} band. However, since the blanketing depends on the metallicity of the progenitor star, we see a different behavior in these bands for the sub-solar metallicity model u18.0. During the hydrogen recombination phase, lasting until $\sim$50 days for this model, there is a plateau-like feature (albeit subluminous) clearly visible in the \textit{U} and \textit{B} bands as well. 

\begin{figure*}
    \centering
    \includegraphics[width=0.8\textwidth]{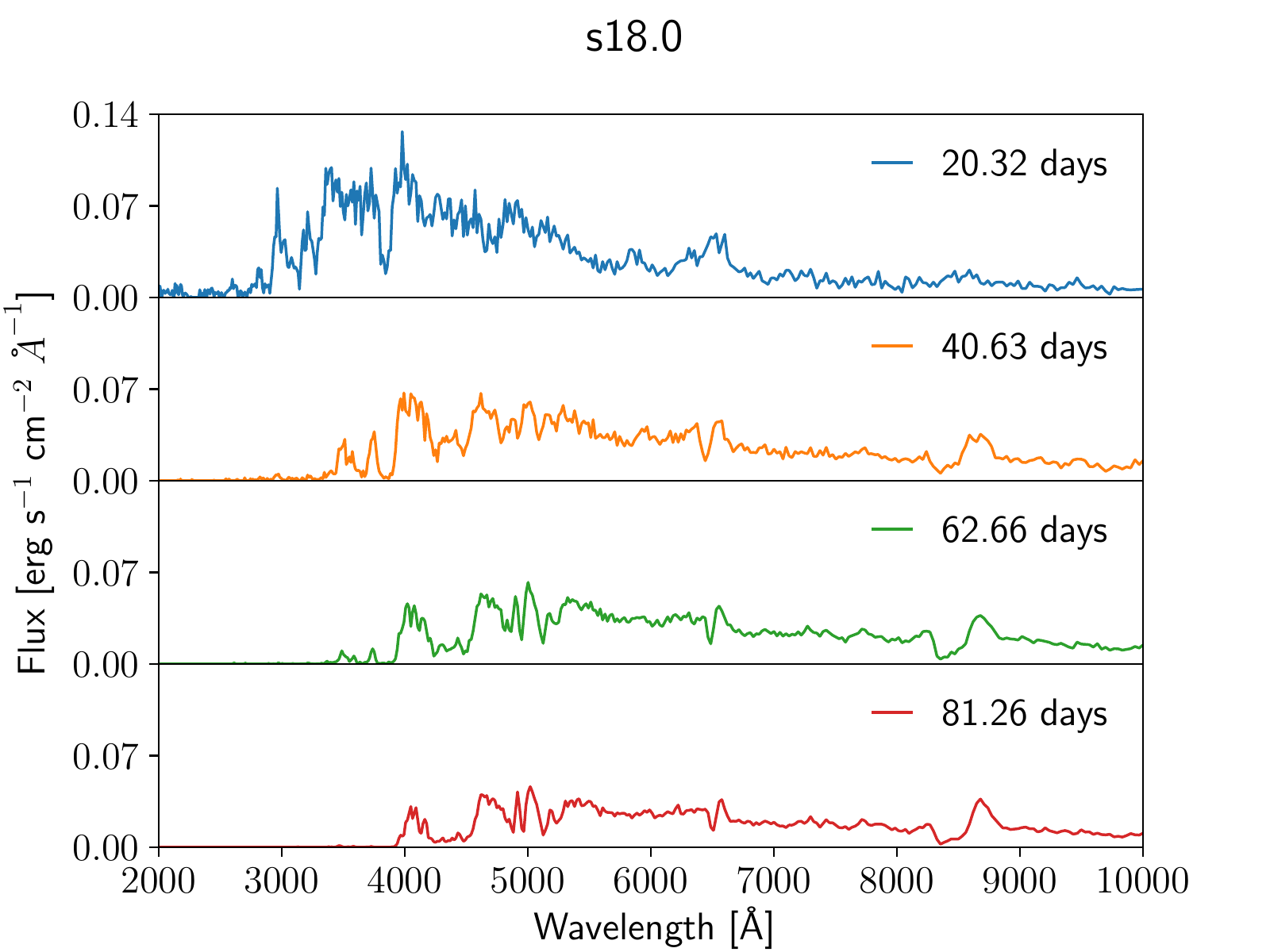}
    \includegraphics[width=0.8\textwidth]{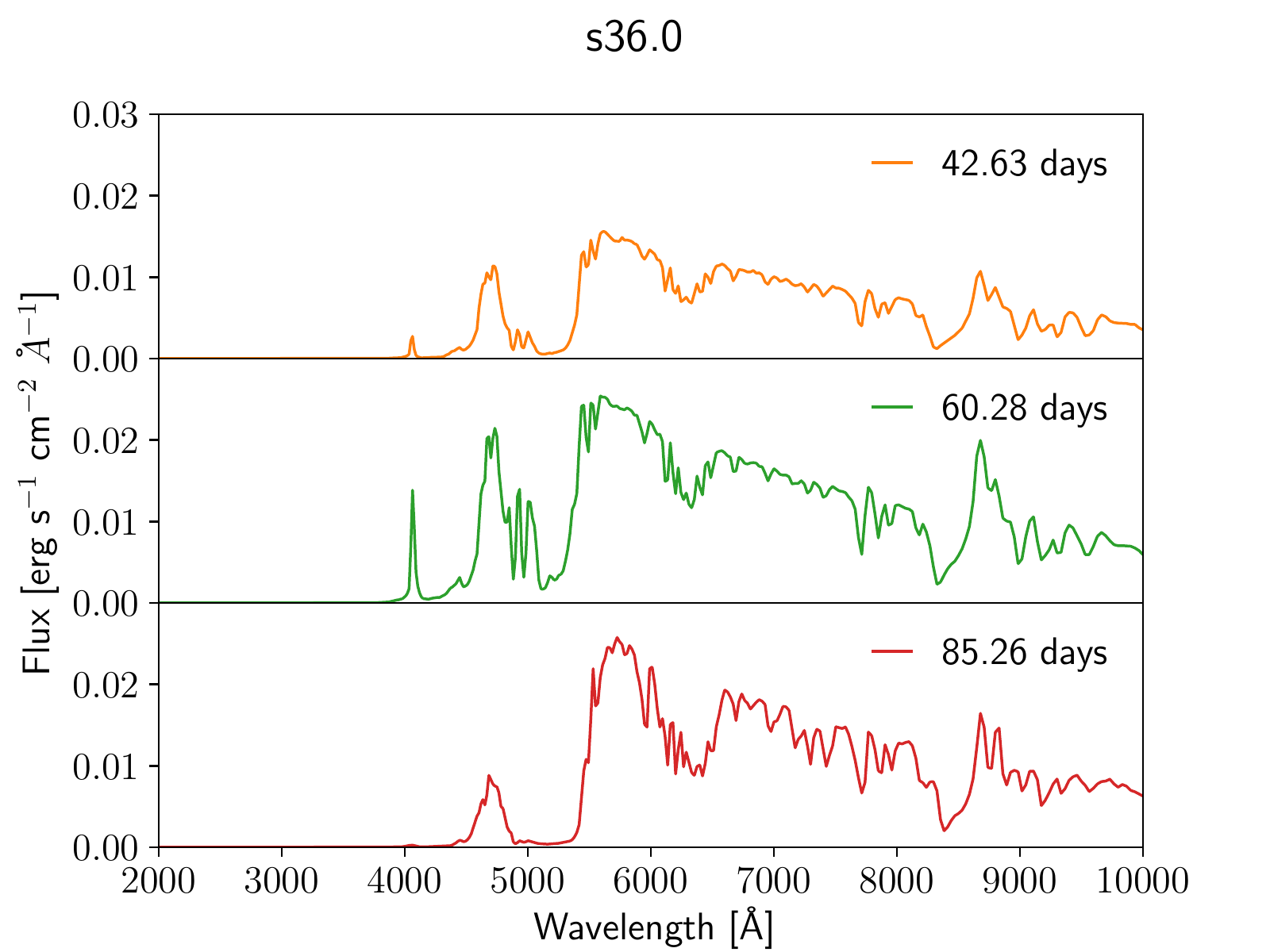}
    \caption{Spectral evolution of solar-metallicity models s18.0 (top) and s36.0 (bottom). The flux is calculated at an observing distance of 10~pc.} 
    \label{fig:s_spectra}
\end{figure*}

\begin{figure*}
    \centering
    \includegraphics[width=0.8\textwidth]{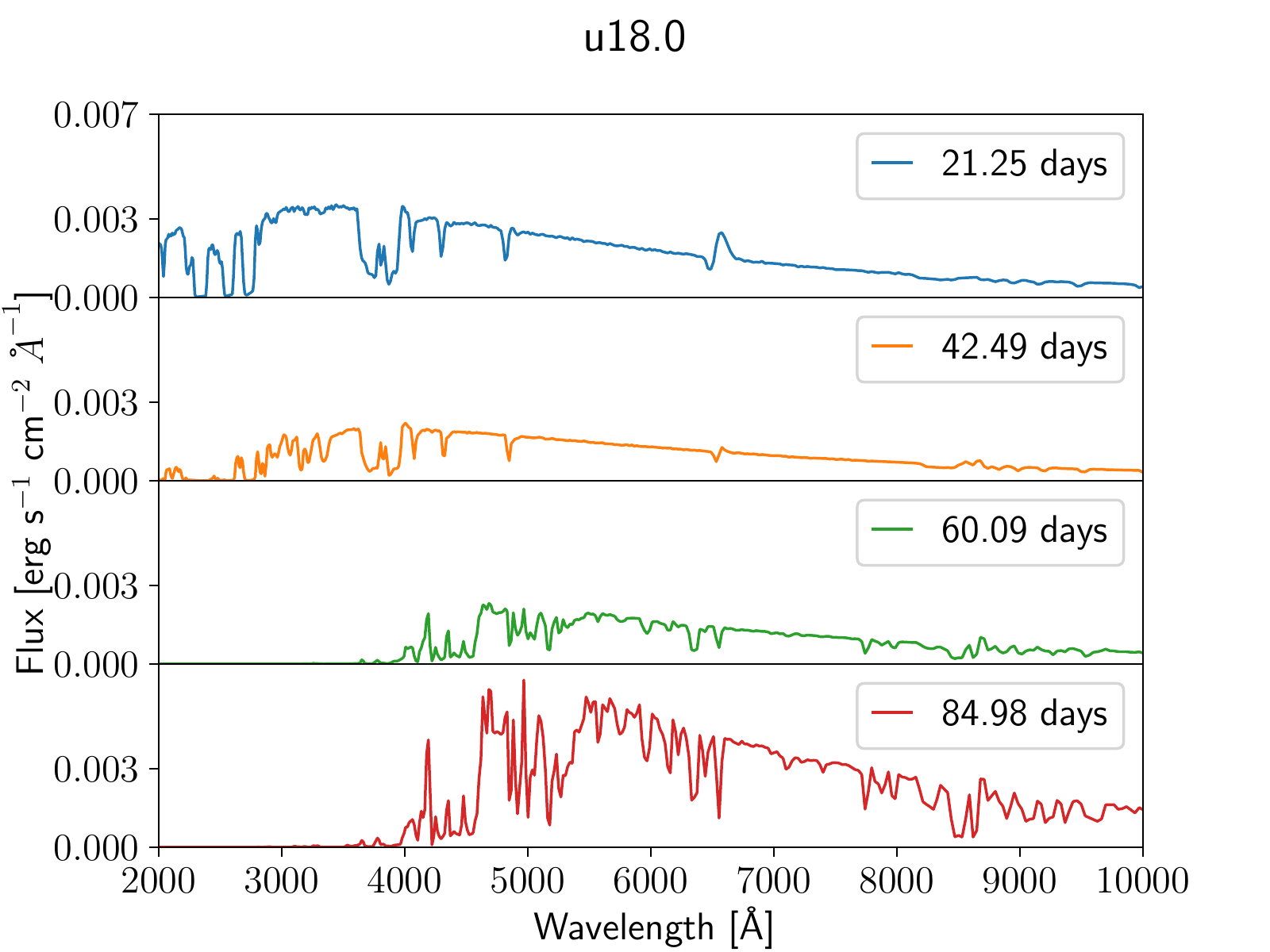}
    \caption{Spectral evolution of sub-solar metallicity model u18.0. The flux is calculated at an observing distance of 10~pc.}
    \label{fig:u_spectra}
\end{figure*}

\begin{figure}
    \centering
    \includegraphics[width=0.48\textwidth]{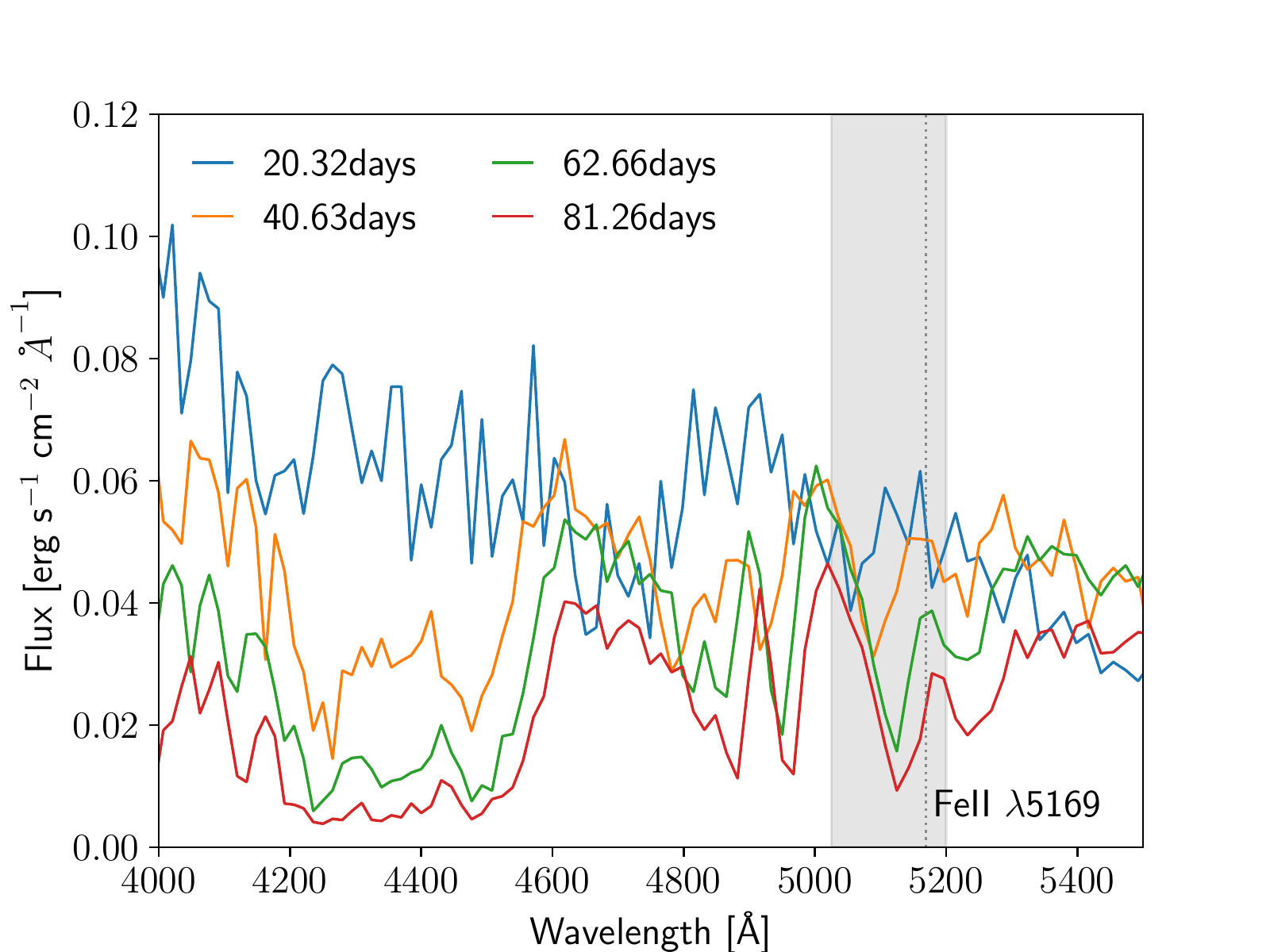}
    \caption{Iron line blanketing in the 4000--5500$\angstrom$ range. The shaded gray region shows the minima associated with the Fe $\scriptstyle\mathrm{II}$ $\lambda$5169.}
    \label{fig:fe2_line}
\end{figure}

The iron group blanketing at shorter wavelengths can also be seen in the spectra of these models, shown in Figures~\ref{fig:s_spectra} and \ref{fig:u_spectra}. The flux at shorter wavelengths decays significantly over time for all models, but the decay is much more rapid for the solar-metallicity models as opposed to the sub-solar metallicity model. The steady decay in flux over time in the 4000 -- 5500 $\angstrom$ spectral range is clearly visible for model s18.0 in the top panel of Figure~\ref{fig:s_spectra}, and is shown in more detail in Figure~\ref{fig:fe2_line}. This happens due to the doppler-shifted Fe $\scriptstyle\mathrm{II}$ and Ti $\scriptstyle\mathrm{II}$ absorption lines. The doppler shift decreases as the photosphere moves inward and the minima of the Fe $\scriptstyle\mathrm{II}$ $\lambda$5169 line moves to longer wavelengths. Models s36.0 (highly stripped) and u18.0 (massive H-envelope) are both much more compact than model s18.0 and behave differently. We do not plot the flux at $\sim$20 days for model s36.0 in Figure~\ref{fig:s_spectra} since it is extremely low. The bolometric luminosities of both models start rising shortly after day 10 and day 50 respectively, primarily due to radioactive heating, and we see an increase in the flux above $\sim$4000$\angstrom$ before it decays once again at late times.

Our late-time spectra do not display strong H$\alpha$ emission although this line is observed for Type II supernovae. This is a result of the assumption of LTE in \texttt{SuperNu}. Non-thermal processes are key to the production of this line as well as other lines like He$\scriptstyle\mathrm{I}$ 10830$\angstrom$. Finally, we note that the spectra presented here have been smoothed (via time-averaging) for the sake of noise reduction, however, we make both the raw and the smoothed spectral data publicly available online.

\subsubsection{Comparison with Observations: SN~1987A and SN~1999em}

A few supernova events, namely SN~1987A and SN~1999em, have a rich dataset of light curves against which predictions and models can be compared. Here we demonstrate how our pipeline and light curve database can be used to better understand an observation. For each event, SN~1987A and SN~1999em, \replaced{we find the models that best fit the bolometric light curve in our database.}{we find in our database the models that most closely resemble the observed bolometric light curves.} If one desires to obtain an ever better match of a chosen synthetic light curve with the observed light curve of a particular supernova, one \replaced{can}{would have to} change \emph{by hand} certain quantities, instead of keeping them at their a priori set default.

When examining an \textit{ensemble} of progenitors and models, it is important to stick to a single, e.g., mixing prescription for \textit{all} models, to enable consistent comparison. However, when examining a \textit{single} model, a variation of e.g.\ the mixing prescription allows us to better understand the relative importance of these quantities and their underlying physical processes within a given event.

We begin with a discussion of SN~1987A. The PUSH model was originally calibrated against the explosion energy and nucleosynthesis for SN1987A using a red supergiant progenitor \citep{push2}. While this progenitor (model s18.8 in our study) reproduces the observed explosion energy and $^{56-58}$Ni yields of SN 1987A, it results in a standard Type IIP light curve, as is expected given its radius and envelope structure which are typical of red supergiants. However, we know that the progenitor of SN1987a was a blue supergiant. We therefore now compare to the blue supergiant model presented in \citet{menon2017}, which also produces a good fit for explosion energy and $^{56-58}$Ni yields in PUSH, as shown in \citet{frohlich2019}. \citet{utrobin2021} show that three-dimensional simulations of the same progenitor can produce bolometric light curves with a reasonable match, but for larger explosion energies than predicted by the existing PUSH calibration.

\begin{figure}
\includegraphics[width=0.48\textwidth]{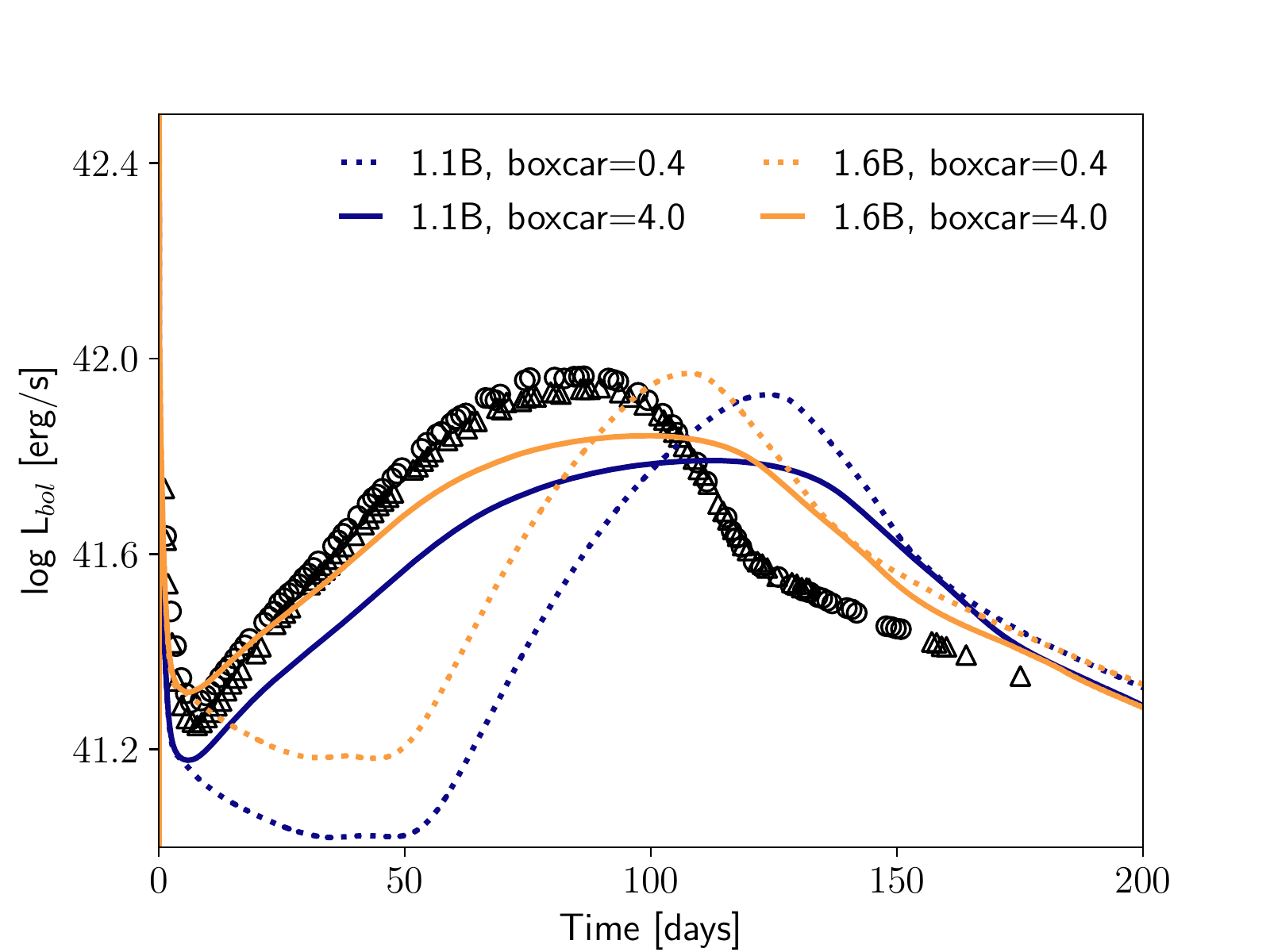}
\caption{Bolometric light curve for SN1987A and the blue supergiant model with different explosion energies (in Bethe) and boxcar widths (given in $M_{\odot}$). The computed light curves are compared with the observed bolometric luminosity of SN 1987A obtained by \citet{Catchpole1987} (open circles) and \citet{Hamuy1988} (open triangles).
\label{fig:87a_compare}
}
\end{figure}

Figure \ref{fig:87a_compare} compares candidate synthetic light curves against the observed one. Here we take the blue supergiant model described above and run it self-consistently (blue dotted line). In addition, we perform calculations with hand-tuned choices (higher explosion energy, orange) and/or enhanced boxcar mixing (solid lines). We find that more vigorous boxcar mixing both reduces time to peak luminosity --- which is needed to reach the SN1987a early peak --- and reduces the peak luminosity. Moreover, even with increased explosion energy, which increases the peak luminosity, boxcar mixing cannot bring our synthetic light curves into agreement with the observed one. Interestingly, the late-time abundances presented in \citet{utrobin2021}, which were produced by a full three-dimensional simulation, cannot be achieved with boxcar mixing. In other words, a more realistic mixing prescription, perhaps inspired by the effects of 3D turbulence may be required to produce a better match for this progenitor.

\begin{figure*}
        \includegraphics[width=0.48\textwidth]{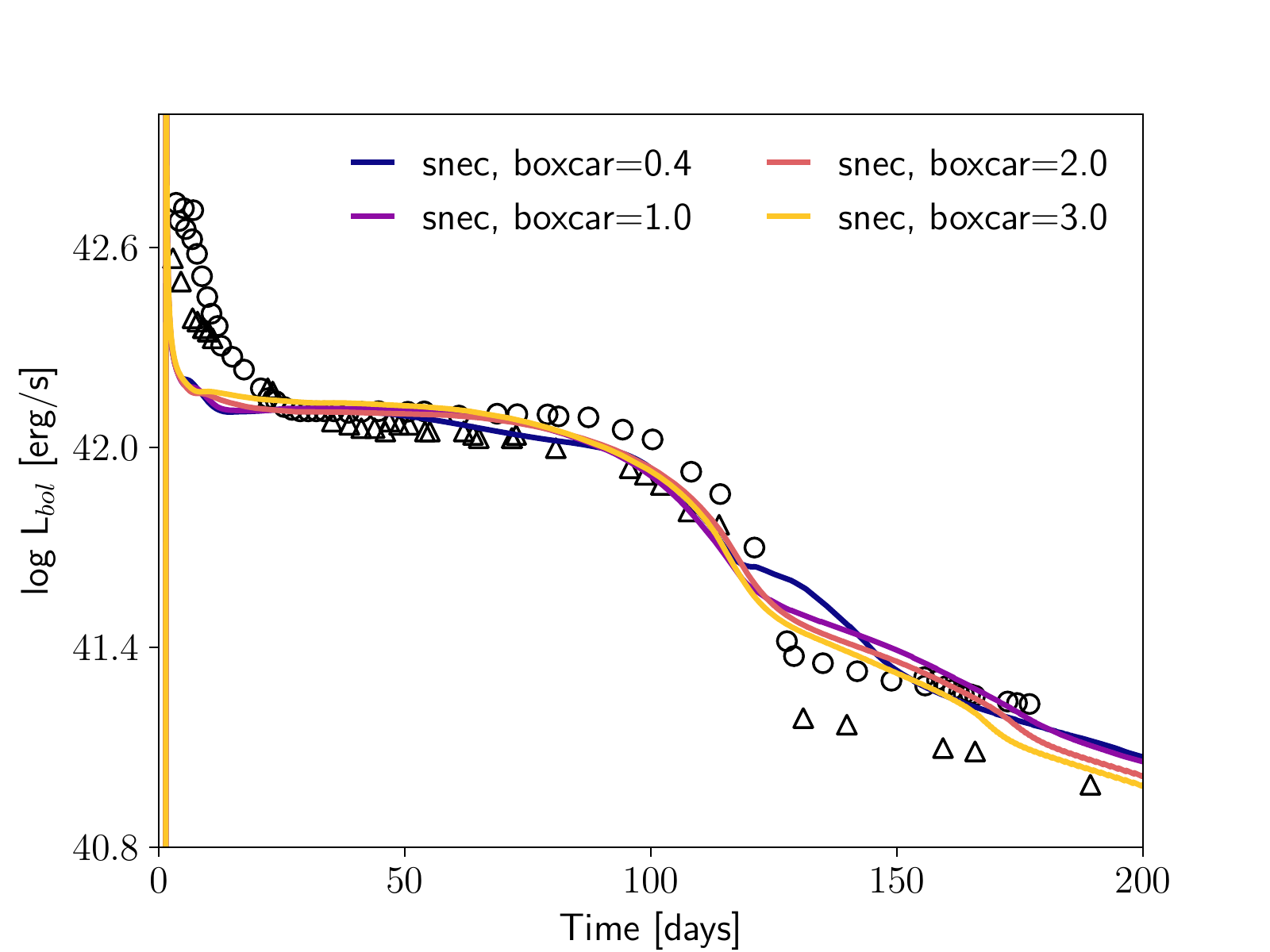}
        \includegraphics[width=0.48\textwidth]{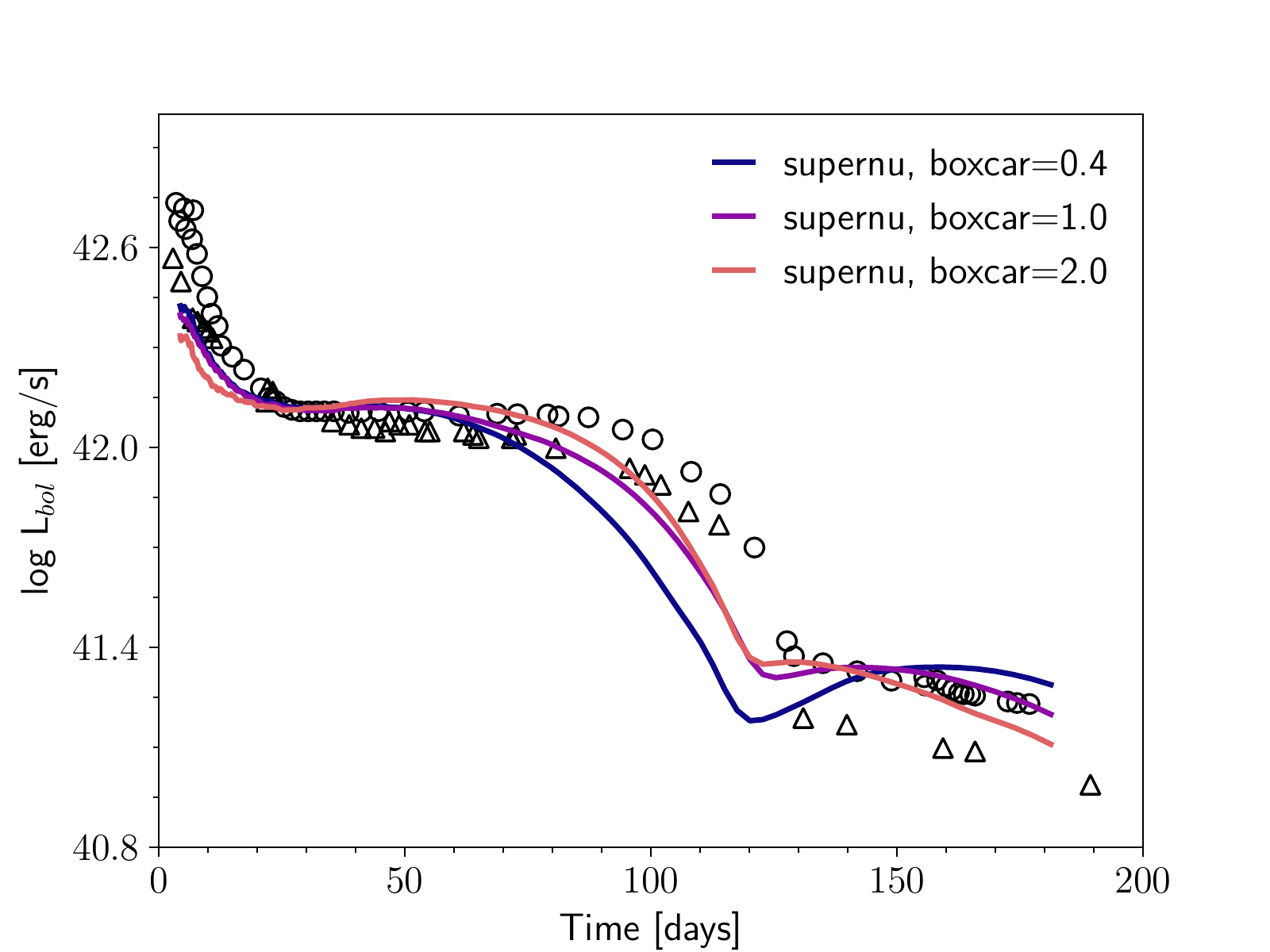}
        \caption{Synthetic light curves for model s12.0 obtained by varying the degree mixing i.e. boxcar width (given in $M_{\odot}$) used in \texttt{SNEC} simulations. The bolometric light curve from \texttt{SNEC} is plotted on the left and the equivalent light curve from \texttt{SuperNu} on the right. The observational data for SN 1999em are plotted as open triangles (provided by V. Utrobin) and open circles (taken from Figure 14 of \citet{Morozova2015}). 
        \label{fig:99em_compare}
        }
\end{figure*}

\begin{figure*}
\begin{center}
        \begin{tabular}{cc}
        \includegraphics[width=0.48\textwidth]{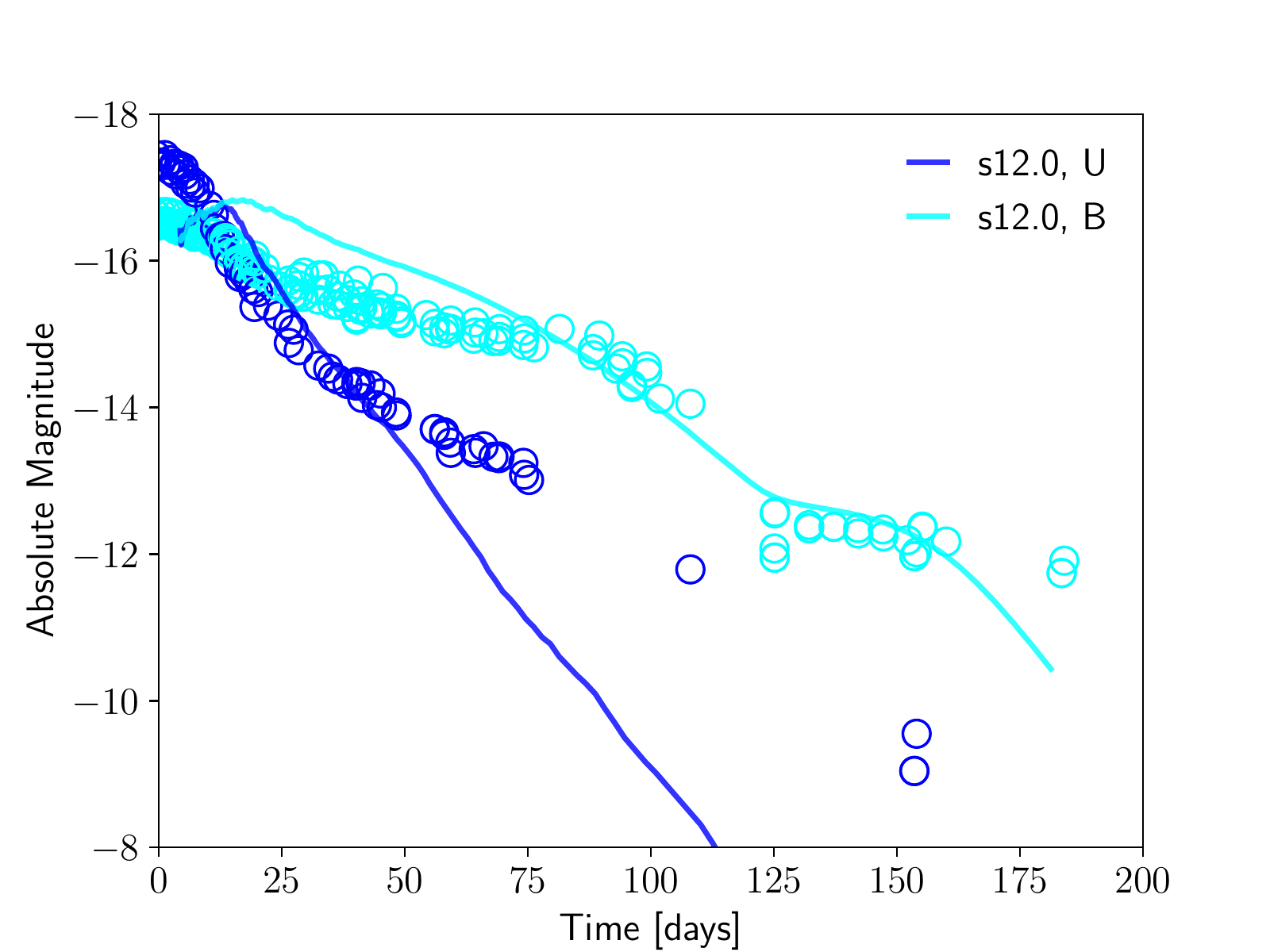}
        \includegraphics[width=0.48\textwidth]{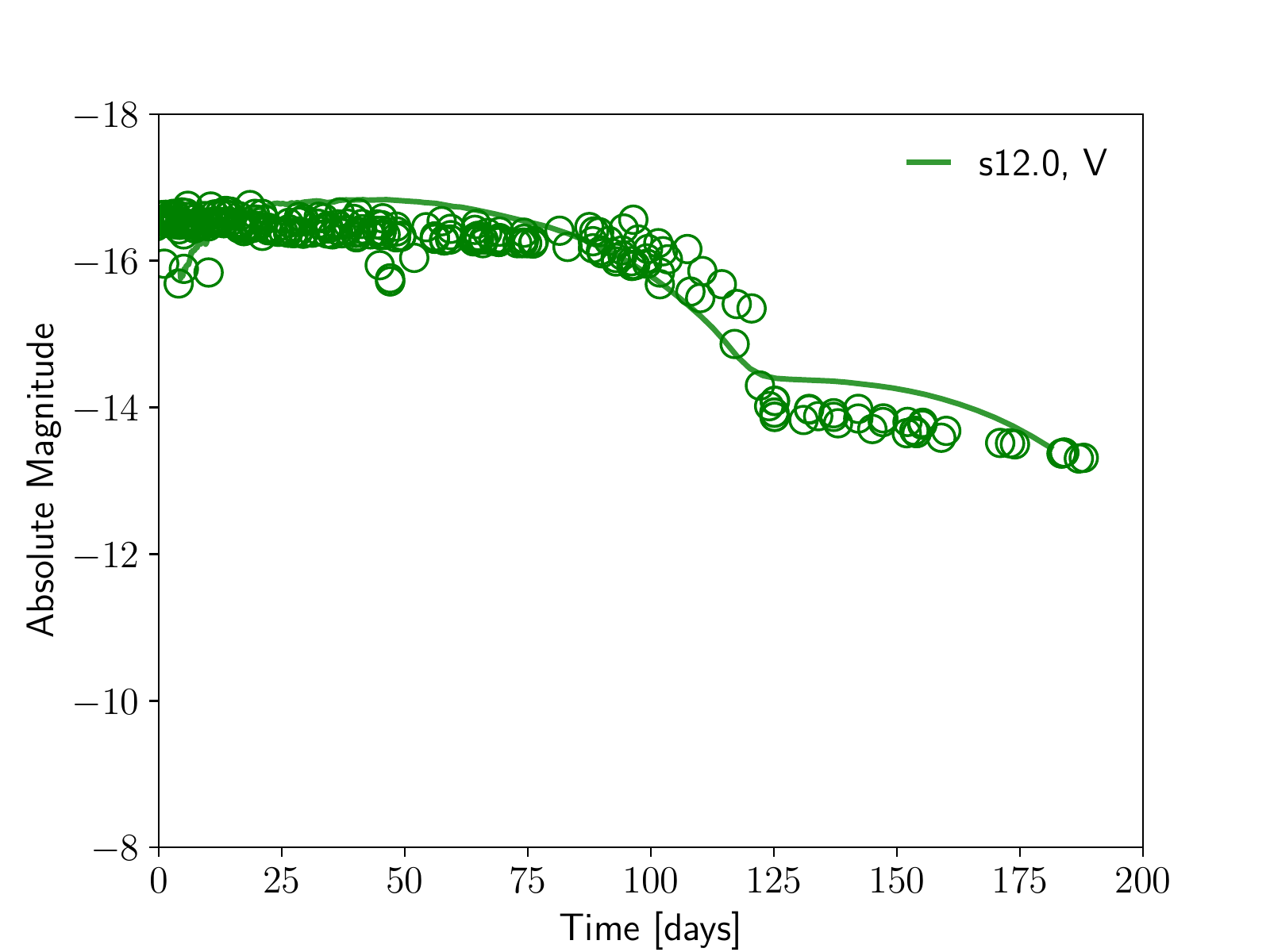}\\
        \includegraphics[width=0.48\textwidth]{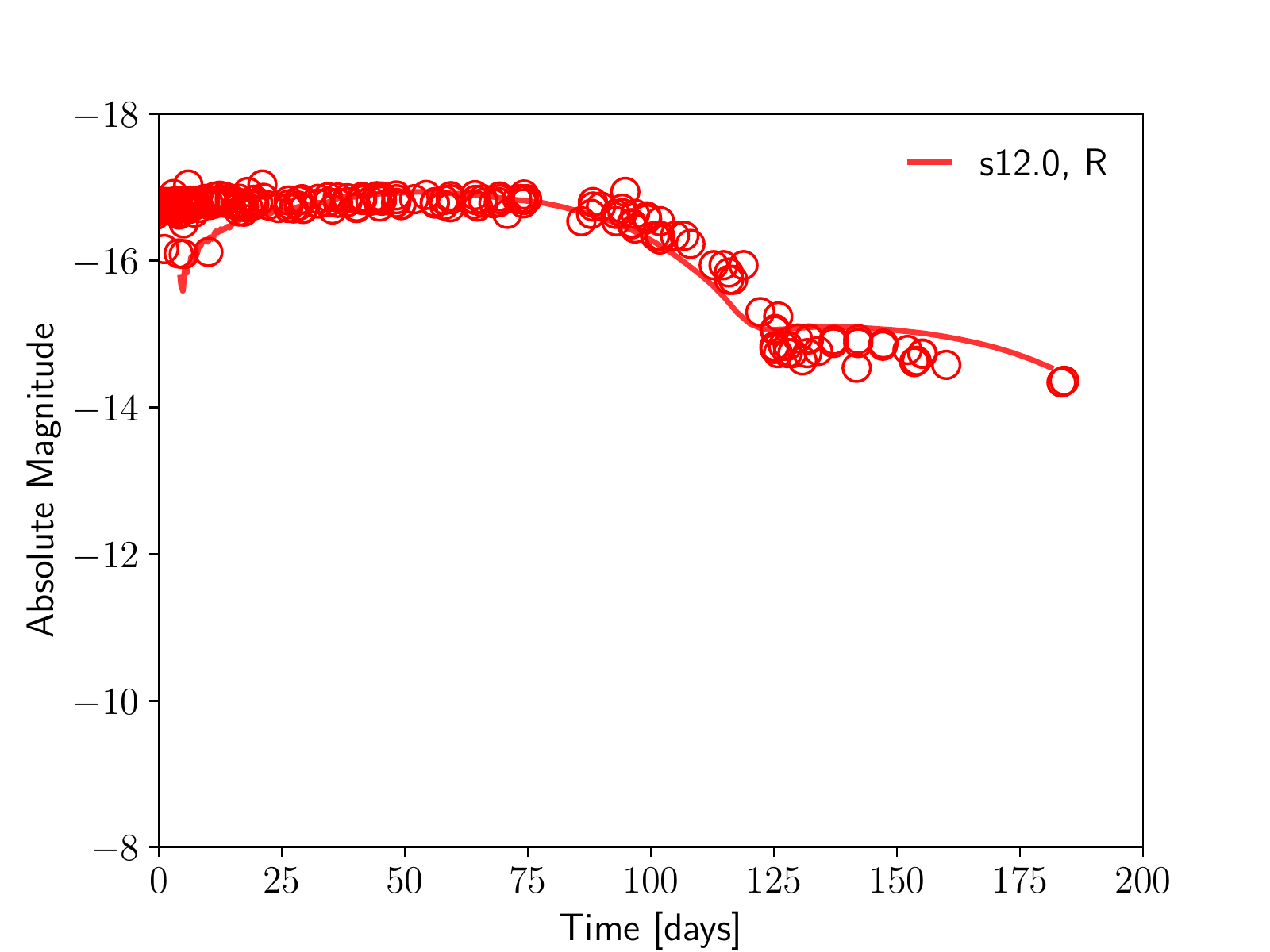}
        \includegraphics[width=0.48\textwidth]{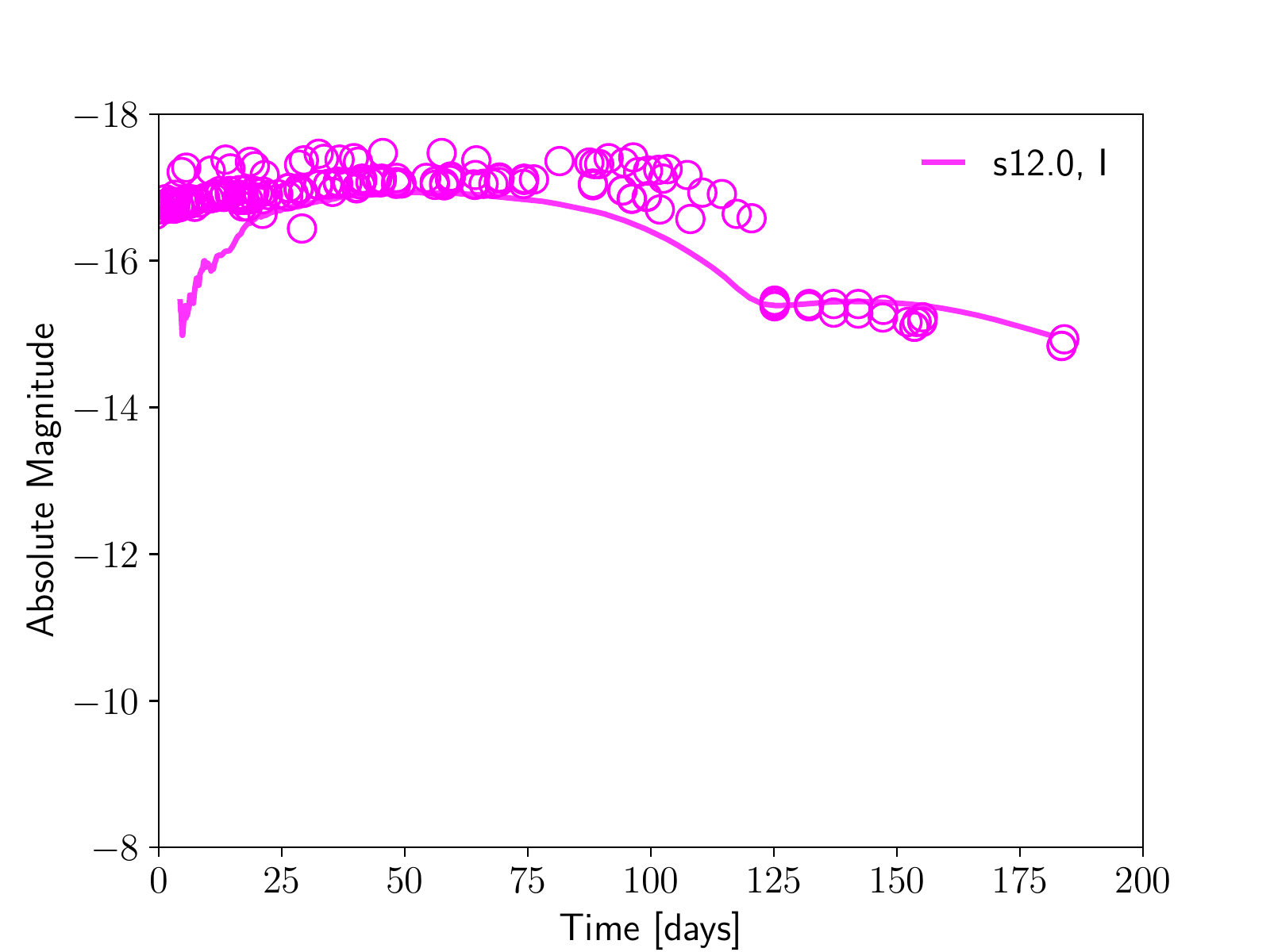}
        \end{tabular}
        \caption{\textit{UBVRI} light curves of model s12.0 with enhanced mixing employing a boxcar width of 1$M_{\odot}$, compared with the corresponding band light curves of SN 1999em (open circles). The apparent magnitude data for SN 1999em are taken from the Open Supernova Catalog \citep{Guillochon2017} and the (approximate) absolute magnitudes are computed adopting a luminosity distance of 11.7 Mpc \citep{Leonard2003} for consistency with the bolometric data (open triangles) in Figure \ref{fig:99em_compare}. The combined dataset is obtained from \citet{Filippenko1999, Stritzinger2002} and \citet{Krisciunas2004}.
        \label{fig:s12_broadbands}
        }
\end{center}
\end{figure*}

In the case of SN~1999em, we found that many low mass Type IIP progenitors in the 11--13 $M_\odot$ range, from both the s- and u-series, were reasonably good \replaced{fits}{matches}. This is consistent with the range of progenitor masses derived for this event in \citet{Smartt2002} and \citet{Smartt2009}. One particularly good \replaced{fit}{match}, our s12.0 model, is presented in Figure \ref{fig:99em_compare}. Again, we perform additional calculations where we vary our boxcar mixing width to obtain --- for SN~1999em --- an improved match with the bolometric luminosity beyond day 50 and a smoother transition to the radioactive tail. The left panel of Figure \ref{fig:99em_compare} shows the effect of increasing the boxcar width. We find that using a 1$M_{\odot}$ boxcar rather than our standard setting of 0.4$M_{\odot}$ removes the unobserved knee-like feature in the light curve as it transitions from the plateau to the tail. Increasing the boxcar width further does not have any significant effect on the light curve morphology although it continues to produce a small shift in the luminosities of the plateau and tail. In the right panel, we show the synthetic light curves produced by \texttt{SuperNu} for our standard mixing prescription, our best fit (employing a boxcar width of 1$M_{\odot}$) and a more extreme case to capture the effects of boxcar mixing. The corresponding magnitudes in different bands are shown in Figure \ref{fig:s12_broadbands}. We find reasonably good agreement across all bands except the \textit{U}-band which decays much faster beyond day 50 than was observed for SN 1999em. The flux in short wavelength bands is very sensitive to the amount and distribution of metals and discrepancies can arise due to any number of reasons ranging from a mismatch between the mixing prescription and true 3D mixing, the LTE approximation used in \texttt{SuperNu}, or the metallicity of the progenitor model itself. It is possible that a lower metallicity model around the same ZAMS mass may produce a better match, however, a detailed exploration of all possible fits to this particular supernova is beyond the scope of this work.

Broadly, the boxcar mixing prescription seems more consistent with Type IIP models than the SN1987A-like. This may imply that mixing is progenitor and engine dependent and that the next iteration of this pipeline should have a more sophisticated mixing prescription to satisfy this constraint.

\added{The basic properties of the pre-explosion models (pre-SN mass, pre-SN radius) and of the explosion (Ni mass, explosion energy, and mass of the neutron star) are presented in Table~\ref{tab:summary_obs}. The table also includes a selection of boxcar widths and the resulting velocities (Ni velocity and velocity of optical depth $\tau =2/3$). The boxcar width was varied for two specific models (b15-7 and s12.0) to outline how \emph{hand-tuning} of parameters might be used to obtain a better fit to a specific supernova. The velocity of $^{56}$Ni in the homologously expanding ejecta is also presented for the different levels of boxcar mixing and at two different mass-fraction cuts. A boxcar width of 4$M_{\odot}$ corresponds to a well-mixed composition for these progenitors. Increasing initial mixing of $^{56}$Ni to higher mass coordinates---and thus larger velocities---leads to more prominent contribution of the radioactive decay luminosity to the total luminosity at early times. At late times, the radioactive decay luminosity becomes smaller since more gamma-rays escape without getting absorbed \citep{Morozova2015}.}

\input{summary_table_obs}

\section{Discussion}
\label{sec:discussion}

Multi-D simulations are a well-suited and necessary tool to investigate the underlying mechanism of CCSNe. Unfortunately, the full solution to the CCSN problem in a self-consistent way is still not converged yet, either within a single code or between groups. The growing set of 2D and 3D CCSN explosion models \citep[see e.g.][]{janka12,burrows13,nakamura15,janka16,bruenn16,Burrows2018} seems to indicate that, while 2D simulations tend to explode more easily, in 3D the resulting explosion energies may be higher, as is shown e.g.\ in  \cite{takiwaki14,lentz15,melson15,Mueller15,janka16,hix16}. Multi-dimensional simulations are also computationally intensive. They are prohibitively expensive if large spatial domains are required and/or if simulations are required to follow the evolution for long timescales beyond the onset of the explosion -- such as is needed for example for the prediction of light curves and spectra. Thus, given the present status, it is still too early to provide complete predictions from multi-D simulations for an extended sample of progenitor stars. 

The path taken in this study bridges the gap between explosion simulations and observations in a \emph{at the present time} computationally feasible way for large samples. The strength of our approach is the predictive power within the framework we use. Specifically, only \emph{one} model is hand-tuned to a particular outcome (matching SN 1987A in explosion energy and $^{56}$Ni yields). This calibration sets the free parameters of the PUSH framework for \emph{all} models \emph{a priori}. The \texttt{Agile} simulations within the PUSH framework then \emph{predict} the explosion energy, mass cut (thus ejecta mass), and the conditions in the ejecta (thus the amount of $^{56}$Ni). These predicted properties are then propagated through out pipeline. For example, the explosion energy predicted from \texttt{Agile} is used in the \texttt{SNEC} simulations without further tuning nor energy injection.

The pipeline used to link massive progenitor models through explosion to electromagnetic counterparts utilizes several software instruments: \texttt{Agile} for the explosion simulations, \texttt{SNEC} to evolve the supernova to homologous expansion, and \texttt{SuperNu} for the late time spectra and light curves. Each of these software instruments has its own approximations, appropriate for the phase of the evolution it is used for. These approximations also present opportunities for future work to improve upon. For the study presented here, we focused on a process that does not involve any hand-tuning along the way. Once the initial calibration of the PUSH framework was complete (see \cite{push1,push2}, the outcomes of the \texttt{Agile} simulations were propagated through the pipeline. Due to known short-comings of the abundance distribution in velocity space from 1D simulations, we employ the boxcar mixing prescription at the point of mapping from \texttt{Agile} to \texttt{SNEC}. We emphasize that we use the standard settings for boxcar mixing as used in the literature and do \emph{not} tune it to achieve a desired outcome in the electromagnetic observables.
In summary, this work resulted in an ensemble of synthetic observables (light curves and spectra) which have been obtained in a self-consistent pineline from explosion to observables. This approach is complementary to hand-tuning parameters for an individual model to obtain the best fit to an observed supernova.

\section{Summary and Conclusions}
\label{sec:summary}

To summarize, we compute bolometric and broadband synthetic light curves
and spectra for a wide variety of progenitors exploded self-consistently in spherical symmetry with the PUSH framework. For this, we map the output from \texttt{Agile} to \texttt{SNEC} and from \texttt{SNEC} to \texttt{SuperNu}. We discuss the subtleties of these mappings in the appendices.

We find that the bolometric light curves can be categorized based on properties of the progenitor. In particular, the mass in the hydrogen envelope and the radius of the star at the end of its life determine the qualitative structure of the light curve. We find roughly three categories of light curve, which we call ``normal Type IIP'', ``stripped envelope-like'', and ``SN1987A-like''. 

The normal Type IIP light curves are seen in models with massive hydrogen envelopes and exhibit an extended plateau powered by hydrogen recombination. The stripped envelope-like light curves mostly emerge from solar-metallicity stars, such as our more massive s-series models, which experience substantial mass loss during their lifetimes. These models exhibit a rapid dip and then rise in bolometric luminosity, powered by nickel decay. The SN1987A-like light curves come from our low-metallicity and zero-metallicity stars, which retain massive hydrogen envelopes but are very compact. 

As a proof of concept, we perform a sensitive-variable analysis both on the normal Type IIP models (with plateaus) and on the models that do not show plateaus (stripped envelope-like and SN1987A-like). As expected,
the most important factor in the quantities we analyzed is the amount of $^{56}$Ni.
Interestingly, the presence of a hydrogen envelope does not equal the presence of an extended plateau. For models that \textit{do} have a plateau, the plateau length appears at least weakly correlated with the hydrogen envelope mass. In the stripped envelope case,
the behavior of models s33.0--s75.0 is weakly correlated with the degree of stripping, as these stars have lost all of their hydrogen and most of their helium. Due to our small sample size, we cannot make strong statements about these weaker correlations.

Our \texttt{SNEC} bolometric light curves broadly agree qualitatively with the other studies in literature that calculate light curves for stellar models without attempting to match a particular observation. Our light curves look quite similar to those presented in \citet{sukhbold16} (who use a piston to induce the explosion) for the normal Type IIP models with ZAMS mass $\lessapprox$ 22 $M_{\odot}$. For stripped-envelope models with ZAMS mass $\gtrapprox$ 30 $M_\odot$, we find longer time scales for rise and decay. To our knowledge, \citet{sukhbold16}
does not show any SN1987A-like models that transition between the two regimes. Our s12.0 model appears similar to the $12 M_{\odot}$ models presented in \citet{Kozyreva.IIp:2019}. 

Overall, our spectra show reasonable agreement with the few synthetic photospheric spectra available in the literature \citep{Kasen2009, Dessart2013}. Our broadband light curves and spectra show the reddening characteristic of iron group blanketing. The Fe $\scriptstyle\mathrm{II}$ and Ti $\scriptstyle\mathrm{II}$ absorption lines are also clearly visible between 4000 and 5000 $\angstrom$. Non-thermal processes are required to produce the H$\alpha$ line. Unfortunately, since \texttt{SuperNu} assumes local thermal equilibrium, this line is absent in our nebular spectra.

Using the examples of SN~1987A and SN~1999em we demonstrate how to use the data of this work in comparisons to observations, and how our data are a complementary approach to hand-tuning input parameters to obtain the best fit between the synthetic and observed light curves for a specific supernova.

Our work constitutes both an analysis of the broad features of electromagnetic counterparts to CCSNe and a database against which observational data can be compared. It also opens the door for more detailed analyses of the spectra and of the collective properties of these electromagnetic signals. Moreover, we present a first-of-its-kind pipeline from a progenitor model, through a self-consistent explosion in spherical symmetry, to synthetic light curves and spectra, applied here to a large number of supernova models. This pipeline allows for an expanded database as more models are produced. Most excitingly, we expect a similar procedure to apply straightforwardly in the multi-dimensional case as these models become more affordable and hence more numerous. 

\begin{acknowledgments}  
We thank Chris Fryer, Valeria U. Hurtado, Patrick Killian, Oleg Korobkin, Nicole Lloyd-Ronning, Albino Perego, Brooke Polak, and Mohira Rassel for useful discussions and we thank Victor Utrobin for providing bolometric luminosities for SN~1987A and SN~1999em. The work at NC State was supported by United States Department of Energy (DOE), Office of Science, Office of  Nuclear Physics under Award DE-FG02-02ER41216. 

The work at Los Alamos was supported by the US DOE Office of Science and the Office of Advanced Scientific Computing Research via the Scientific Discovery through Advanced Computing (SciDAC4) program and Grant DE-SC0018297 and through the Los Alamos National Laboratory (LANL). Additional funding was provided by the Laboratory Directed Research and Development Program, the Center for Space and Earth Science, and the Center for Nonlinear Studies at LANL under project numbers 20190021DR, 20180475DR (TS), and 20170508DR. This research used resources provided by the LANL Institutional Computing Program. LANL is operated by Triad National Security, LLC, for the National Nuclear Security Administration of U.S. DOE under Contract No. 89233218CNA000001. LANL has cleared this article for unlimited release, LA-UR-20-26099.

We are grateful to the countless developers contributing to software
projects on which we relied in this work, including Agile \citep{Liebendoerfer.Agile}, CFNET \citep{cf06a}, 
SNEC \citep{Morozova2015}, SuperNu \citep{Wollaeger2014}, Python \citep{rossumPythonWhitePaper}, the GNU compiler
\citep{stallman2009using}, numpy and scipy\citep{numpy,scipyLib}, and Matplotlib \citep{matplotlib}. We also thank the OPAL opacity project for making opacity tables readily available. This research made extensive use of the SAO/NASA Astrophysics Data System (ADS).
\end{acknowledgments}

\software{Agile \citep{Liebendoerfer.Agile}, CFNET \citep{cf06a}, 
SNEC \citep{Morozova2015}, SuperNu \citep{Wollaeger2014}, Python \citep{rossumPythonWhitePaper}, the GNU compiler
\citep{stallman2009using}, numpy and scipy\citep{numpy,scipyLib}, and Matplotlib \citep{matplotlib}}.

\clearpage

\bibliography{main}{}
\bibliographystyle{aasjournal}

\appendix

\section{Boxcar smoothing}
\label{app:mixing}
\begin{figure*}
\begin{center}
        \begin{tabular}{cc}
\includegraphics[width=0.48\textwidth]{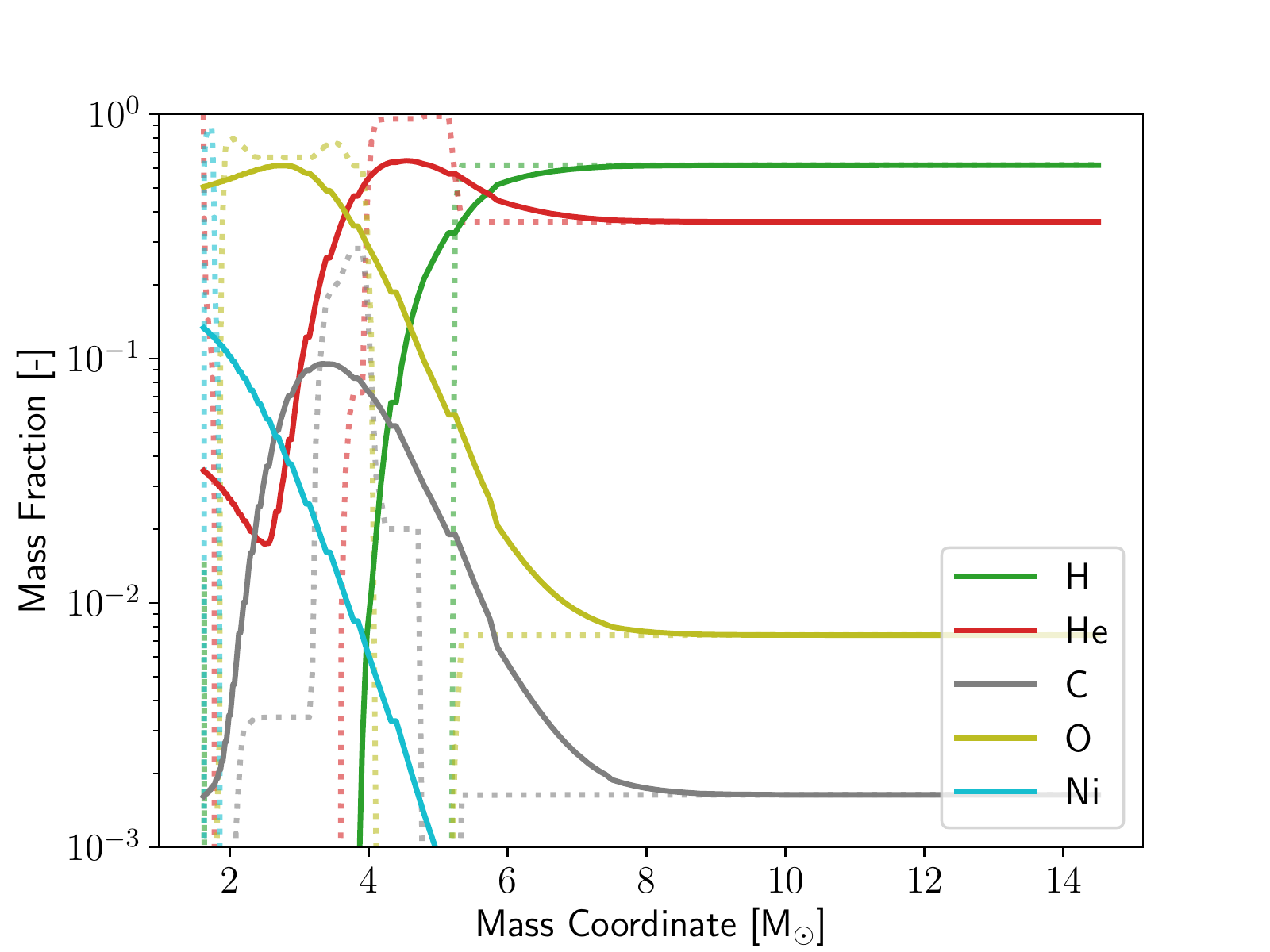}
\includegraphics[width=0.48\textwidth]{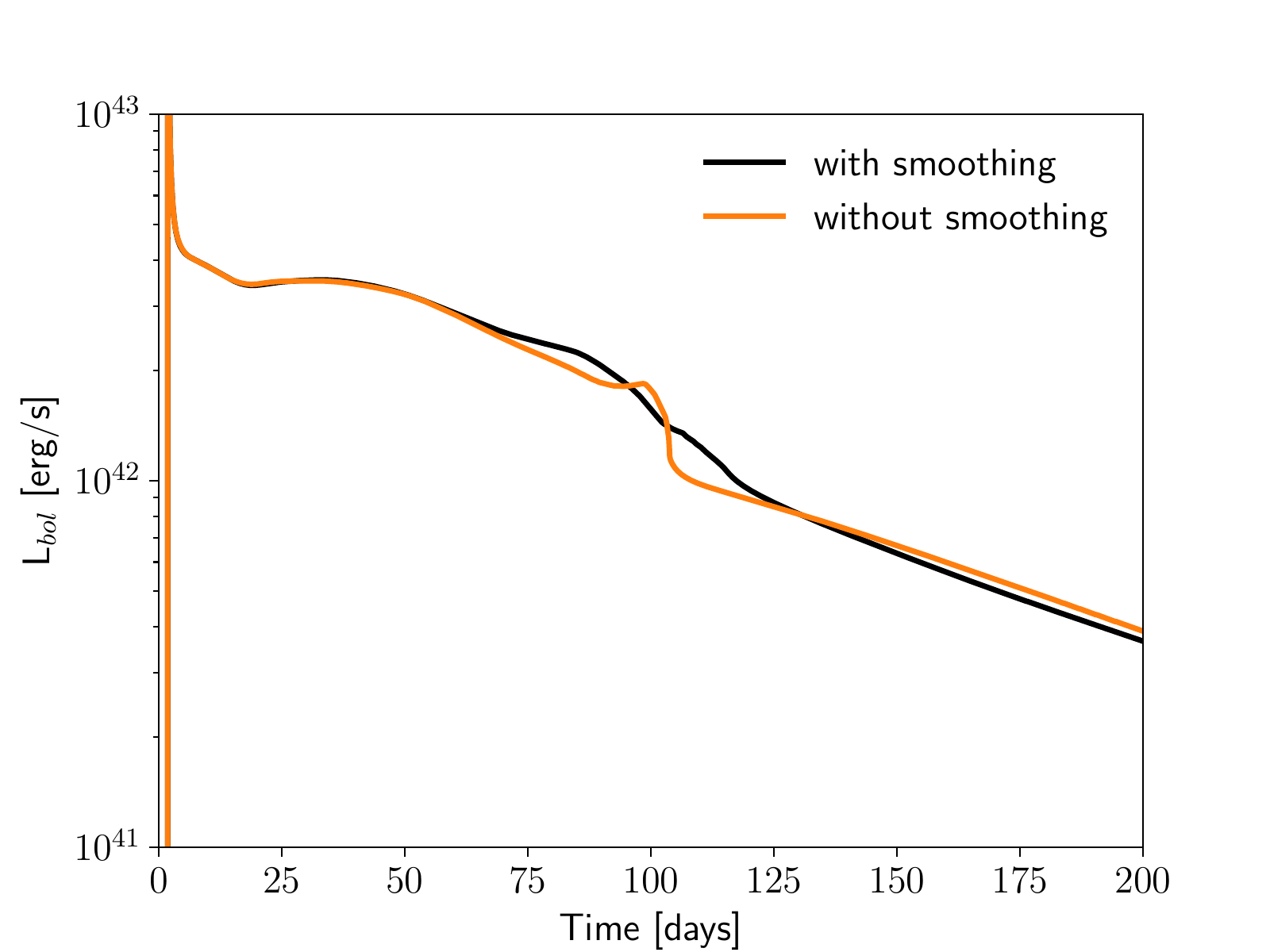}
        \end{tabular}
\caption{The unmixed and mixed compositions (left) and corresponding light curves (right) for model s18.0. The dotted lines in the left panel correspond to the unmixed post-explosion composition while the solid lines correspond to the composition obtained after boxcar smoothing in \texttt{SNEC}.
\label{fig:s18_mixing_lcs}
}
\end{center}
\end{figure*}

In Figure~\ref{fig:s18_mixing_lcs}, we show how artificial mixing affects the \texttt{SNEC} light curves for model s18.0. The panel on the left shows the composition of this model before and after the application of boxcar averaging in \texttt{SNEC}. The boxcar width is set to 0.4$M_{\odot}$ and it is run four times through the profile. 
We find that the predicted light curve, shown in the right panel of Figure~\ref{fig:s18_mixing_lcs}, changes in the same way as reported by \cite{Morozova2015}: the abrupt drop at the end of the plateau is replaced by a more gradual decrease in bolometric luminosity and a small knee appears at $\sim$100 days.

In the unmixed case, the composition interfaces are steep and discontinuous. Once the photosphere reaches the helium layer, helium recombination occurs rapidly and the photosphere sweeps through the layer very quickly. This leads to an abrupt decrease in the photospheric radius which translates into a drop in the bolometric luminosity. In the mixed case, the helium layer contains some amount of hydrogen, oxygen as well as $^{56}$Ni. Note that the $^{56}$Ni is mixed out to $\sim$5$M_{\odot}$ as opposed to being centrally distributed below $\sim$2$M_{\odot}$ prior to smoothing. The photosphere moves through the helium layer at a relatively slower rate, also uncovering energy input from radioactive heating as it moves inward in mass coordinate. This additional luminosity is responsible for the small knee-like feature in the light curve as it gradually transitions from the plateau to the radioactive tail.

\section{SNEC Convergence tests}
\label{app:snec_convergence}

\begin{figure*}
\begin{center}
        \begin{tabular}{cc}
        \includegraphics[width=0.48\textwidth]{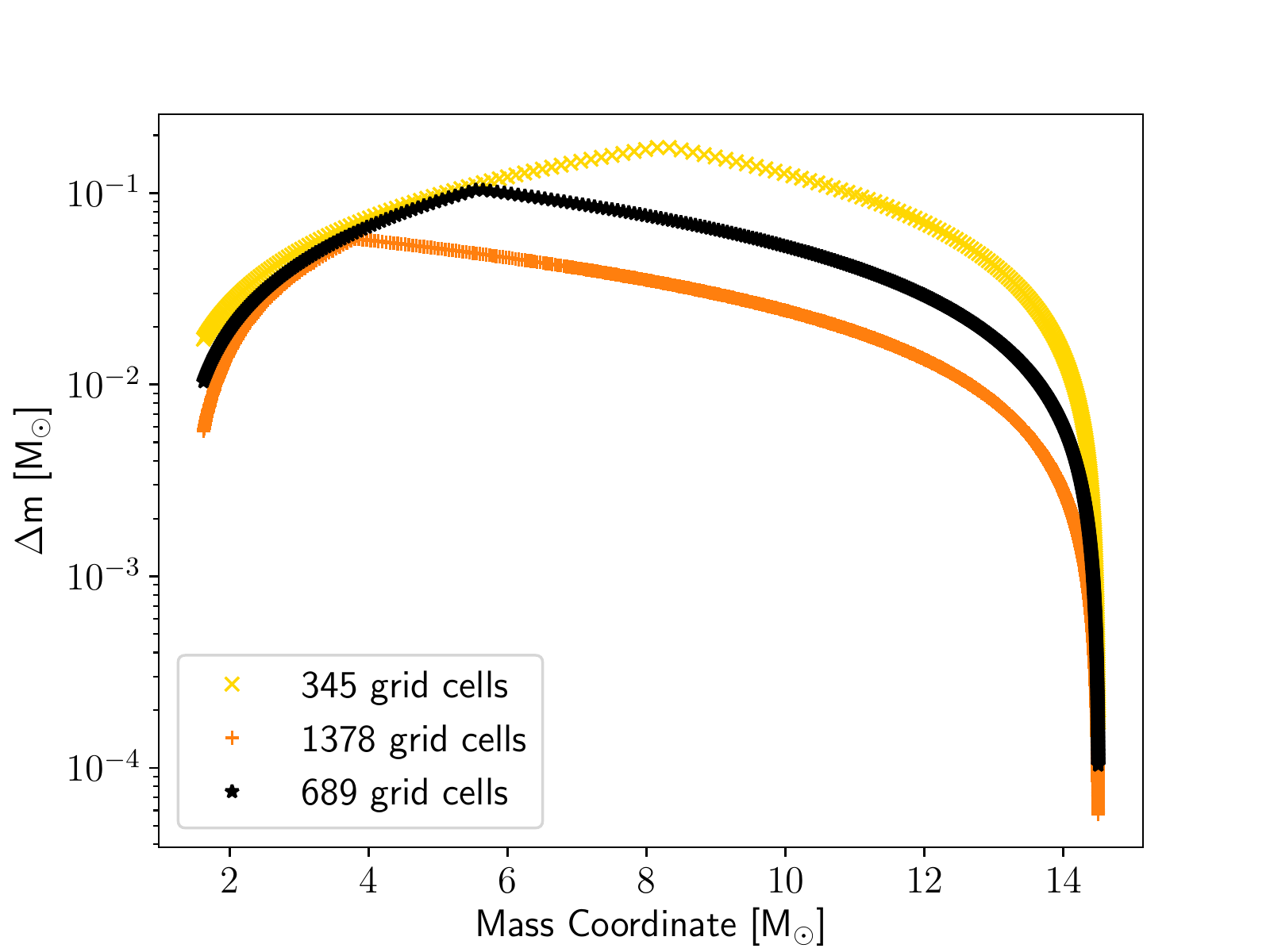}
        \includegraphics[width=0.48\textwidth]{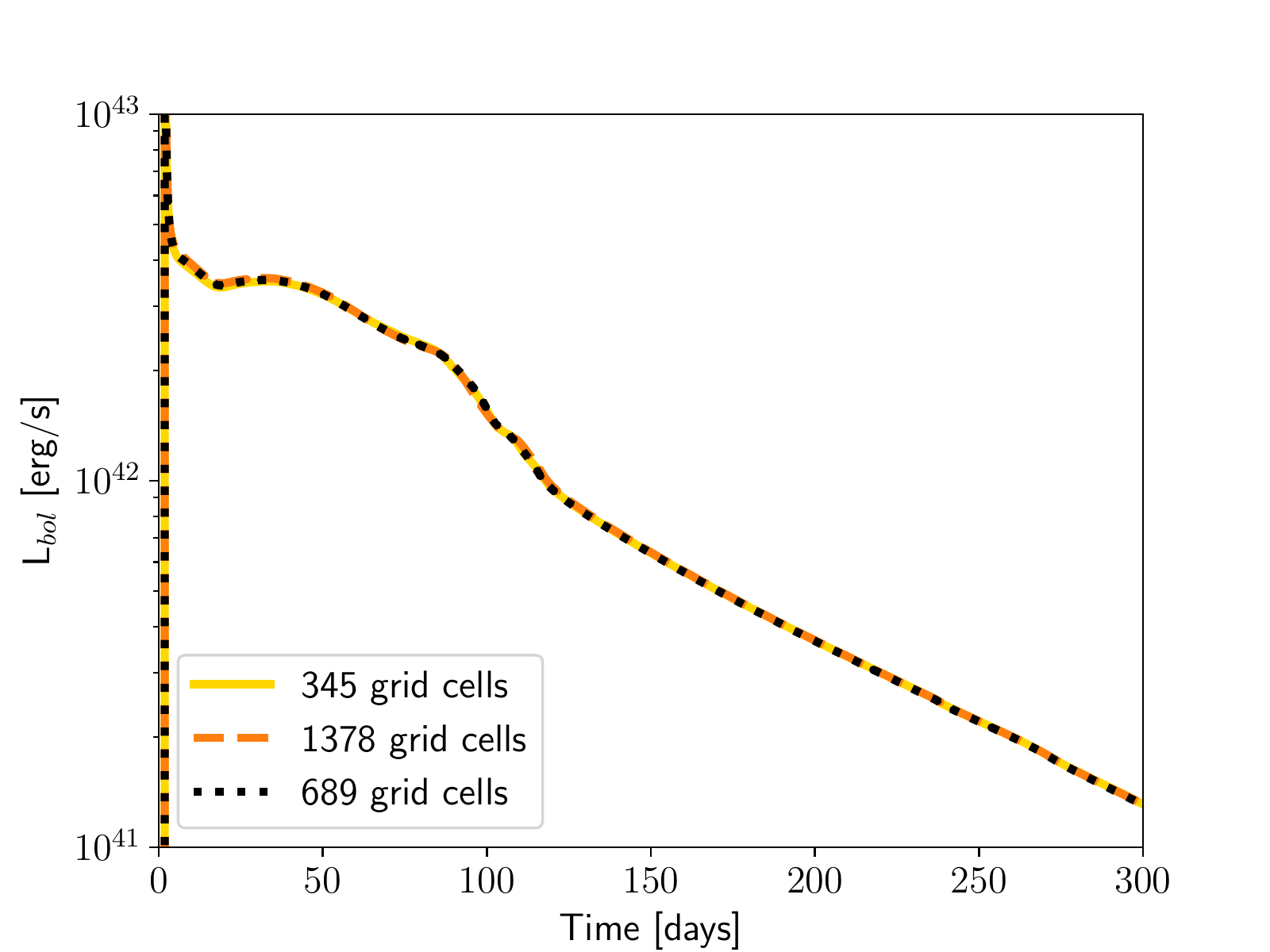}
        \end{tabular}
        \caption{Convergence tests for model s18.0 using different number of grid cells in \texttt{SNEC}, which corresponds to differences in the mass-resolution as shown in the panel on the left. The corresponding light curves are converged and shown on the right.
        \label{fig:snec_converge}
        }
\end{center}
\end{figure*}

\texttt{SNEC} maps the input model onto a new grid that focuses resolution on the inner layers and the stellar surface. Given a desired number of grid cells, the corresponding mass grid is generated by coarsening the resolution (according to geometric progression) between the innermost cell and the coarsest cell, and refining it in the same manner between the coarsest cell and the surface cell. The grid pattern this produces is shown for model s18.0, using 345, 689 and 1378 total cells, in the left panel of Figure~\ref{fig:snec_converge}. The corresponding light curves are shown in the panel on the right and lie on top of each other, indicating that our results are numerically converged. 

The light curve presented in this paper for model s18.0 corresponds to the 689 grid cell run. This results in a mass resolution of the order of $\sim$10$^{-2} M_{\odot}$ for the innermost cell, $\sim$10$^{-1} M_{\odot}$ for the coarsest cell (at a mass coordinate of $\sim$6$M_{\odot}$) and $\sim$10$^{-4} M_{\odot}$ at the surface. For every model, we set the default number of grid cells in \texttt{SNEC} equal to the total number of cells obtained by combining the \texttt{Agile} simulations with the excised portion of the progenitor model. This typically corresponds to 600--800 cells but can be as low as 225 for some of the most stripped models. However, for all models, we find the resulting grid to have a similar or better mass resolution as shown above for model s18.0. We also ran our models at half of our chosen resolution and found the light curves to be unchanged. 

\section{SNEC Opacities}
\label{app:opacities}

The opacity tables included with \texttt{SNEC} are a combination of OPAL Type II opacity tables and tables from \cite{Ferguson2005}\footnote{\url{https://www.wichita.edu/academics/fairmount_college_of_liberal_arts_and_sciences/physics/Research/opacity.php}}, valid for solar compositions. To simulate the sub-solar and zero metallicity models, we need to supply similar tables valid at the corresponding metallicities. OPAL tables at different metallicities can be generated upon request \footnote{\url{https://opalopacity.llnl.gov/type2inp.html}}. However, while we could obtain both sets of tables for zero metallicity, neither table is available for the extremely low metallicity (Z=10$^{-4}Z_{\odot}$) of the u-series models. 

We combine the two sets of zero-metallicity opacity tables to create five separate input tables for \texttt{SNEC}, each corresponding to a certain mass fraction of hydrogen (X = 0, 0.03, 0.1, 0.35, 0.7). Preference is given to \cite{Ferguson2005} values wherever the tables overlap ($ 10^{3.75} \rm{K} < T < 10^{4.5} \rm{K}$). There are regions of low temperature and density where opacities are not available. As was done in \cite{Morozova2015}, we set these opacities to the nearest available values at the same temperature. Along with the tables, the \texttt{envelope\_metallicity} parameter must be changed in the \texttt{SNEC} source code.

\begin{figure}
\begin{centering}
\includegraphics[width=0.48\textwidth]{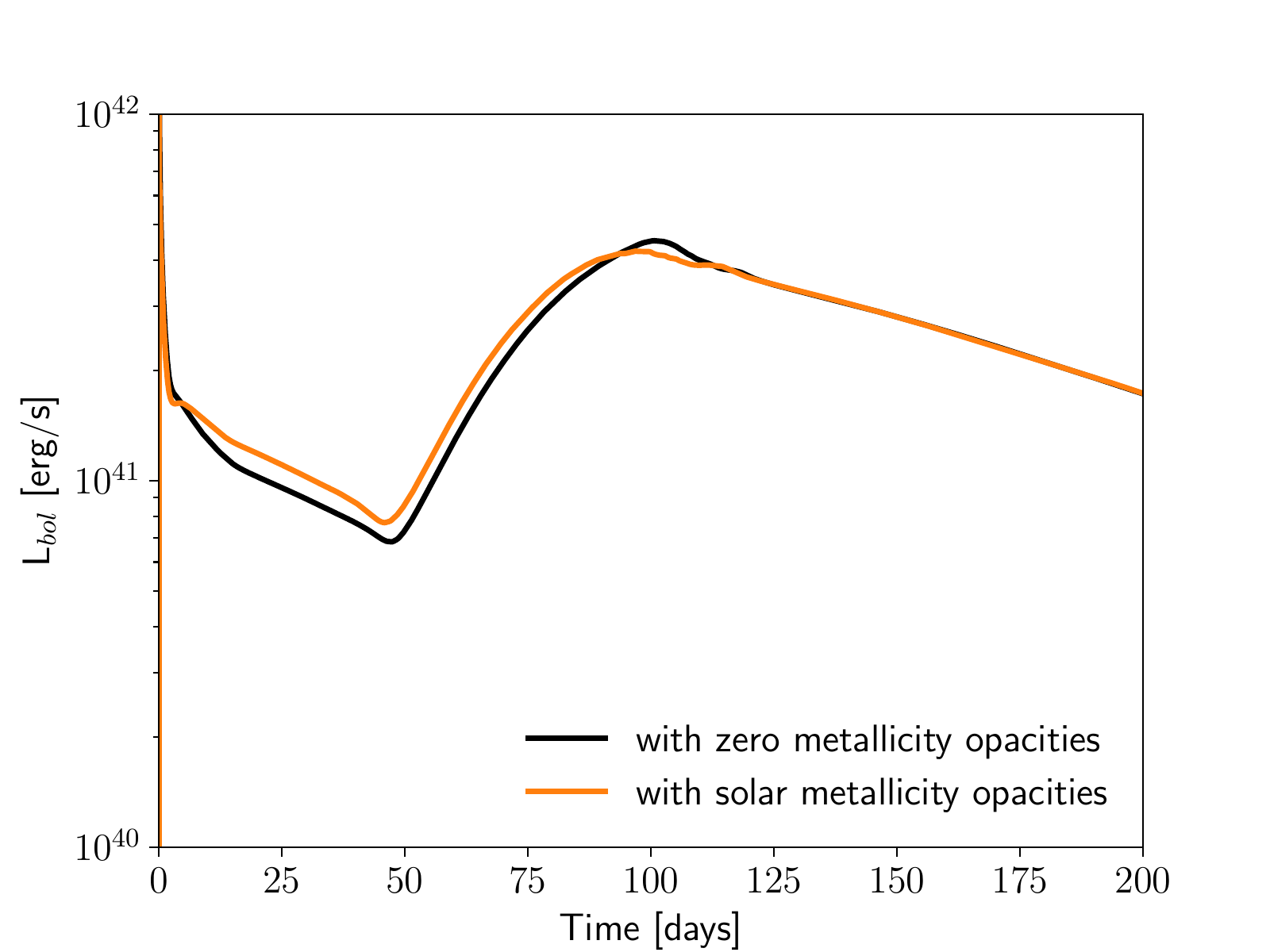}
\caption{The \text{SNEC} bolometric light curve for model u18.0 from simulations employing solar and zero metallicity opacities.
\label{fig:u02_lcs_opactest}
}
\end{centering}
\end{figure}

Since the metallicity of the u-series models is almost zero, we use this new set of zero-metallicity tables to simulate both the u-series and the z-series models. To estimate the extent to which this approximation affects our light curve predictions, we simulate model u18.0 using both the solar and zero metallicity tables. The results are shown in Figure~\ref{fig:u02_lcs_opactest}. We find that the light curves show minor differences and the peak is shifted to slightly earlier times in the solar metallicity case. However, note that this represents a bound on the maximum uncertainty introduced by opacities between $Z=0$ and $Z = Z_{\odot}=0.02$. We therefore expect the light curves to be minimally affected by the our use of $Z=0$ opacities in place of $Z=10^{-4}Z_{\odot}$ opacities for the u-series models. 

\section{Peculiar Light Curves}
\label{app:weird}

\begin{figure}
\begin{centering}
\includegraphics[width=0.48\textwidth]{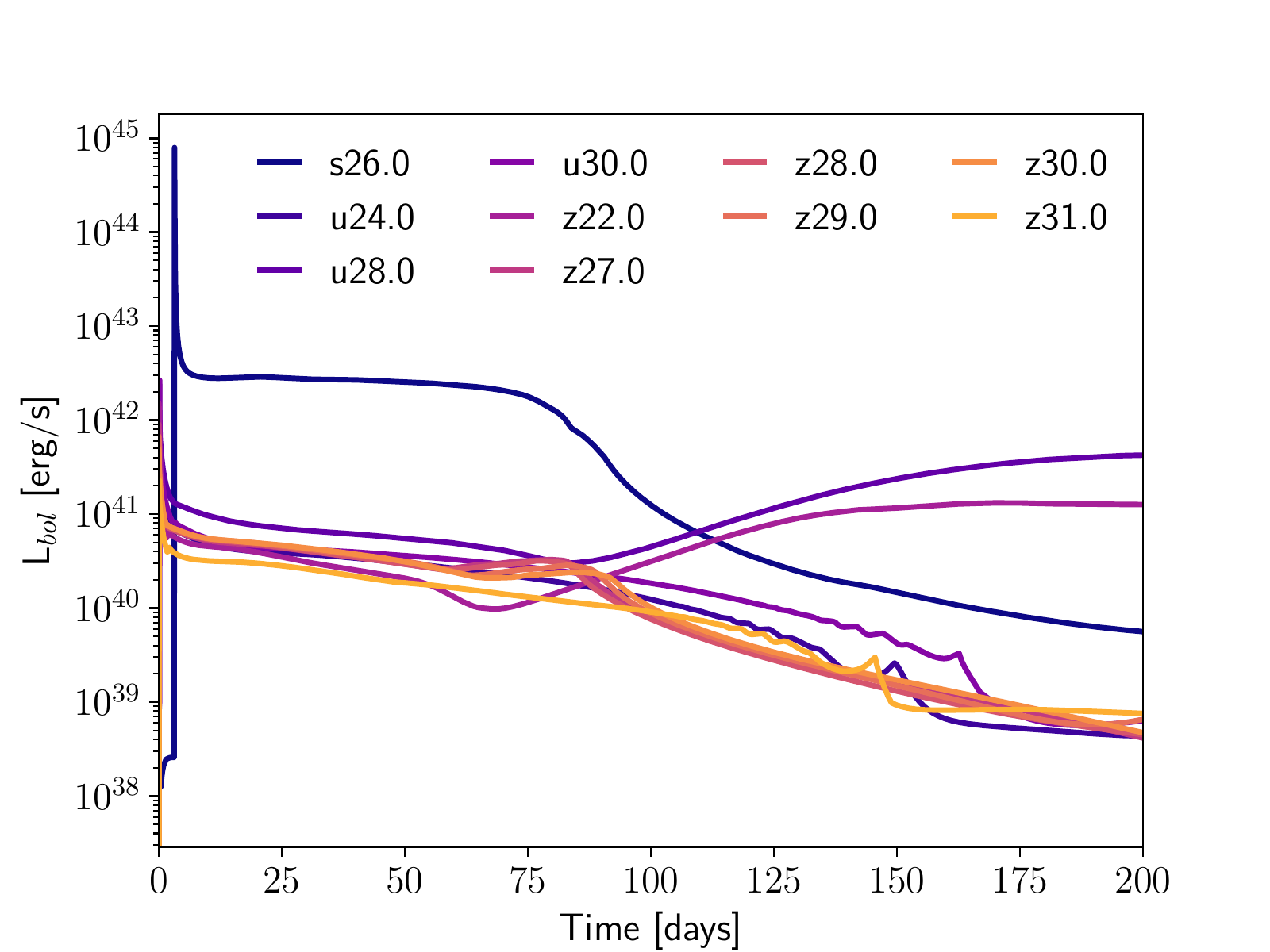}
\caption{The \texttt{SNEC} bolometric light curves for ten of our input models. These light curves are presented separately since they look unusual and different from other models in the same series. 
\label{fig:appndx_lcs}
}
\end{centering}
\end{figure}

We show the peculiar light curves obtained for ten of our models in Figure~\ref{fig:appndx_lcs}. For these models, the explosion energy estimated by \texttt{SNEC} differs significantly from that predicted by \texttt{Agile}. This could be due to uncertainties introduced by mapping between the two codes, or due to the fact that the explosion energy is still increasing at the end of the \texttt{Agile} simulations for these models, when the shock leaves our grid.

As a check, we extended the end time of the \texttt{Agile} simulations for some of these models and found that the two estimates of explosion energy begin to converge. This improves the resulting light curves but they continue to look unusual. We also experimented with allowing \texttt{SNEC} to deposit $\sim$0.2 Bethe of thermal bomb energy, for example, for model s26.0. With the extra energy input, the light curve of this model changes to fit well with those shown in the middle panel of Figure~\ref{fig:s02_lcs}. This suggests that mapping from \texttt{Agile} simulations at a later time when the explosion energy has saturated may produce normal light curves for these models. However, this would require including more mass in the hydrodynamical simulations to follow the shock evolution for a longer time, and is beyond the scope of this work. 

\section{Active Subspace Analysis}
\label{app:uq}

Active subspace analysis is a form of sensitivity analysis and dimension reduction that allows us to identify the most influential model parameters. The analysis identifies a set of important \textit{directions} in the parameter space that best predict a response. This method has been successfully used on computational models in other fields \citep[e.g.,][]{Constantine.as-jets, Jefferson.as-hydrology,  Vohra.as-chemistry}.

We apply a linear model as described in Algorithm 1.3 of \citet{as_textbook_constantine} to identify the single most important important direction in the parameter space. This allows us to reduce the dimension of the parameter space to one. Following the nomenclature of \citet{Jefferson.as-hydrology}, we call this direction the ``active variable weight'' $\textbf{w}$. When we project our parameters onto the active variable weight, we calculate the ``active variable'' $\textbf{w}^T \textbf{x}_i$, where the $\textbf{x}_i$ are vectors representing a position in our parameter space.

\begin{figure}
    \includegraphics[width=0.48\textwidth]{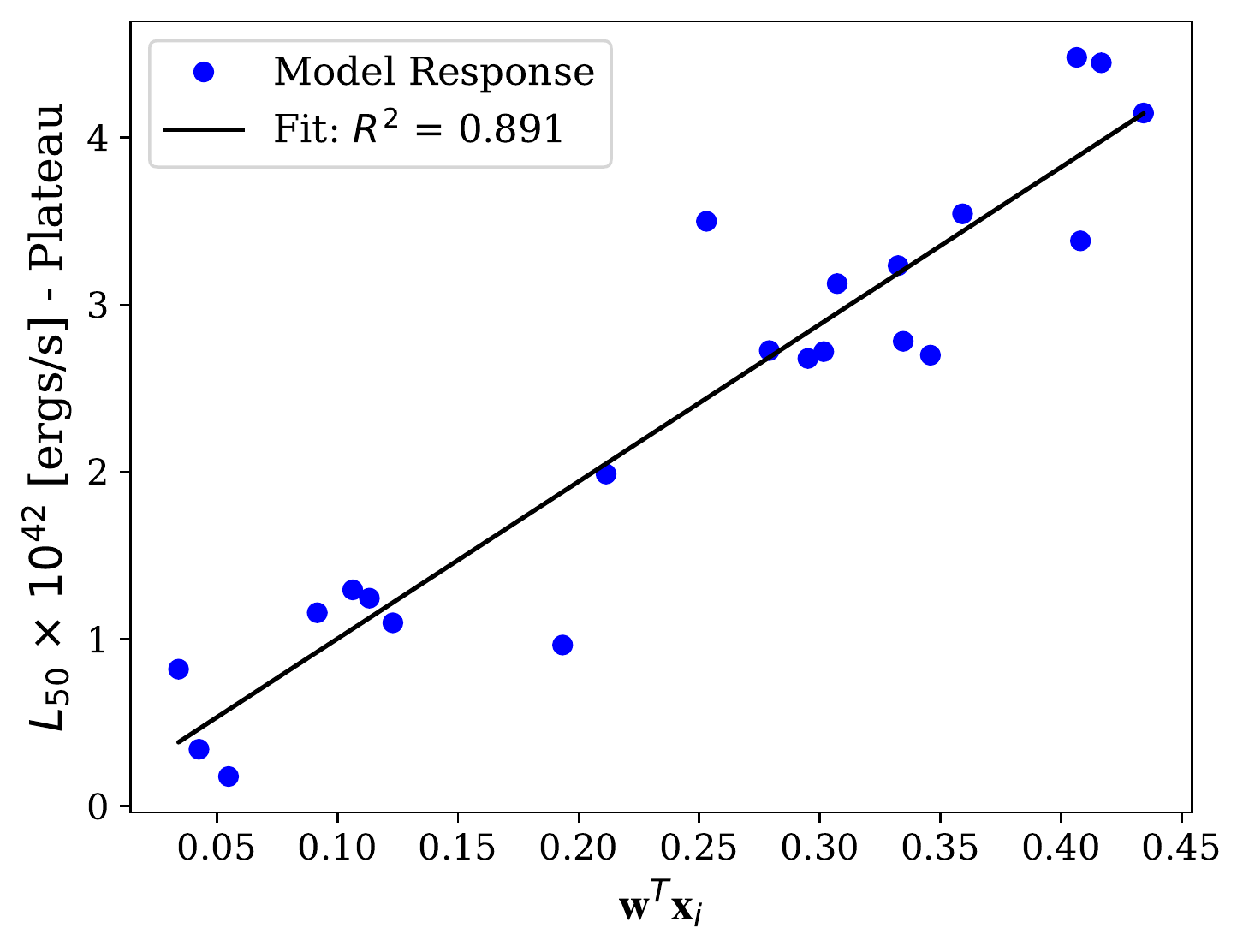}
    \includegraphics[width=0.48\textwidth]{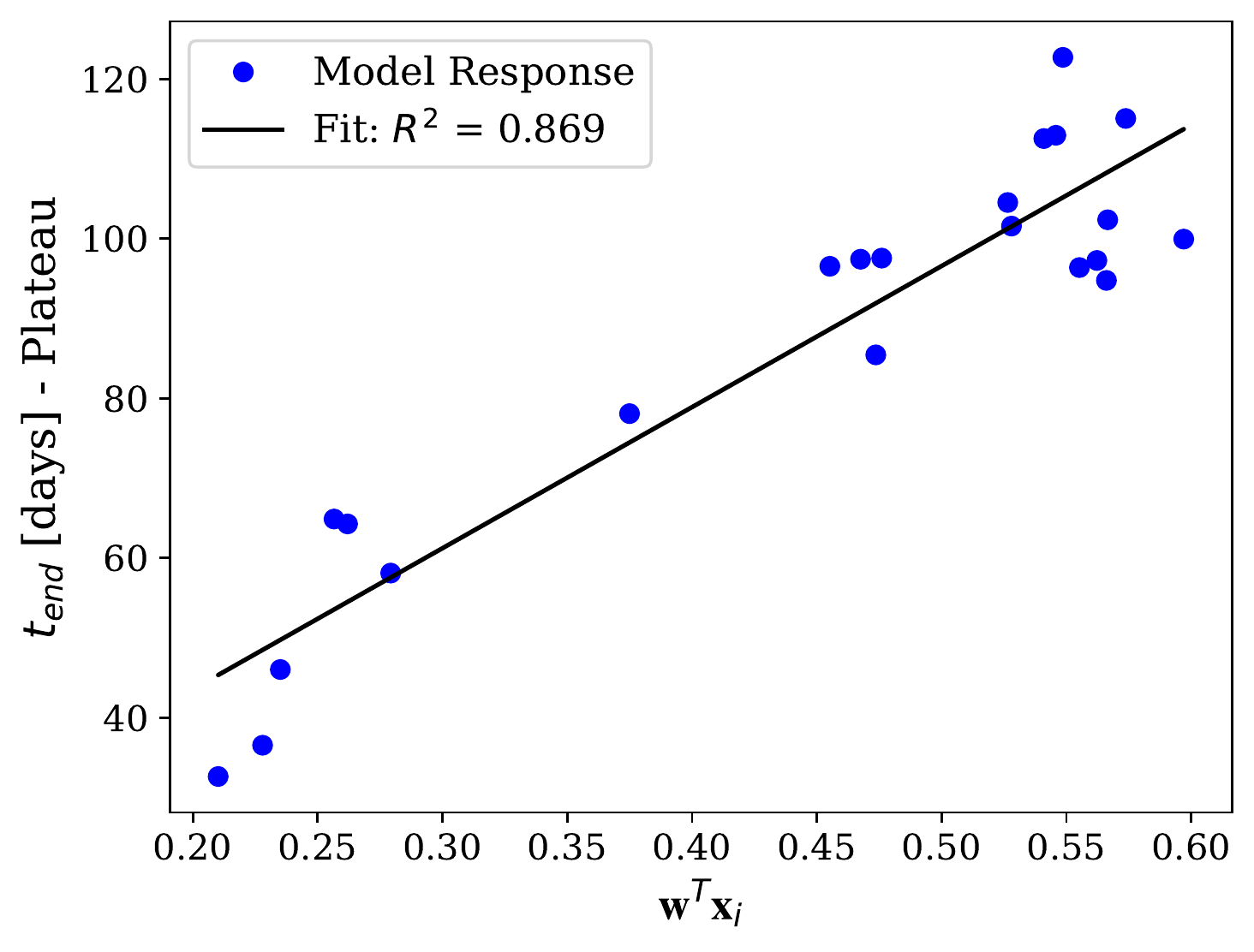} \\
    \includegraphics[width=0.48\textwidth]{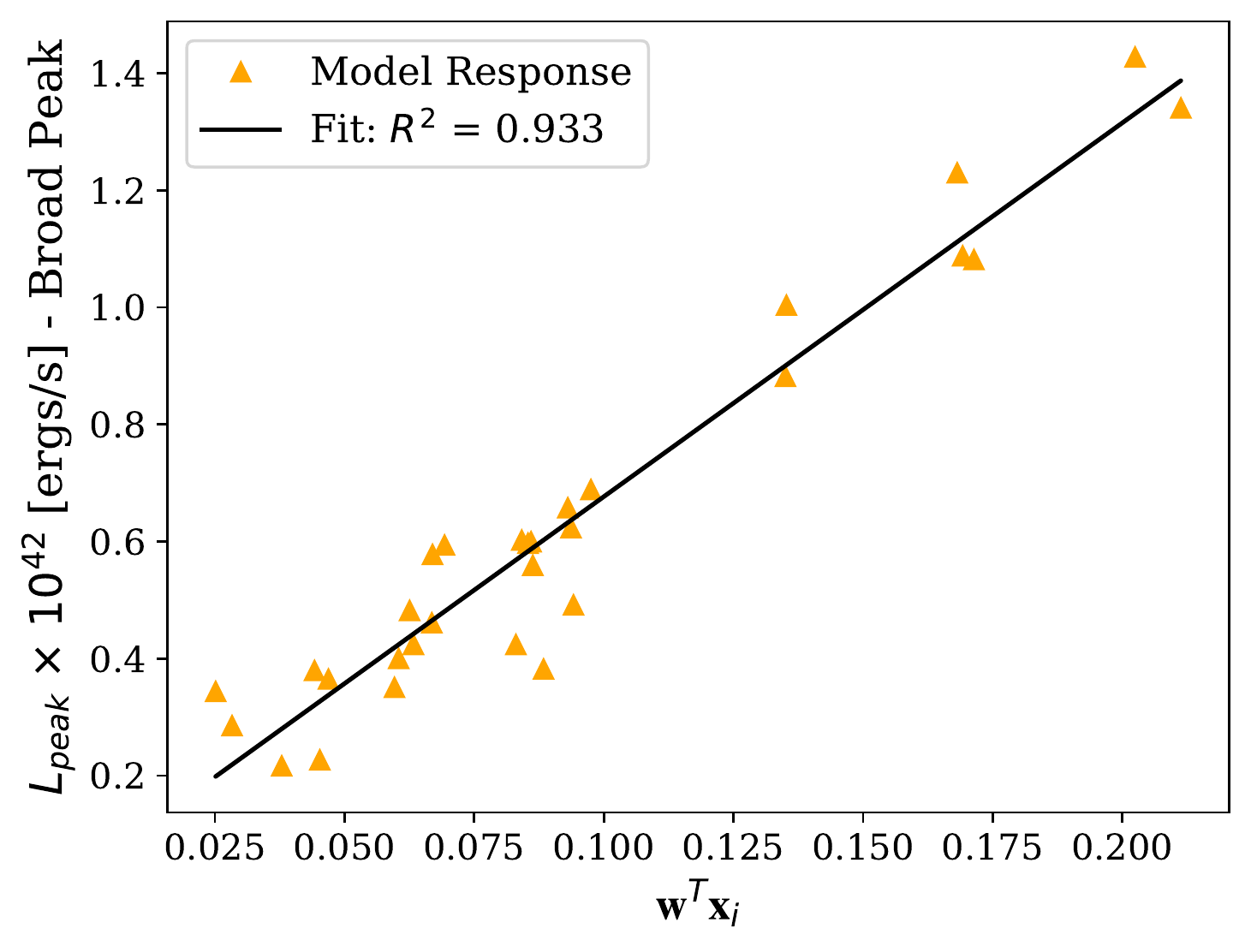}
    \includegraphics[width=0.48\textwidth]{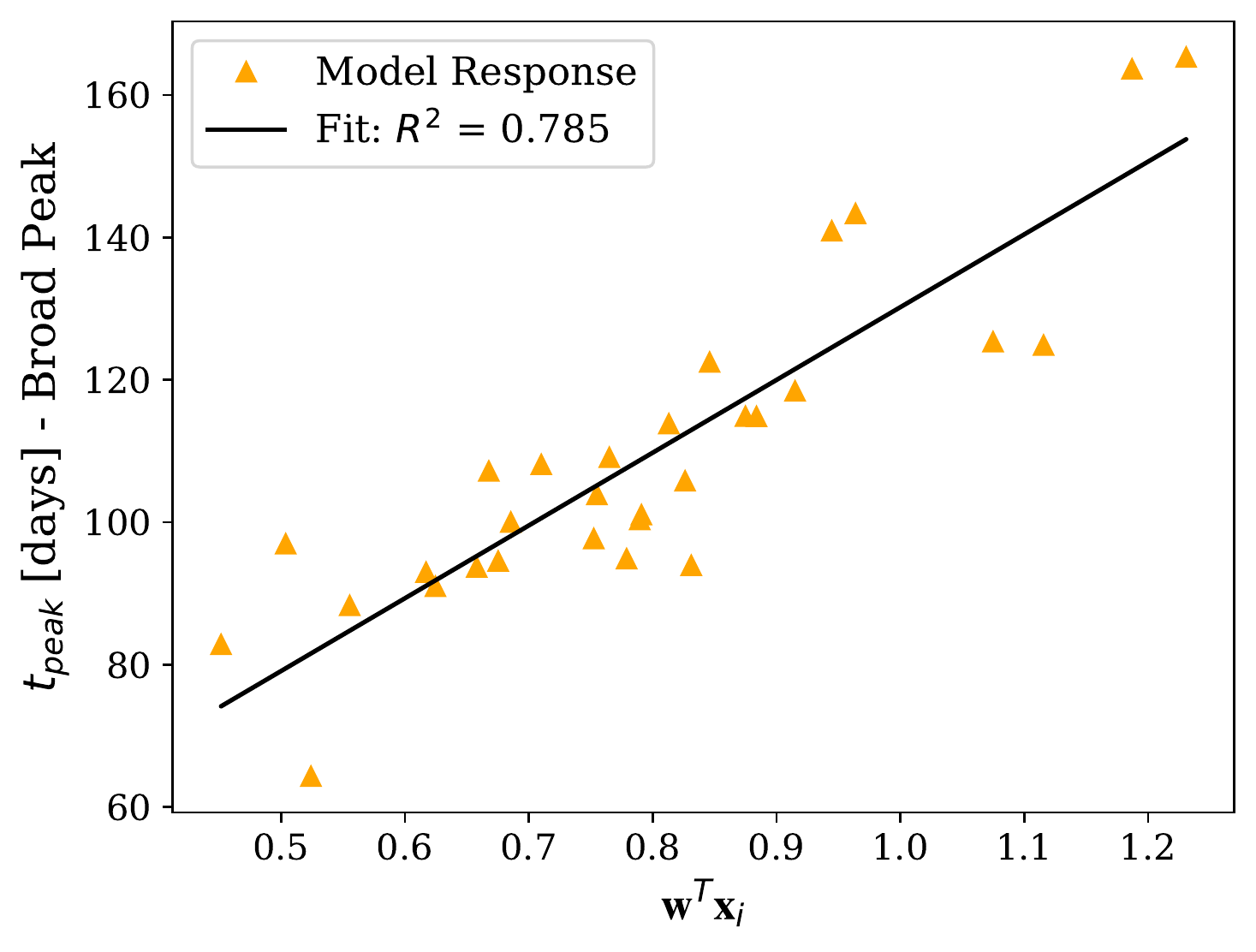}

    \caption{Sufficient summary plots for the luminosity at 50 days ($L_{50}$; top left) and plateau end time ($t_\mathrm{end}$; top right) for plateau models, and for peak luminosity ($L_\mathrm{peak}$; bottom left) and time to peak ($t_\mathrm{peak}$; bottom right) for broad peak models. The linear fit and the $R^2$ value for are also shown.
    \label{fig:as_linear}
    }
\end{figure}

We applied this active subspace analysis to two groups of light curves, those with a plateau (Type IIP) and those with a broad peak (stripped-envelope like and SN1987A-like). For each group, we have chosen two relevant response variables: Luminosity at 50 days ($L_{50}$) and plateau end time ($t_\mathrm{end}$) for plateau models, and peak luminosity ($L_\mathrm{peak}$) and time to peak ($t_\mathrm{peak}$) for broad peak models. In Figure \ref{fig:as_linear}, we find linear trends (with some scatter) between all of the response variables and their associated active variables. This indicates that a linear subspace analysis is appropriate for constructing the active subspace for all of these response variables on their respective model subsets. We report the linear fits in Table \ref{tab:as_slope_intercepts}.

\begin{table}
    \begin{center}
    \caption{Slope and intercepts of the linear fits in the linear subspace analysis.
    \label{tab:as_slope_intercepts}
        }
    \begin{tabular}{lccccc}
        \tableline \tableline
        Response & Slope & Intercept \\
        \tableline
        $L_{50}$ & $9.406 \times 10^{42}$ & $6.089 \times 10^{40}$ \\
        $t_{\mathrm{end}}$ & 176.767 & 8.185 \\
        \tableline
        $L_\mathrm{peak}$ & $6.384 \times 10^{42}$ & $3.828 \times 10^{40}$ \\
        $t_\mathrm{peak}$ & 102.210 & 27.987 \\
        \tableline
    \end{tabular}
    \end{center}
    \tablecomments{The first part of the table refers to the light curve sample with plateaus, and the second part refers to the sample of broad peak light curves.}
\end{table}

\section{SuperNu Convergence tests}
\label{app:supernu_convergence}

A number of choices have to be made when mapping from \texttt{SNEC} to \texttt{SuperNu}. The time of mapping affects the resulting light curve, in addition to the number of energy groups, the time step resolution and the number of MC particles generated per time step. The earlier the mapping time, the higher is the luminosity of the radioactive tail calculated by \texttt{SuperNu} during the nebular phase. This is not unexpected since \texttt{SNEC} and \texttt{SuperNu} treat radioactive heating and radiative transfer very differently. However, there is a physically-motivated way of choosing the mapping time i.e. mapping at the earliest time possible while making sure the outflow is homologous, as required by \texttt{SuperNu}. 

\begin{figure*}
\begin{center}
        \begin{tabular}{cc}
        \includegraphics[width=0.48\textwidth]{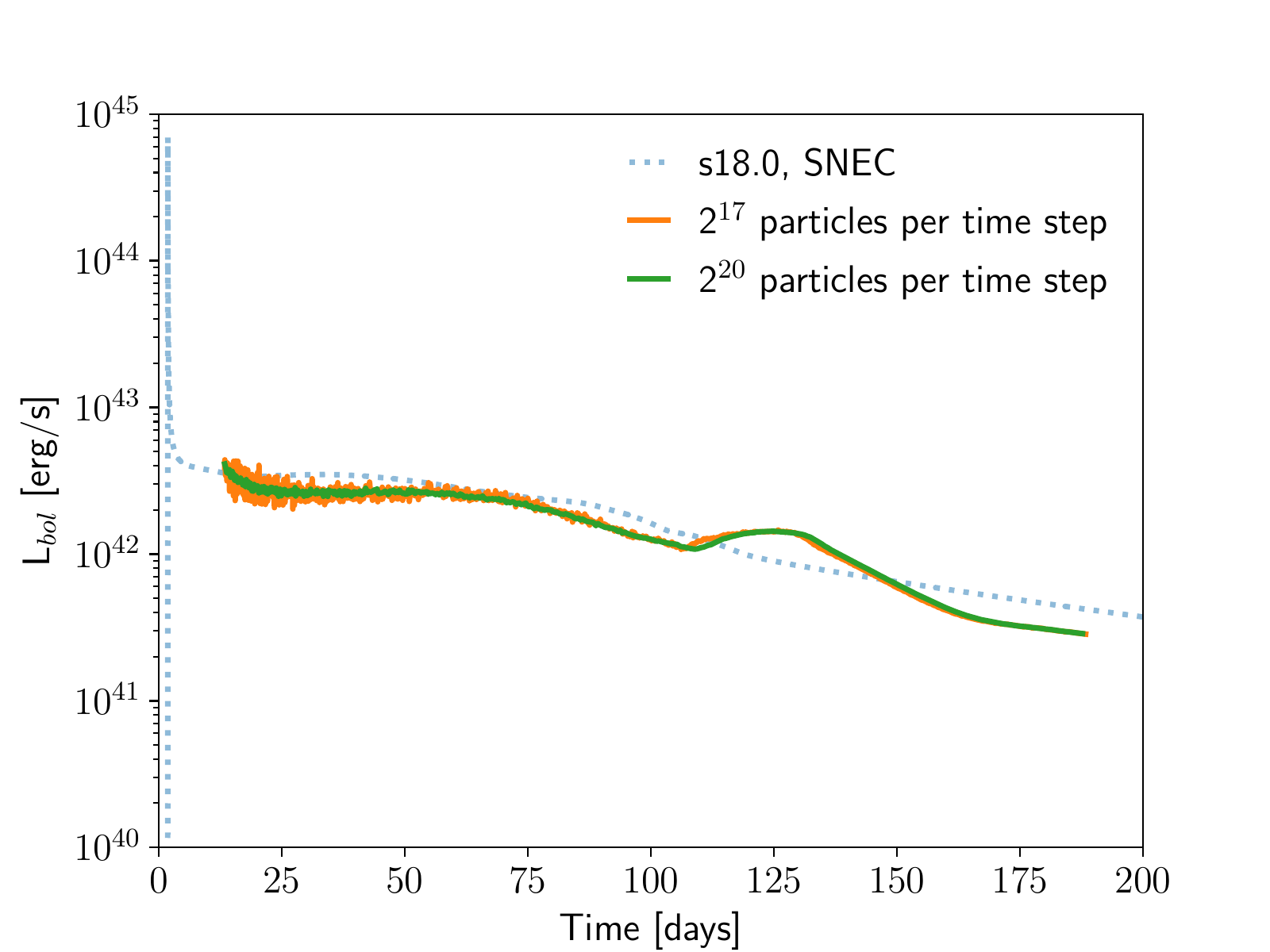}
        \includegraphics[width=0.48\textwidth]{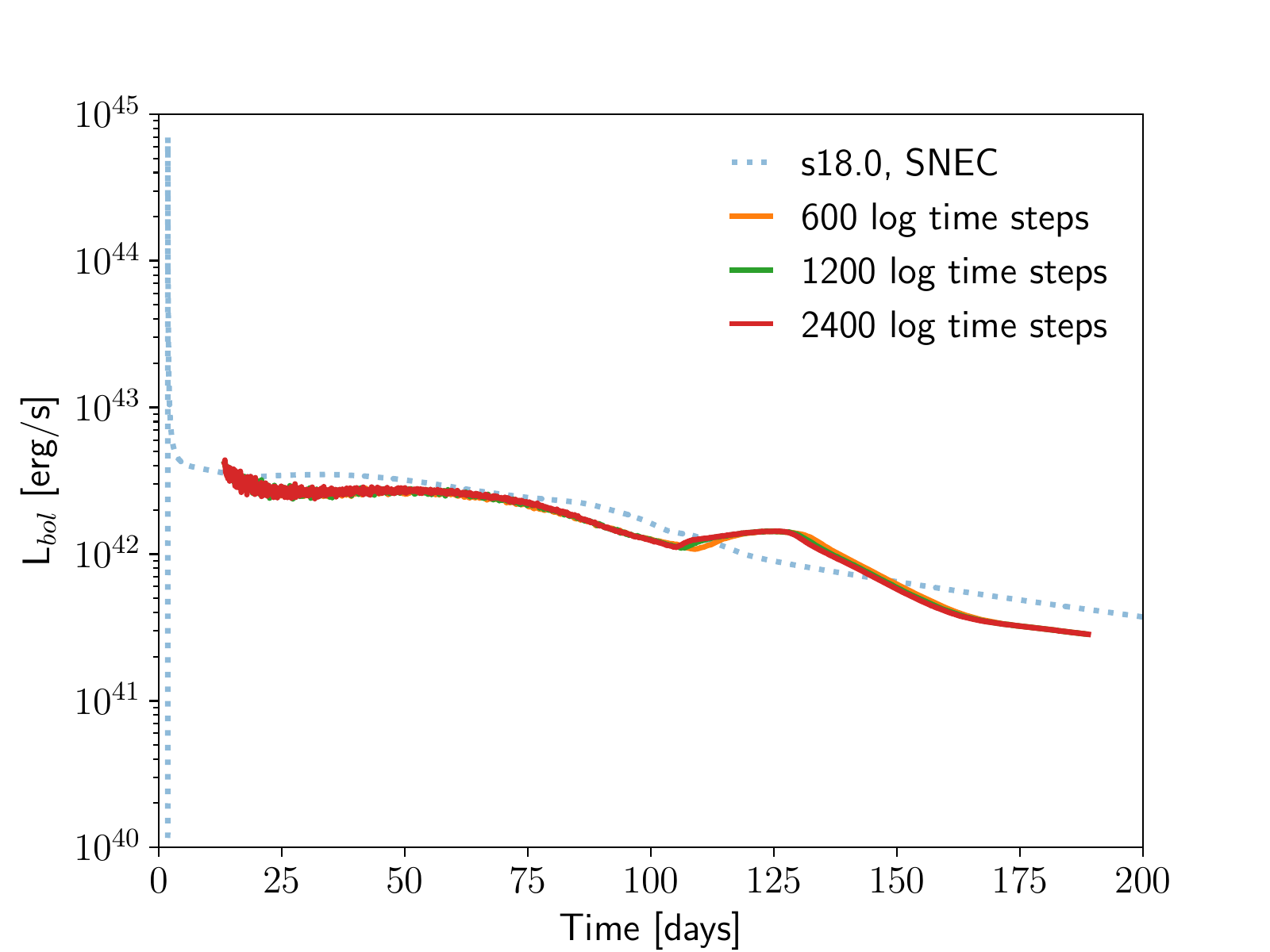}
        \end{tabular}
        \caption{SuperNu Convergence tests for model s18.0 for MC particle count (left) and timestep resolution (right).
        \label{fig:supernu_converge}
        }
\end{center}
\end{figure*}

We also varied the number of groups, time steps, and particles per time step in \texttt{SuperNu} until our light curves and spectra were converged. The number of time steps and MC particle count are not fully independent parameters with respect to increasing the fidelity of our simulations. Smaller time steps require a corresponding increase in the MC particle count per time step to counteract the increased MC noise. In Figure~\ref{fig:supernu_converge}, we show how varying each parameter independently affects the light curve computed by \texttt{SuperNu} for model s18.0. The number of MC particles is increased by a factor of eight on the left, and the total number of time steps is doubled on the right (corresponds to halving the time step size). As seen in the Figure, increasing the number of MC particles reduces the level of MC noise, especially in the early time bolometric light curve, however, the luminosity of the plateau and the time of transition to the radioactive tail do not change noticeably. Similarly, on the right, starting with 600 total time steps (in log space) and doubling them does not change the level of the plateau and has only a minor effect on the location of the bump feature at $\sim$125 days. We choose 1200 total time steps and $2^{20}$ particles per time step as our fiducial resolution. Models with smaller radii do not require the same degree of resolution and we halve the number of total time steps wherever appropriate for the sake of computational efficiency. 

\section{Sample Data Tables}
\label{app:data}

\input{sample_table_lum}
\input{sample_table_spectra}

\listofchanges

\end{document}

%% file: summary_table_plateau.tex
\begin{table*}
        \begin{center}
                \caption{Progenitor, explosion, and light curve properties for plateau-type light curves
                \label{tab:summary_plateau}
        }
                \begin{tabular}{lcccccccc}
                        \tableline \tableline
                        Model & $M_{\rm{pre-SN}}$ &  $M_{\rm{NS}}$ & $R_{\rm{pre-SN}}$ & $E_{\rm{expl}}$ & $M_{\rm{Ni}}$ & $v_{\mathrm{Ni}}$ & $L_{50}$ & $t_{\rm{plateau}}$\\
                        (-) & ($M_{\odot}$) & ($M_{\odot}$) & ($R_{\odot}$) & (10$^{51}$ erg) & ($M_{\odot}$) & (10$^3$ km s$^{-1}$) & (erg s$^{-1}$) & (days) \\
                        \tableline
                        s11.0   &   10.61   &   1.41     &   587.09      &   0.57  &   2.38E-02   &  1.70  &    1.10e+42  &   104.52 \\
                        s12.0   &   10.93   &   1.47     &   636.63      &   0.62  &   3.69E-02   &  1.78  &    1.24e+42  &   112.53\\
                         s13.0   &   11.36   &   1.49     &   709.74      &   1.00  &   4.38E-02   &  2.10  &    1.99e+42  &    97.55\\
                        s14.0   &   11.97   &   1.66     &   789.44      &   1.33  &   6.85E-02   &  2.40  &    2.73e+42  &    96.53\\
                        s15.0   &   12.64   &   1.71     &   842.98      &   1.53  &   9.22E-02   &  2.33  &    3.13e+42  &    97.42\\
                        s16.0   &   13.25   &   1.54     &   912.92      &   1.24  &   7.54E-02   &  1.74  &    2.68e+42  &   101.57\\
                        s17.0   &   13.84   &   1.54     &   957.82      &   1.23  &   7.93E-02   &  1.43  &    2.72e+42  &    96.37\\
                        s18.0   &   14.50   &   1.62     &   1010.52     &   1.46  &   1.12E-01   &  1.50  &    3.23e+42  &   102.36  \\
                        s18.8   &   15.05   &   1.55     &   1042.64     &   1.21  &   6.90E-02   &  1.20  &    2.78e+42  &    99.94\\
                        s19.0   &   15.04   &   1.78     &   1040.71     &   1.58  &   1.12E-01   &  1.41  &    3.54e+42  &    97.26\\
                        s20.0   &   14.74   &   1.51     &   1122.93     &   1.03  &   6.24E-02   &  1.31  &    2.70e+42  &    94.76\\
                        s21.0   &   13.00   &   1.82     &   1246.03     &   1.47  &   1.35E-01   &  2.11  &    4.15e+42  &    78.08\\
                        s22.0   &   14.42   &   1.57     &   1255.79     &   1.15  &   6.91E-02   &  1.59  &    3.38e+42  &    85.44\\
                        s27.0   &   12.45   &   1.67     &   1481.38     &   0.99  &   8.43E-02   &  1.86  &    4.45e+42  &    64.87\\
                        s28.0   &   12.67   &   1.66     &   1515.46     &   0.98  &   8.71E-02   &  1.84  &    4.48e+42  &    64.26\\
                        s29.0   &   12.62   &   1.61     &   1330.68     &   1.06  &   9.81E-02   &  2.14  &    3.50e+42  &    58.10\\
                        s30.0   &   12.25   &   1.71     &   1223.57     &   1.27  &   1.33E-01   &  2.68  &    9.64e+41  &    46.02\\
                        u11.0   &   11.00   &   1.51     &   321.79      &   0.75  &   4.30E-02   &  2.03  &    8.19e+41  &   122.70\\
                        u12.0   &   12.00   &   1.57     &   349.42      &   1.13  &   6.10E-02   &  2.27  &    1.16e+42  &   112.94\\
                        u13.0   &   13.00   &   1.68     &   353.46      &   1.35  &   7.21E-02   &  2.25  &    1.29e+42  &   115.05\\
                        \tableline
                        \tableline
                \end{tabular}
        \end{center}
        \tablecomments{$v_{\mathrm{Ni}}$ is the velocity of $^{56}$Ni in the homologously expanding ejecta specified at the location where the mass-fraction drops below 10$^{-4}$.
        }
\end{table*}

%% file: summary_table_peak.tex
\begin{table*}
        \begin{center}
                \caption{Progenitor, explosion, and light curve properties for broad-peak light curves
                \label{tab:summary_peak}
                }
                \begin{tabular}{lcccccccc}
                        \tableline \tableline
                        Model & $M_{\rm{pre-SN}}$ &  $M_{\rm{NS}}$ & $R_{\rm{pre-SN}}$ & $E_{\rm{expl}}$ & $M_{\rm{Ni}}$ & $v_{\mathrm{Ni}}$ & $L_{\rm{peak}}$ & $t_{\rm{peak}}$\\
                        (-) & ($M_{\odot}$) & ($M_{\odot}$) & ($R_{\odot}$) & (10$^{51}$ erg) & ($M_{\odot}$) & (10$^3$ km s$^{-1}$) & (erg s$^{-1}$) & (days)  \\
                        \tableline
                        s31.0   &   11.72   &   1.58     &   993.70      &   0.95  &   8.24E-02   &  2.04  &        5.13e+41  &  116.56\\
                        s32.0   &   12.00   &   1.68     &   1111.92     &   1.09  &   1.00E-01   &  2.35  &       6.06e+41  &  111.70\\
                        s33.0   &   11.44   &   1.74     &   3.08        &   1.08  &   1.07E-01   &  2.20  &      6.23e+41  &  113.79\\
                        s34.0   &   11.78   &   1.80     &   1.15        &   1.01  &   1.15E-01   &  1.97  &      6.57e+41  &  122.45\\
                        s35.0   &   10.64   &   1.69     &   1.30        &   1.04  &   9.67E-02   &  2.43  &     5.96e+41  &  109.07\\
                          s36.0   &   10.31   &   1.71     &   2.04        &   1.07  &   1.04E-01   &  2.65    &  6.88e+41  &  103.83\\
                        s37.0   &    9.72   &   1.83     &   1.27        &   1.29  &   1.58E-01   &  3.23   &  1.09e+42  &   94.84\\
                        s38.0   &    9.27   &   1.66     &   1.64        &   0.99  &   8.97E-02   &  2.61   &  5.99e+41  &   99.98\\
                         s40.0   &    8.75   &   1.82     &   1.11        &   0.93  &   8.79E-02   &  2.81   &  6.02e+41  &   93.64\\
                        s75.0   &    6.36   &   1.62     &   0.94        &   0.84  &   2.65E-02   &  4.85  &  2.16e+41  &   64.28\\
                        u14.0   &   14.00   &   1.67     &   102.90      &   1.52  &   1.08E-01   &  2.40  &  1.34e+42  &   90.94\\
                        u15.0   &   15.00   &   1.68     &   51.23       &   1.34  &   7.67E-02   &  1.94  &  8.82e+41  &   94.51\\
                        u16.0   &   16.00   &   1.77     &   37.67       &   1.66  &   1.36E-01   &  2.04    &  1.43e+42  &  100.33\\
                        u17.0   &   17.00   &   1.50     &   35.16       &   1.17  &   5.57E-02   &  1.37   &  5.59e+41  &   97.65\\
                        u18.0   &   18.00   &   1.56     &   33.33       &   1.02  &   5.05E-02   &  1.15  &  4.61e+41  &  101.02\\
                        u19.0   &   19.00   &   1.58     &   38.46       &   1.01  &   4.61E-02   &  0.99   &  3.50e+41  &   93.90\\
                        u20.0   &   20.00   &   2.03     &   44.72       &   0.77  &   1.38E-01   &  0.83  &  1.00e+42  &  143.32\\
                        u25.0   &   25.00   &   1.85     &   44.67       &   0.86  &   6.53E-02   &  0.65   &  3.65e+41  &  124.83\\
                        u26.0   &   26.00   &   1.90     &   42.58       &   0.75  &   9.35E-02   &  0.44   &  4.24e+41  &  163.63\\
                        u27.0   &   27.00   &   1.89     &   43.97       &   0.75  &   9.40E-02   &  0.41  &  4.00e+41  &  165.28\\
                        z11.0   &   11.00   &   1.49     &   19.81       &   0.52  &   2.02E-02   &  1.71  &  2.85e+41  &   82.80\\
                        z12.0   &   12.00   &   1.41     &   12.60       &   0.49  &   2.64E-02   &  1.51 &  3.44e+41  &   96.93\\
                        z13.0   &   13.00   &   1.54     &   10.93       &   0.80  &   3.12E-02   &  1.78  &  3.79e+41  &   88.27\\
                        z14.0   &   14.00   &   1.59     &   10.58       &   1.02  &   4.70E-02   &  1.87  &  5.93e+41  &   92.93\\
                        z15.0   &   15.00   &   1.49     &   15.59       &   0.88  &   5.53E-02   &  1.54   &  5.77e+41  &  107.14\\
                        z16.0   &   16.00   &   1.50     &   11.17       &   0.92  &   5.41E-02   &  1.44 &  4.82e+41  &  108.08\\
                        z17.0   &   17.00   &   1.76     &   10.47       &   1.60  &   1.22E-01   &  1.92  &  1.23e+42  &  105.76\\
                        z18.0   &   18.00   &   1.78     &   9.34        &   1.54  &   1.34E-01   &  1.69   &  1.08e+42  &  114.81\\
                        z19.0   &   19.00   &   1.61     &   10.57       &   1.20  &   8.17E-02   &  1.17  &  4.91e+41  &  114.83\\
                        z20.0   &   20.00   &   1.60     &   15.74       &   1.12  &   7.77E-02   &  0.89 &  4.24e+41  &  118.39\\
                        z21.0   &   21.00   &   1.50     &   11.70       &   0.88  &   6.13E-02   &  0.58  &  2.27e+41  &  140.88\\
                        z23.0   &   23.00   &   1.70     &   12.13       &   1.15  &   9.69E-02   &  0.55 &  3.82e+41  &  125.32\\
                        \tableline
                        \tableline
                \end{tabular}
        \end{center}
        \tablecomments{$v_{\mathrm{Ni}}$ is the velocity of $^{56}$Ni in the homologously expanding ejecta specified at the location where the mass-fraction drops below 10$^{-4}$.
        }
\end{table*}

%% file: summary_table_obs.tex
\begin{table*}
        \begin{center}
                \caption{Models resembling SN1987A and SN1999em
                \label{tab:summary_obs}
        }
                \begin{tabular}{lccccccccc}
                        \tableline \tableline
                        Model & $M_{\rm{pre-SN}}$  & $R_{\rm{pre-SN}}$ & $M_{\rm{Ni}}$  & $M_{\rm{NS}}$  & $E_{\rm{expl}}$  & Boxcar Width & $v^1_{\mathrm{Ni}}$ & $v^2_{\mathrm{Ni}}$ & $v_{\mathrm{\tau=2/3}}$\\
                        (-) & ($M_{\odot}$) &  ($R_{\odot}$) & ($M_{\odot}$) & ($M_{\odot}$) &   (10$^{51}$ erg) & ($M_{\odot}$) & (10$^3$ km s$^{-1}$) & (10$^3$ km s$^{-1}$) & (10$^3$ km s$^{-1}$)\\
                        \tableline
                         b15-7 & 21.06 & 36.99 & 0.072 & 1.64 & 1.10 & 0.4 & 1.24 & 1.42 & 18.94\\
                           &   &   &  &   1.64 & 1.10 & 4.0 & 2.19 & - & 18.68\\
                           &   &   &   & 1.60 & 1.62 & 0.4 & 1.44 & 1.67 & 24.24\\
                          &   &   &   &   1.60 & 1.62 & 4.0 & 2.21 & - & 24.10\\
                        s12.0   &   10.93   &     636.63    &  0.037  &   1.47   & 0.62   & 0.4 & 1.31 & 1.78 & 8.02\\
                         &     &       &         &    &         & 1.0 & 2.19 & - & 8.03\\
                         &     &       &         &    &         & 2.0 & 9.57 & - & 8.04\\
                         &     &       &         &    &         & 3.0 & 9.53 & -  & 8.00\\
                        \tableline
                        \tableline
                \end{tabular}
        \end{center}
        \tablecomments{$v^1_{\mathrm{Ni}}$ and $v^2_{\mathrm{Ni}}$ represent the velocity of $^{56}$Ni in the homologously expanding ejecta specified at the location where the mass-fraction drops below 10$^{-3}$ and  10$^{-4}$ respectively. 
        }
\end{table*}

%% file: sample_table_lum.tex
\begin{table*}
        \begin{center}
                \caption{Sample data table of bolometric luminosities from \texttt{SuperNu}
                \label{tab:sample_data_lumin}
        }
        \begin{tabular}{ccc}
             \#1: iteration & 2: time [days] & 3: luminosity [erg/s] \\
             0  & 14.3652315 & 4.18126e+42 \\
             1  & 14.6797891 & 3.82635e+42 \\
             2  & 15.0012346 & 3.55915e+42\\
             .  & . &  . \\
             .  & . &  . \\
             .  & . &  . \\
             115 & 173.4359708 & 3.30533e+41\\
             116 & 177.2337240 & 3.19157e+41\\
             117 & 181.1146372 & 3.08052e+41\\
        \end{tabular}
        \end{center}
        \tablecomments{These data have been smoothed over time and are available as machine-readable tables included with this paper. We also provide upon request similar files containing the raw data from our simulations. Additionally, bolometric luminosities from \texttt{SNEC} simulations  are available for all models in our online database.
        }
\end{table*}

%% file: sample_table_spectra.tex
\begin{table*}
        \begin{center}
                \caption{Sample data table of spectral evolution from \texttt{SuperNu}
                \label{tab:sample_data_spectra}
        }
        \begin{tabular}{ccc}
             \# it=     4  time[d]= 20.316&&\\
             1.0000000e-05  & 1.0035000e-05 & 2.99311e-04 \\
             1.0035000e-05 &  1.0070000e-05 & 0.00000e+00 \\
             1.0070000e-05 & 1.0105000e-05 & 0.00000e+00 \\
             .  & . &  . \\
             .  & . &  . \\
             .  & . &  . \\
             3.1669000e-04 & 3.1779000e-04 & 8.35088e-06\\
             3.1779000e-04 & 3.1889000e-04 & 5.48667e-06\\
             3.1889000e-04 & 3.2000000e-04 &  1.54620e-04\\
             &&\\
             &&\\
             \# it=     5  time[d]= 22.154&&\\
             1.0000000e-05 & 1.0035000e-05 & 2.30068e-04\\
             1.0035000e-05 & 1.0070000e-05 &  0.00000e+00\\
             1.0070000e-05 & 1.0105000e-05 & 0.00000e+00\\
             .  & . &  . \\
             .  & . &  . \\
             .  & . &  . \\
             3.1669000e-04 & 3.1779000e-04 & 9.32209e-06\\
             3.1779000e-04 & 3.1889000e-04 &  7.11555e-06\\
             3.1889000e-04 & 3.2000000e-04 & 1.83080e-05\\
        \end{tabular}
        \end{center}
        \tablecomments{The spectra are binned into a specified number of wavelength bins and provided as comment-separated blocks, each block corresponding to a particular time. The first two columns of each block specify the lower and upper edges of the wavelength bin (in cm) and the third column gives the flux (in erg s$^{-1}$ cm$^{-2}$ \angstrom$^{-1}$) in that bin. These data have been smoothed over time and are available as machine-readable tables included with this paper. We also provide upon request similar files containing the raw data from our simulations.
        }
\end{table*}